\documentclass[12pt]{article}
\usepackage[utf8]{inputenc}
\usepackage{amsthm,amsmath,amssymb}
\usepackage{mathrsfs}
\usepackage[margin=.7in]{geometry}
\usepackage{pgfplots}
    \pgfplotsset{compat=newest}
\usepackage{tikz}
\usepackage{booktabs}
\usepackage{threeparttable}
\usepackage{tikz-3dplot}
\usetikzlibrary{petri,patterns,tikzmark}
\usepgflibrary{patterns}
\usepackage{pgfplotstable}
\usepackage{filecontents}
\usepackage{pgfcalendar}
\usetikzlibrary{pgfplots.dateplot}
\usepackage{setspace}
\usepackage{ragged2e}
\usepackage{graphicx}
\usepackage{subcaption,placeins}
\usepackage{xcolor}
\usepackage{comment}
\usepackage{hyperref}
\usepackage{bm}
\usepackage{enumerate}
\hypersetup{
    colorlinks=true,
    urlcolor=blue,
    citecolor=blue,
    linkcolor=blue
}
\PassOptionsToPackage{hyphens}{url}\usepackage{hyperref}
\PassOptionsToPackage{sloppy}{url}\usepackage{hyperref}
\usepackage{cleveref}
\usepackage{ctable}
\usepackage{indentfirst}
\usepackage{makecell}

\usepackage[bibstyle=numeric, citestyle=authoryear, natbib=true, doi=false, url=true, backend=biber, maxbibnames=10, maxcitenames=4, uniquelist=false, uniquename=false, sorting=nyt]{biblatex}
\renewbibmacro{in:}{}
\DeclareNameAlias{default}{family-given/given-family}

\newcommand{\Cov}{\mathbb{C}ov}
\newcommand{\DTT}{\mathrm{DTT}}
\newcommand{\QTT}{\mathrm{QTT}}
\newcommand{\E}{\mathbb{E}}
\newcommand{\F}{\mathrm{F}}
\newcommand{\Prob}{\mathbb{P}}
\newcommand{\T}{\mathcal{T}}
\newcommand{\Y}{\mathcal{Y}}
\newcommand{\indicator}[1]{\mathbbm{1}\{#1\}}
\newcommand{\indicatorbig}[1]{\mathbbm{1}\big\{#1\big\}}
\newcommand{\indicatorBig}[1]{\mathbbm{1}\Big\{#1\Big\}}

\newtheorem{assumption}{Assumption}
\newtheorem{definition}{Definition}
\newtheorem{proposition}{Proposition}
\newtheorem{lemma}{Lemma}
\newtheorem{example}{Example}
\newtheorem{result}{Result}
\newtheorem{theorem}{Theorem}
\newtheorem{corollary}{Corollary}
\newtheorem{condition}{Condition}
\newtheorem{remark}{Remark}
\newtheorem{algorithm}{Algorithm}
\newtheorem{inneruassumption}{Assumption}
\newenvironment{namedassumption}[1]
  {\renewcommand\theinneruassumption{#1}\inneruassumption}
  {\endinneruassumption}

\crefname{assumption}{Assumption}{Assumptions}
\crefname{lemma}{Lemma}{Lemmas}
\crefname{inneruassumption}{Assumption}{Assumptions}

\newcommand\independent{\protect\mathpalette{\protect\independenT}{\perp}}
    \def\independenT#1#2{\mathrel{\setbox0\hbox{$#1#2$}%
    \copy0\kern-\wd0\mkern4mu\box0}} 
\newcommand{\mathbbm}[1]{\text{\usefont{U}{bbm}{m}{n}#1}} 

\title{Quantile and Distribution Treatment Effects on the Treated with Possibly Non-Continuous Outcomes}

\author{Nelly Djuazon\footnote{Email: nellydj14@gmail.com. Nutrition, Diets and Health Unit, International Food Policy Research Institute, Almadies, Parcelles 22, Zone 10, Lot 227, Dakar, Senegal.} \and Emmanuel Selorm Tsyawo\footnote{Corresponding author. Email: estsyawo@gmail.com. Department of Economics, Finance and Legal Studies, Culverhouse College of Business, University of Alabama.}}

\begin{document}  

\maketitle

\abstract{
	\noindent Applied Difference-in-Differences studies often involve outcomes that are discrete, mixed, censored, or otherwise non-continuously distributed, while policy questions frequently concern distributional effects rather than mean effects alone. This paper develops a distributional DiD framework for identifying and conducting uniform inference on distribution and quantile treatment effects on the treated in such settings under stated identifying and regularity conditions. Identification is based on distributional parallel trends and no-anticipation assumptions, illustrated through an economic model of crime that generates count-valued untreated potential outcomes. The identification and asymptotic theory accommodate staggered treatment adoption and a general sampling scheme encompassing repeated cross-sections, unbalanced panels, rotating panels, and balanced panels. The paper also proposes a test of functional over-identifying restrictions as a diagnostic for the identifying assumptions and working-CDF specification. An empirical application to the effect of police on crime illustrates the practical relevance of the approach and shows how distributional effects can be interpreted as event-probability effects for count outcomes.
}

\bigskip

\bigskip

\bigskip

\bigskip

\bigskip

\bigskip


\bigskip


\vspace{250pt}

\onehalfspacing

\pagebreak
\begin{refsection}
\section{Introduction}

Applied researchers are often interested in policy evaluations involving outcomes with mass points, counts, bunching, censoring, or mixed support, but such outcomes create distinct inferential challenges. Policies of interest may also have heterogeneous effects beyond mean parameters such as the Average Treatment Effect on the Treated (ATT), which can mask important distributional variation. The quantile treatment effect on the treated (QTT) and the distribution treatment effect on the treated (DTT), for example, are useful causal parameters for studying treatment-effect heterogeneity across the distribution, either separately or jointly. For instance, legal minimum wage policies are expected to affect low-wage workers rather than those earning significantly above the minimum wage \citep{ghanem2025evaluating,cengiz-Arindrajit-2019effect}. \citet{almond-hoynes-schanzenbach-2011-inside} reports that the largest birth-weight gains from the Food Stamp Program in the United States occur among infants with the lowest birth weights. Moreover, \citet{callaway2021bounds} finds that the average effects of job displacement on workers' earnings are a mix of large negative effects and more modest impacts. \citet{callaway-li-oka-2018} demonstrates statistically significant effects of minimum wage increases on earnings, particularly at the lower end of the distribution.

Non-continuously distributed outcome variables are prevalent in empirical work and arise for a variety of reasons. In some settings, discreteness is intrinsic to the outcome of interest, e.g., the count of car thefts \citep{diTella-2004}, attitudes to gun control measured on a three-point ordinal scale \citep{newman-hartman-2019-mass,yamauchi2020difference}, or the frequency of marijuana consumption recorded on a discrete scale \citep{gutknecht2024generalized}. In other cases, apparent non-continuity reflects features of data construction or measurement, such as compositional changes in parties' vote shares \citep{Boussim-2025}. Finally, discontinuities may be induced by policy interventions that generate bunching or mass points in the outcome distribution, as in the case of minimum wage laws \citep{ghanem2025evaluating,cengiz-Arindrajit-2019effect} or price controls \citep{ruback1982effect}. 

Despite the prevalence of non-continuously distributed outcomes in empirical work, the distributional DiD literature has historically lacked a general pathway from point identification to valid inference in such settings. To illustrate, with only two periods of data and under the distributional difference-in-differences assumption of \citet{callaway-li-2019}, the QTT is only partially identified (\citealp{fan-yu-2012}). The resulting bounds—see, for instance, \citet{fan-yu-2012}—are often too wide to be informative in empirical applications (\citealp{callaway-li-2019}). Moreover, standard asymptotic arguments underpinning existing QTT estimators rely on continuity of the outcome distribution; e.g., \citet[Assumption 3.2]{callaway-li-2019}. While \citet{ghanem2025evaluating} provides partial identification results for non-continuous outcomes, it is not immediately clear how to complement these with valid uniform inference that (typically) requires Hadamard differentiability of the quantile map. Finally, practical fixes such as jittering, i.e., smoothing discontinuities in the distribution function, can be unsatisfactory; see, e.g., Section C of the supplement of \citet{chernozhukov2019generic}. 

The recent works of \citet{kim-wooldridge-2024-difference,fernandez-meier-vuuren-vella-2024distribution,li-lin-2024-identification} are the closest to the current paper. These papers study distributional and quantile treatment effects in DiD settings, using restrictions closely related to distributional parallel trends. \citet{fernandez-meier-vuuren-vella-2024distribution} develops a distribution-regression DiD framework based on a transformed distributional parallel-trends restriction, with particular emphasis on covariate adjustment and extensions to multiple outcomes. \citet{kim-wooldridge-2024-difference} proposes a QTT estimator based on a common time-effect restriction on untreated outcome CDFs, allows for covariates, and establishes uniform inference, including simultaneous confidence bands that remain valid for discrete outcomes. \citet{li-lin-2024-identification} extends the QTT identification of \citet{callaway-li-oka-2018} to staggered adoption settings.

This paper complements these contributions by providing a route from point identification to uniform inference for possibly non-continuous outcomes. Specifically, it establishes uniform inference results for DFs, DTT, QFs, and QTT under staggered treatment adoption designs and a general sampling scheme that encompasses balanced panels, unbalanced panels, rotating panels, and repeated cross-sections. To provide a microfoundation for count-valued potential outcomes in the running application, this paper derives such outcomes from a stochastic rational choice model applied to the economics of crime. As a specification check, this paper proposes a \emph{functional over-identifying restrictions test} in a DiD setting with staggered treatment adoption. The test nests over-identifying restrictions from multiple control groups and pre-trend restrictions from multiple pre-treatment periods. It can therefore be viewed as a functional extension of the DiD-based over-identifying restrictions tests proposed in \citet{marcus-santanna-2020,koumou-tsyawo-2024difference}. An empirical application to the effect of police presence on car theft illustrates the usefulness of the DTT representation: the estimates point to an economically meaningful increase in the probability that treated blocks experience no theft, while the corresponding QTT evidence is less statistically decisive.

Operationally, the paper works on the DF scale. It first models and identifies the counterfactual DF of untreated potential outcomes for the treated group, then constructs uniform confidence bands for the observed and counterfactual DFs and for DTT. Uniform inference for QFs and QTT is obtained by applying the band-inversion arguments of \citet{chernozhukov2019generic} to these DF bands. This DF-first route is useful in the present setting because valid uniform bands for DFs are available under discreteness, whereas direct inference on QFs is complicated by non-continuity of the outcome distribution.

The remainder of the paper is organised as follows. \Cref{Sect:Identif} presents the identification results alongside the illustrative economic model of crime. \Cref{Sect:Asymptotic_Theory} provides the estimator and its asymptotic properties. \Cref{Subsect:spec_test} introduces the functional over-identifying restrictions test and studies its asymptotic behaviour. \Cref{Sect:Empirical_Appl} provides an empirical application illustrating the practical relevance of the proposed approach, and \Cref{Sect:Conclusion} concludes. The supplementary appendix contains proofs of all theoretical results, a covariate extension, and simulation studies.

\section{The Model}\label{Sect:Identif}
This section introduces the model framework in a multi-group, multi-period setting. It defines the causal functional parameters of interest and outlines the identification strategy and main results. The section also presents illustrative models for untreated potential outcomes and concludes with a micro-founded model of crime that generates count-valued potential outcomes.
\subsection{Setup}
Consider a staggered treatment adoption design. Let $\T$ and $T$ denote, respectively, the number of pre-treatment and post-treatment periods relative to the first treated group, so that the study contains $T+\T+1<\infty$ calendar periods. Time is indexed by
\(
t \in [-\T:T] := [-\T] \cup [T],
\) where $[-\T]:=\{-\T,\ldots,-1,0\}$ and $[T]:=\{1,\ldots,T\}$ denote, respectively, calendar periods before and after treatment begins for the first treated group. Units are partitioned into groups according to their treatment timing. Let $G \in \mathcal{G}$ denote the group index, where
\(
\mathcal{G} \subseteq ([T] \cup \{\infty\}).
\)
Group $g \in \mathcal{G}\setminus \{\infty\} $ consists of units first treated in period $g$. Treatment state is assumed to be absorbing, so that once treated, units remain treated in all subsequent periods. The case $G=\infty$ corresponds to the never-treated group, which may or may not be available in a given empirical setting.

Under the staggered treatment assignment adopted in this paper, the indexing is defined as follows: 
(a) treated groups are indexed by $g \in \mathcal{G} \setminus \{\infty\}$; 
(b) time periods satisfy $t \in [g:T]$ for post-treatment periods and $s \in [-\T:(g-1)]$ for pre-treatment periods for a fixed treated group $g \in \mathcal{G}\setminus\{\infty\}$; and (c) the admissible control groups correspond to the not-yet-treated and never-treated groups, i.e., $h \in \mathcal{G} \setminus [1:t]$ for a given post-treatment period $t$ of treated group $g\leq t$. For not-yet-treated controls, admissible index combinations satisfy the temporal ordering
\[
-\T \le s < g \le t < h;
\]
the case $h=\infty$ denotes never-treated controls and is included in the same admissible-control set. Define the admissible index set over groups and periods by
\[
\mathcal I :=
\Big\{(g,t,h,s):
g\in\mathcal G\setminus\{\infty\},\;
t\in[g:T],\;
h\in\mathcal{G}\setminus [1:t],\;
s\in[-\T:(g-1)]
\Big\}.
\]

$Y_t$ denotes the observed outcome at period $t \in [-\T:T]$, taking values in $\overline{\mathbb{R}}, \ \overline{\mathbb{R}}:= \mathbb{R} \cup \{-\infty,\infty\} $.\footnote{The binary-support case is not the focus of this paper, as DiD methods for binary outcomes are already treated in \citet{puhani2012treatment,wooldridge-2023-simple}.} Let $\overline{\Y}\subseteq \overline{\mathbb{R}}$ denote the full support of the outcome and let $\Y \subseteq \overline{\Y}$ be a compact index set of interest, i.e., the set of outcome thresholds over which estimation and inference are conducted. $Y_t(0)$ and $Y_t(1)$, respectively, denote the untreated and treated potential outcomes at period $t$. For each $(g,t) \in \mathcal{G} \times [-\T:T]$, define the distribution function \( \F_{gt}(y) := \Prob(Y_t \le y \mid G=g)\). Further, define the potential outcome DFs \(
\F_{Y_t(e) \mid G_g=1}(y) := \Prob\big(Y_t(e) \le y \mid G=g\big),
\quad e \in \{0,1\}, \) where $G_g := \indicator{G=g}$. For example, $\F_{Y_1(0) \mid G_g=1}(y)$ denotes the DF of the untreated potential outcome at period $t=1$ for group $g$.

\subsection{Parameters of interest}

With the data structure and potential outcome notation in place, the causal objects of interest can be defined as functionals of the treated and counterfactual outcome distributions. The focus is on distributional effects for treated groups in post-treatment periods, indexed by \((g,t)\in \mathcal{G}\setminus\{\infty\}\times[g:T]\). The two leading parameters are DTT and QTT.

The Distribution Treatment Effect on the Treated (DTT) for group $g$ at period $t$ is defined as
\begin{equation}
    \DTT_{gt}(y) = \F_{Y_t(1) \mid G_g=1}(y) - \F_{Y_t(0) \mid G_g=1}(y), \quad y\in\Y.\label{eqn:DTT}
\end{equation} $\DTT(y)$ is the effect on the likelihood that the outcome of the treated group is no larger than $y$. Equivalently, it is the treatment-induced change in the share of treated units whose outcome is at most $y$. While QTT is well-explored and easily interpretable, e.g., \citet{athey-imbens-2006,callaway-li-2019,kim-wooldridge-2024-difference}, DTT has received less direct attention. In the present setting, however, this probability-scale interpretation is particularly useful for count-valued outcomes, where effects on event probabilities, such as the probability of no car theft, are economically meaningful. The interpretation of $\DTT(y)$ applies a natural functional generalisation of the Average Treatment Effect on the Treated (ATT) for binary outcomes -- see \citet{wooldridge-2023-simple,puhani2012treatment}. Although labelled DTT, the parameter of interest in \citet{gutknecht2024generalized} applies to the probability mass function of a discrete ordered outcome, whereas \eqref{eqn:DTT} applies to the cumulative distribution function (CDF). Thus, the framework of \citet{gutknecht2024generalized} rules out continuous or mixed outcomes, unlike the current paper.

To make the interpretation of $\DTT$ more concrete, consider the following empirical running example from \citet{diTella-2004}, which studies the effect of increased police presence on crime.

\begin{example}[DTT -- Police presence and car theft]\label{Ex:Interpretation_DTT}
The treatment indicator $G_g$ denotes increased police presence, the outcome $Y$ is the number of car thefts, and the unit of observation is the city block. The distribution function $\F_{Y_t(1)\mid G_g=1}(y)$ in \eqref{eqn:DTT} gives the probability that at most $y$ vehicles are stolen under increased police presence for blocks that were actually treated, whereas $\F_{Y_t(0)\mid G_g=1}(y)$ gives the corresponding counterfactual probability in the absence of treatment. A positive value of $\DTT(y)$ indicates that increased police presence raises the likelihood of observing no more than $y$ thefts, i.e., induces a leftward shift of the outcome distribution at threshold $y$. Thus, when $y$ indexes low crime levels, a positive $\DTT(y)$ increases the share of treated blocks experiencing low crime. Equivalently, it reduces the share of treated blocks exceeding that theft threshold. If $\DTT(y) \geq 0$ for all $y\in\overline{\Y}$, and $\DTT(y) > 0$ for some $y\in\overline{\Y}$, then increased police presence induces first-order stochastic dominance, implying a uniform shift of the distribution toward lower theft counts. When the inequalities are assessed on a restricted index set $\Y\subset\overline{\Y}$, they provide evidence of this dominance pattern over the thresholds in $\Y$. If $\DTT(y)$ changes sign across $y$, the distributions may cross, indicating heterogeneous effects across the distribution rather than a uniform improvement.
\end{example}

To define the quantile-based counterpart, let the QF associated with a generic DF be its left-inverse.
\begin{definition}[Quantile Function]\label{Def:QF}
Given a DF $y\mapsto \F(y)$ with support $\overline{\Y}$, the corresponding QF is
\( \displaystyle 
    \F^{\leftarrow}(\tau)
    :=
    \inf\{y\in\overline{\Y}: \F(y)\geq \tau\},
    \qquad \tau\in(0,1).
\)
\end{definition}
\noindent The quantile treatment effect on the treated (QTT) for group $g$ at period $t$ compares the treated and counterfactual QFs:
\begin{equation}
    \QTT_{gt}(\tau) = \F_{Y_t(1)\mid G_g=1}^{\leftarrow}(\tau) - \F_{Y_t(0)\mid G_g=1}^{\leftarrow}(\tau),
    \qquad \tau\in(0,1).
    \label{eqn:QTT}
\end{equation}
While DTT captures distributional changes at fixed outcome thresholds, QTT measures treatment effects at fixed quantile ranks, thereby providing a complementary perspective on heterogeneity across the outcome distribution. In the running example, this gives the following interpretation.

\begin{example}[QTT -- Police presence and car theft]\label{Ex:Interpretation_QTT}
$\F_{Y_t(1)\mid G_g=1}^{\leftarrow}(\tau)$ denotes the $\tau$-th quantile of the distribution of car theft counts for blocks with increased police presence, while $\F_{Y_t(0)\mid G_g=1}^{\leftarrow}(\tau)$ denotes the corresponding counterfactual quantile that would have prevailed for those same blocks in the absence of increased police presence. The quantile treatment effect on the treated, $\QTT(\tau)$, therefore measures the effect of treatment on the $\tau$-th quantile of the theft distribution. A negative value of $\QTT(\tau)$ indicates that increased police presence reduces car thefts at quantile $\tau$, corresponding to a downward shift of the distribution at that point. When $\QTT(\tau)$ varies across $\tau$, it reveals heterogeneity in treatment effects across the distribution.
\end{example}

Beyond $\DTT(y)$ and $\QTT(\tau)$, the identified distribution functions $\F_{Y_t(1)\mid G_g=1}$ and $\F_{Y_t(0)\mid G_g=1}$ deliver a range of policy-relevant causal parameters beyond average effects typically identified in standard DiD designs. First, they identify probabilities of key events. In the running example, these event probabilities are especially interpretable because they translate distributional effects into the probability that a treated block experiences no car theft. In particular, the probability that a block is crime-free is given by
\(
\Prob(Y_t(d)=0 \mid G_g=1) = \F_{Y_t(d)\mid G_g=1}(0),
\)
so that the causal effect on the probability that a treated block is crime-free is
\(
\F_{Y_t(1)\mid G_g=1}(0) - \F_{Y_t(0)\mid G_g=1}(0) = \DTT(0).
\)
Similarly, the probability of any theft is
\(
\Prob(Y_t(d)>0 \mid G_g=1) = 1 - \F_{Y_t(d)\mid G_g=1}(0),
\)
and the corresponding treatment effect is
\(
\Prob\big(Y_t(1)>0\mid G_g=1\big) - \Prob\big(Y_t(0)>0\mid G_g=1\big) = -\DTT(0).
\)
For integer count outcomes, the event $Y_t(d)>0$ is equivalent to $Y_t(d)\ge 1$. This formulation is useful in the empirical application, where the outcome can be recorded on a fractional support after aggregation. Second, the distribution functions identify inequality in crime across blocks. Let $\mathrm{Gini}(d)$ denote the Gini coefficient of $Y_t(d)$ for treated blocks. When $\E[Y_t(d)\mid G_g=1]>0$, it can be written as a functional of the distribution as
\(\displaystyle 
\mathrm{Gini}(d)
= \frac{1}{2\,\E[Y_t(d)\mid G_g=1]}
\iint |y - y'| \, d\F_{Y_t(d)\mid G_g=1}(y)\, d\F_{Y_t(d)\mid G_g=1}(y'),
\)
where the mean itself is given by
\(\displaystyle 
\E[Y_t(d)\mid G_g=1] = \int y \, d\F_{Y_t(d)\mid G_g=1}(y).
\)
In the discrete count setting, this reduces to
\[
\E[Y_t(d)\mid G_g=1] = \sum_{y=0}^\infty y \, \Prob(Y_t(d)=y \mid G_g=1),
\]
and
\[
\mathrm{Gini}(d)
= \frac{1}{2\,\E[Y_t(d)\mid G_g=1]}
\sum_{y=0}^\infty \sum_{y'=0}^\infty |y - y'| \, \Prob(Y_t(d)=y \mid G_g=1)\, \Prob(Y_t(d)=y' \mid G_g=1).
\]
The causal effect on inequality is therefore
\(
\mathrm{Gini}(1) - \mathrm{Gini}(0),
\)
which captures how increased police presence affects the concentration of crime across blocks. Together, these distributional parameters characterise how treatment affects both the probability of crossing policy-relevant crime thresholds and the shape of the crime distribution, features that are not recoverable from mean comparisons alone.

\subsection{Identification}
The fundamental problem of identifying DTT and QTT is that, unlike $\F_{Y_t(1) \mid G_g=1}(y)$ which is identified from the data sampling process, the distribution of the untreated potential outcome of treated units in the post-treatment period, i.e., $\F_{Y_t(0) \mid G_g=1}(y)$, is not identified without assumptions. Thus, identification assumptions are needed much like in the wider distributional treatment effects literature, e.g., \citet{athey-imbens-2006,callaway-li-2019,bonhomme-sauder-2011,gutknecht2024generalized,wooldridge-2023-simple,kim-wooldridge-2024-difference}. To this end, distributional parallel trends and distributional no-anticipation assumptions are leveraged to identify $\F_{Y_t(0) \mid G_g=1}(y)$.

The identifying restrictions used below have two components. First, for each threshold $y$, untreated outcome distributions are assumed to evolve in parallel across treated and comparison groups after applying the link transformation $\Phi^{-1}$. Second, pre-treatment outcomes for treated groups are assumed not to be affected by future treatment. The latent-index representation below provides one way to rationalise these restrictions while allowing $Y_t(0)$ to be discrete, mixed, or continuous.

Let $\dot{Y}_t(0;y)$ be a local underlying continuously distributed latent variable (on a possibly enlarged probability space) such that $ \indicator{Y_t(0)\leq y} = \indicator{\dot{Y}_t(0;y)\leq 0}$.\footnote{\Cref{rem:latent_Y_exist} demonstrates the existence of such a latent variable in the context of the running example.} Fix $y\in\Y$, and for each $t\in [-\T:T] $, assume the $y$-indexed latent untreated potential outcome satisfies 
\begin{equation}\label{eqn:latent_Y0}
\dot{Y}_t(0;y)
=
U(0)-\delta_{Gt}(y),
\qquad
\delta_{g\nu}(y)
=
\alpha(y)+\beta_g(y)+\gamma_\nu(y),
\quad (g,\nu)\in\mathcal{G}\times[-\T:T].
\end{equation}
where $U(0)$ is a common innovation process, i.e., it does not depend on $y$. That is, for any $y,y'\in\Y$, the latent indices share the same stochastic component $U(0)$ and differ only through the deterministic shifts $\alpha(\cdot)$, $\beta_{g'}(\cdot)$, and $\gamma_\nu(\cdot)$. In the running example, one can see $\dot{Y}_t(0;y)$ as an unobservable index capturing the (counterfactual) underlying tendency or vulnerability to crime on a block in the absence of \emph{increased police presence}. This structural interpretation allows one to model the probability of observing a certain level of crime as a threshold-crossing event. Observe that the above restriction is imposed only on the untreated potential outcome, and the DF of the post-treatment treated potential outcome is left unrestricted. Relative to a fully saturated specification $\delta_{g\nu}(y)=\alpha(y)+\beta_g(y)+\gamma_\nu(y)+\kappa_{g\nu}(y)$, the restriction in \eqref{eqn:latent_Y0} sets the group-time interaction $\kappa_{g\nu}(y)$ to zero. Thus, group heterogeneity and common time shocks are allowed to vary flexibly with $y$, but untreated latent outcomes are ruled out from having group-specific time shocks; see \citet[Assumption 1]{fernandez-meier-vuuren-vella-2024distribution}. From \eqref{eqn:latent_Y0} and the foregoing, the DF of the untreated potential outcome of group $g$ at period $t$ is given by
\begin{align}
\F_{Y_t(0) \mid G_g=1}(y) =& \Prob(Y_t(0) \leq y \mid G_g = 1) =\Prob(\dot{Y}_t(0;y) \leq 0 \mid G_g=1) \nonumber \\
& = \Phi\big(\alpha(y) + \beta_g(y) + \gamma_t(y) \big) \label{eqn:F_C}
\end{align}
\noindent for a strictly increasing function $\Phi: \mathbb{R} \rightarrow [0,1]$.

The approach to imposing identifying restrictions on latent untreated potential outcomes has antecedents in the literature, e.g., \citet{wooldridge-2023-simple,yamauchi2020difference,gutknecht2024generalized}. This formulation differs from ordered-discrete threshold-crossing approaches because the latent index is introduced locally for each threshold $y$, rather than as a single global latent variable generating a finite set of ordered outcomes. This allows the framework to accommodate discrete, mixed, and continuous outcomes while avoiding global distributional assumptions on $Y_t(0) \mid G_g=1$. Since the DF of post-treatment treated potential outcomes of the treated group is identified from the data sampling process, i.e., $ \F_{Y_t(1) \mid G_g=1}(y) := \Prob(Y_t(1) \leq y \mid G_g=1) = \Prob(Y_t \leq y \mid G_g=1) = \F_{gt}(y) $, it remains to identify the counterfactual DF, namely $\F_{Y_t(0) \mid G_g=1}(y)$. 

Consider the following distributional parallel trends assumption, stated for admissible quadruples $(g,t,h,s)\in\mathcal I$.
\begin{assumption}[Distributional Parallel Trends]\label{ass:PT}
For every $y\in\Y$, every $(g,t,h,s)\in \mathcal I $, and every $(d,\nu)\in(\{0,1\}\times \{s,t\}) $, the following restriction and representation hold:
\begin{align*}
\F_{Y_{\nu}(0)\mid G_g=d,\;G_g+G_h=1}(y)
=
\Phi\Big(\alpha_s^h(y)
+ d\,\beta_s^{gh}(y)
+ \indicator{\nu=t}\,\gamma_{st}^h(y)\Big),
\end{align*}
 where $\Phi:\mathbb{R} \mapsto [0,1] $ is a known strictly increasing function with inverse $\Phi^{-1}: (0,1) \mapsto \mathbb{R} $.
\end{assumption}

\noindent \Cref{ass:PT} is imposed on the untreated potential outcome only -- the DF of the treated potential outcome in the post-treatment period is unrestricted. $\Phi(\cdot)$ in \Cref{ass:PT} is a known or user-specified working CDF; in practice, one can use known standard CDFs such as the normal or logistic. By flexibly modelling each point $y$ on $\Y$, the choice of $\Phi(\cdot)$ should be viewed as a working CDF, i.e., it need not hold \emph{stricto sensu}, since the approximation of the DF is local rather than global; see, e.g., \citet[Sect 2.4]{fernandez-meier-vuuren-vella-2024distribution}. The standard uniform CDF and identity link are useful benchmark specifications but do not satisfy the strict proper-link conditions in \Cref{ass:PT}: the former is flat outside $[0,1]$, while the latter is not a proper CDF on $\mathbb{R}$. The identity link $\Phi(x)=x$ corresponds to imposing parallel trends directly on the DF scale as in distributional DiD restrictions such as \citet{havnes-mogstad-2015-universal}, \citet[eqn.~5]{roth-santanna-2023parallel}, and \citet{kim-wooldridge-2024-difference}. The simulations in \Cref{App:Sect_Sim} therefore report results for the standard normal, standard uniform, and identity benchmark specifications, and document the finite-sample sensitivity of the procedure to this working-link choice.

No-anticipation assumptions often encountered in the DiD literature have their analogue in the current setting.
\begin{assumption}[Distributional No-Anticipation]\label{ass:NA} 
For every $y\in\Y$, every $g\in\mathcal{G}\setminus\{\infty\} $ and any $s\in [-\T:(g-1)] $, \( \displaystyle \F_{Y_s(1) \mid G_g=1}(y) = \F_{Y_s(0) \mid G_g=1}(y). \)
\end{assumption}
\noindent \Cref{ass:NA} says the DF of both treated and untreated potential outcomes of the treated group in pre-treatment periods is the same -- this rules out anticipatory effects of treatment at any point of the distribution of the outcome. In the running example, \Cref{ass:NA} holds for periods preceding the terrorist attack as treatment, i.e., increased police presence, is driven by an unanticipated shock. However, the time lag between the terrorist attack and increased police presence on certain blocks may be contaminated by anticipatory effects. It suffices to avoid using these time periods; see \citet[Figure 1]{diTella-2004}.


With \Cref{ass:PT,ass:NA} in hand, one can proceed with the identification of the counterfactual DF $\F_{Y_t(0) \mid G_g=1}(y)$.
\begin{theorem}[Identification]\label{Thm:Identif}
Suppose the maintained sampling setup holds and \Cref{ass:PT,ass:NA} hold. Then, for every 
$y \in \Y$ and every $(g,t,h,s)\in \mathcal I $, the counterfactual distribution 
$\F_{Y_t(0) \mid G_g=1}(y)$ is identified:
\begin{equation}\label{eqn:DF_Counterfactual}
\F_{Y_t(0) \mid G_g=1}(y)= \F_{Y_t(0)\mid G_g=1,G_g+G_h=1}(y)
= 
\Phi\!\Big(
\Phi^{-1}\big(\F_{gs}(y)\big)
+
\Phi^{-1}\big(\F_{ht}(y)\big)
-
\Phi^{-1}\big(\F_{hs}(y)\big)
\Big).
\end{equation}
\end{theorem} 
\noindent It follows from \Cref{ass:PT} that, within the pairwise comparison sample $G_g+G_h=1$, $\F_{Y_s(0) \mid G_g=d,G_g+G_h=1}(y) = \Phi\big( \alpha_s^h(y)+d\beta_s^{gh}(y) \big) $ and $\F_{Y_t(0) \mid G_g=d,G_g+G_h=1}(y) = \Phi(\alpha_s^h(y)+d\beta_s^{gh}(y) +\gamma_{st}^h(y)) $. Combining the above with the invertibility of $\Phi(\cdot)$, $\gamma_{st}^h(y) = \Phi^{-1}\big(\F_{Y_t(0) \mid G_g=d,G_g+G_h=1}(y)\big) - \Phi^{-1}\big(\F_{Y_s(0) \mid G_g=d,G_g+G_h=1}(y)\big)$ for each $d\in\{0,1\}$. The latter implies a parallel trends assumption on a non-linear transformation of $\F_{Y_\nu(0) \mid G_g=d,G_g+G_h=1}(y), \ \nu\in\{s,t\} $ for every $y\in\Y$ -- cf. \citet[eqn. 2.8]{wooldridge-2023-simple}. Indeed, under \Cref{ass:NA}, \eqref{eqn:DF_Counterfactual} has the equivalent expression for all $y\in\Y$
\begin{equation}\label{eqn:NL_PT}
    \begin{split}
        &\Phi^{-1}\big(\F_{Y_t(0) \mid G_g=1,G_g+G_h=1}(y)\big) - \Phi^{-1}\big(\F_{Y_s(0) \mid G_g=1,G_g+G_h=1}(y)\big) \\
    = &\Phi^{-1}\big(\F_{Y_t(0) \mid G_g=0,G_g+G_h=1}(y)\big) - \Phi^{-1}\big(\F_{Y_s(0) \mid G_g=0,G_g+G_h=1}(y)\big).
    \end{split}
\end{equation}
\noindent The above equation is a distributional parallel trends condition on a strictly monotonic transformation of the DF of untreated potential outcomes for the treated group $g$ and comparison group $h$ within the pairwise comparison sample, for all $y\in\Y$.\footnote{Support restrictions on the distribution of $Y_t\mid G=g$, $g\in\mathcal G\setminus\{\infty\}$, are not required in \Cref{Thm:Identif} because the link function $ \Phi(\cdot) $ in \Cref{ass:PT} is restricted to proper CDFs; see \citet[pp.~5--6]{fernandez-meier-vuuren-vella-2024distribution}.} 

With the availability of multiple control groups, multiple pre-treatment periods, or both, \Cref{Thm:Identif} implies over-identifying restrictions for all admissible $ (h,s) \in \big(\mathcal{G}\setminus [1:t]\big) \times [-\T:g-1]$. This is exploited in \Cref{Subsect:spec_test} to develop a functional over-identifying restrictions test of identification and specification.

\subsection{Examples}
The following examples illustrate simple benchmark specifications and data-generating processes related to \Cref{ass:PT}. They are meant as toy constructions: the first clarifies the link with existing distributional DiD restrictions, while the remaining examples show how continuous, discrete, and censored outcomes can be generated from latent continuously distributed random variables.

\begin{example}\label{Ex:G_Unif}
  Consider the standard-uniform benchmark link, $\Phi(x) = x\indicator{0\leq x\leq 1} + \indicator{x>1} $. Although this link is not strictly increasing on $\mathbb{R}$, substituting it into the counterfactual formula in \eqref{eqn:DF_Counterfactual} yields the bounded benchmark
\begin{equation}\label{eqn:DF_G_Unif}
\F_{Y_t(0) \mid G_g=1}(y) = \Pi_{ht}(y)\indicator{0\leq \Pi_{ht}(y)\leq 1} + \indicator{\Pi_{ht}(y) > 1}
\end{equation}
where $\Pi_{ht}(y): = \F_{gt}(y) + \F_{ht}(y) - \F_{hs}(y), \ y\in\Y$.
\end{example}

\noindent Without the support restrictions imposed by the standard uniform link in \eqref{eqn:DF_G_Unif}, the expression $\Pi_{ht}(y)$ corresponds to the identity-link distributional DiD counterfactual; see \citet{havnes-mogstad-2015-universal}, \citet[eqn.~5]{roth-santanna-2023parallel}, and \citet{kim-wooldridge-2024-difference}. This identity-link benchmark is not guaranteed to be a DF unless $\Pi_{ht}(y)\in[0,1]$ and monotonicity in $y$ hold. Thus, $\Phi(\cdot)$ in the proper-CDF formulation \eqref{eqn:DF_Counterfactual} plays the role of enforcing the $[0,1]$ bound on the counterfactual DF, while the identity case remains useful for comparison and for specification tests based on direct DF-scale restrictions.

\begin{example}\label{ex:Latent_Loc_Shift}
Consider the model: $ \displaystyle Y_t(0) = U(0) - \alpha - \sum_{g'\in\mathcal{G}}G_{g'} \beta_{g'} - \gamma_t,$ where $U(0) $ is continuously distributed. Then, $\F_{Y_t(0) \mid G_g=1}(y)=\Phi\big(\alpha(y) + \beta_g + \gamma_t \big), $ where $\alpha(y) = y+\alpha $.
\end{example}

\noindent \Cref{ex:Latent_Loc_Shift} is the familiar continuous-outcome case: the latent variable coincides with the observed untreated outcome, and \Cref{ass:PT} imposes a pure location shift in the DF at each $y\in\Y$.

\begin{example}\label{ex:Ordered_Logit}
Consider a discrete untreated potential outcome $Y_t(0)$ with support $ \overline{\Y}:= \{y_0,y_1,\ldots,y_K\}$. Suppose there exists a latent index
$ \displaystyle \dot{Y}_t(0) = U(0) - \alpha - \sum_{g'\in\mathcal{G}}G_{g'} \beta_{g'} - \gamma_t,$
and cut-points $\tilde y_{-1}=-\infty<\tilde y_0<\cdots<\tilde y_K=+\infty$ such that \(
Y_t(0)=y_k \quad\Longleftrightarrow\quad \tilde y_{k-1}<\dot{Y}_t(0)\le \tilde y_k, \text{ and } k\in \{0\} \cup [K].
\)
Then, for $y\in (y_{k-1},y_k]$ (with $y=y_0$ if $k=0$),
\(
\F_{Y_t(0)\mid G_g=1}(y)
=
\Prob(Y_t(0) \le y_k \mid G_g=1)
=
\Prob(\dot{Y}_t(0) \le \tilde y_k \mid G_g=1)
=\Phi(\alpha_k + \beta_g + \gamma_t).
\)
\end{example}

\noindent In \Cref{ex:Ordered_Logit}, the observed outcome is discrete, but the distribution still satisfies \Cref{ass:PT} at each \( y \in \Y\subseteq\overline{\Y} \), with the threshold-specific intercept \(\alpha_k\) varying over the support.

\begin{example}\label{ex:Censored}
Consider the latent index in \Cref{ex:Ordered_Logit} and define the censored count-valued untreated potential outcome
\[
Y_t(0)=\max\big\{\lceil \dot{Y}_t(0)\rceil,\,0\big\}.
\]
Then $Y_t(0)$ is discrete with $\overline{\Y}:=\{0,1,2,\ldots\}$, while $\dot{Y}_t(0)$ remains the generating latent variable. For any $y\in\Y\subseteq\overline{\Y}$, \( \F_{Y_t(0)\mid G_g=1}(y)=\Phi(\alpha(y)+\beta_g+\gamma_t) \), where $\alpha(y):=\alpha+y$.
\end{example}
\noindent The last two examples show that \Cref{ass:PT} can hold even when the observed untreated outcome is non-continuous and differs from the latent variable. The next subsection provides an economic model that gives this latent construction a concrete interpretation in the running crime application.

\subsection{A Stochastic Random Utility Model of Crime}

This section develops a micro-founded model of crime based on a random utility framework, with a Roy-type occupational choice interpretation. The model serves two purposes: (i) to illustrate how count-valued potential outcomes can arise from economic primitives, and (ii) to give an economic interpretation to assignment mechanisms under which \Cref{ass:PT} is plausible.

The economic model of crime-utility originates with \citet{Becker-1968}, who conceptualises criminal behaviour as the outcome of rational decision-making under uncertainty, whereby individuals compare the expected benefits and costs of engaging in illegal activities. \citet{ehrlich-1973-participation} extends this framework by embedding criminal participation within a broader occupational choice problem. Departing from these deterministic expected-utility formulations, this section adopts a stochastic specification in which realised outcomes depend on idiosyncratic shocks in addition to \emph{ex ante} incentives. Specifically, while the decision to offend is formed \emph{ex ante} via utility maximisation, the realisation of crime is subject to unforeseen factors at the time of execution, such as random opportunities, environmental conditions, or situational frictions. In keeping with the running empirical application of \citet{diTella-2004}, the analysis focuses on a setting with a single treated group and a single control group observed over multiple pre- and post-treatment periods. Potential outcomes are discrete-valued block aggregates generated by the crime decisions of individuals under uncertainty. 

Let $w^L(\cdot)$ and $w^C(\cdot)$ denote an individual's net earnings from engaging in legal and crime sectors, respectively. For example, one can understand $w^L(\cdot)$ and $ w^C(\cdot)$ as increasing in ability and decreasing in transportation costs from one's place of residence. There are $ M \in \mathbb{N} $ individuals (potential offenders) with access to a space of $ N \in \mathbb{N} $ blocks. In this subsection, blocks are indexed by $j\in[N]$, in keeping with the unit-level notation used elsewhere in the paper, and potential offenders are indexed by $i\in[M]$. For an individual $i$ on block $j$ at period $t \in [-\T:T] $, the utility from each choice includes an idiosyncratic shock:
\begin{align*}
U_{jit}^L(e) =& w^L(e;Z_{jt},X_{it},\zeta_i) + \mathcal{E}_{jit}^L \qquad \text{and} \\
U_{jit}^C(e) =& (1-\varpi_{jit}(e))w^C(e;Z_{jt},X_{it},\zeta_i) - \varpi_{jit}(e) \cdot f_i + \mathcal{E}_{jit}^C
\end{align*} where $e \in \{0,1\} $ is a binary argument for the status quo versus \emph{increased level of law enforcement deterrence}; $Z_{jt}$ captures block characteristics such as population density, employment opportunities, the \emph{status quo} level of law enforcement deterrence, and the degree of self-protection against crime and victimisation; $X_{it}$ captures individual $i$'s observable characteristics such as level of education, available time for work in both sectors, and block of residence; and $\zeta_i$ represents individual skill or ability.\footnote{The interaction of $Z_{jt}$ and $X_{it}$ through $w^C(\cdot)$ and $w^L(\cdot)$ captures transaction costs in the respective sectors, e.g., transportation costs to individual $i$ in conducting business in block $j$ in the legal sector or the costs of accomplices' services, accessories, or transportation in the illegal sector, e.g., \citet{ehrlich-1973-participation}.} $\varpi_{jit}(e):= \varpi(e;Z_{jt},X_{it},\zeta_i) $ captures individual $i$'s (subjective) probability of apprehension in block $j$ at period $t$ under state $e$, and $f_{i}$ is disutility imposed in case of apprehension, e.g., fines, legal fees, or the opportunity cost of prison time. 

The idiosyncratic utility shocks $\mathcal{E}_{jit}^C$ and $\mathcal{E}_{jit}^L$ capture unobserved factors that influence the utility from choosing crime or legal work, e.g., unanticipated changes in the work environment (for both sectors) such as weather shocks, socio-cultural influences in the legal sector, personal risk preferences, peer effects, or opportunity-specific risks in the crime sector. The stochastic decision rule (as a function of indicator $e$) for a rational economic agent \( i \) is to commit a crime on block \( j \) at period $t$ if \( U_{jit}^C > U_{jit}^L \) :
\begin{align*}
    Y_{jit}(e):& = \indicator{ U_{jit}^C(e) > U_{jit}^L(e) } \\
    =& \indicatorBig{ \underbrace{\mathcal{E}_{jit}^C - \mathcal{E}_{jit}^L}_{\widetilde{\mathcal{E}}_{jit}} >
    \begin{aligned}
        &w^L(e;Z_{jt},X_{it},\zeta_i) - w^C(e;Z_{jt},X_{it},\zeta_i) \\
        &\quad + \varpi_{jit}(e)\big(f_i + w^C(e;Z_{jt},X_{it},\zeta_i) \big)
    \end{aligned} } \\
    =&:\indicatorbig{ \widetilde{\mathcal{E}}_{jit} > \varphi(e; Z_{jt},X_{it},\zeta_i,f_i ) }.
\end{align*}
$ Y_{jit}(0) $ and $ Y_{jit}(1) $ are individual $i$'s binary untreated and treated potential outcomes, respectively, on block $j$ at period $t$. Specifically, $ Y_{jit}(1) = 1 $ if $i$ would \emph{commit a crime} if there were \emph{increased police presence}, and $ Y_{jit}(0) = 1 $ if $i$ would \emph{commit a crime} if there were \emph{no increased police presence} on block $j$ at period $t$. 

The following economic primitives suffice for the count-valued potential outcomes to be well-defined.
\begin{condition}\label{Ass:economic_model}
\quad
\begin{enumerate}[(a)]
    \item There is an infinite number of economic agents, formally the limit of the economy as $M\rightarrow \infty $.
    \item For each block $j\in [N]$ and period $t \in [-\T:T] $, only finitely many crimes can be committed, almost surely: \( \displaystyle \max_{e \in \{0,1\} } \sum_{i=1}^\infty Y_{jit}(e) \, < \, \infty \ a.s. \)
\end{enumerate}
\end{condition}

\noindent \Cref{Ass:economic_model}(a) is a useful approximation device for removing the dependence of potential outcomes on $M$ and for integrating out individual-level characteristics over blocks. \Cref{Ass:economic_model}(b) is intuitive. A sufficient condition is that there is only a finite number of possibilities or opportunities for crime on any given block at any period.

Aggregating potential outcomes across individuals, one obtains the following representation of the block $j$ potential outcome at period $t$:
\begin{align*}
    Y_{jt}(e):= \lim_{M \rightarrow \infty} \sum_{i=1}^M Y_{jit}(e) =  \lim_{M \rightarrow \infty} \sum_{i=1}^M \indicatorbig{ \widetilde{\mathcal{E}}_{jit} > \varphi(e; Z_{jt},X_{it},\zeta_i,f_i ) },
\end{align*} thereby generating count-valued block-level potential outcomes $ Y_{jt}(e), \ e \in \{0,1\} $.   Under \Cref{Ass:economic_model}, potential outcomes \( Y_{jt}(e), \, e\in \{0,1\} \) are defined on $\mathbb{N}$ for each $(j,t)$.

The following result characterises the distribution of potential outcomes under given conditions. 
\begin{proposition}[Characterising the distribution of Potential Outcomes]\label{prop:Pois_Pot_outcome}
In addition to \Cref{Ass:economic_model}, suppose (i) \( \big\{Y_{jit}(e) \big\}_{i=1}^M \) are $i.i.d.$ across agents given $\mathcal{X}_{jt}$ where $\mathcal{X}_{jt}$ is the sigma-field capturing common information, e.g., $Z_{jt}$ and (ii) \( \Prob\big( Y_{jit}(e) = 1 \mid \mathcal{X}_{jt} \big) = p_{Mt}(e) \ a.s. \) for all $i=1,\ldots,M$. 
\begin{enumerate}[(a)]
    \item Then
    \[
        \sum_{i=1}^M Y_{jit}(e) \mid \mathcal{X}_{jt}
        \sim
        \mathrm{Binomial}\big(M,p_{Mt}(e)\big).
    \]
    Further, under the rate condition
    \[
        \lim_{M\rightarrow \infty} Mp_{Mt}(e) = \lambda_t(e),
        \qquad
        \max_{e\in \{0,1\}} \lambda_t(e) < \infty,
    \]
    the potential outcome $ Y_{jt}(e) \mid \mathcal{X}_{jt} $ converges in distribution to $ \mathrm{Pois}\big(\lambda_t(e)\big) $ for each period $t$.

    \item The identification result in \Cref{Thm:Identif} holds if, for $G_{gj}:=\indicator{G_j=g}$, either (i) \( Y_{jt}(0) \mid G_{gj} = d \,  \sim \, \mathrm{Pois}\big(\lambda_{d\cdot}(0)\big) \) for each $d\in\{0,1\}$ or (ii) \( Y_{j\nu}(0) \mid G_{gj} = d \,  \sim \, \mathrm{Pois}\big(\lambda_{\cdot \nu}(0)\big) \) for each $\nu\in\{s,t\}$, where the dot indicates invariance in the omitted index.
\end{enumerate}
\end{proposition}
\begin{proof}
    For part (a), conditional independence across agents gives a Binomial count, and the stated rare-event rate condition gives the Poisson limit. For part (b), the Poisson distribution is fully determined by its mean, so group-invariant or time-invariant untreated Poisson means imply the transformed parallel trends representation in \eqref{eqn:NL_PT}. The conclusion then follows from \Cref{Thm:Identif}.
\end{proof}
\noindent \Cref{prop:Pois_Pot_outcome}(a) above demonstrates that under fairly non-parametric assumptions on economic agents' preferences and sampling of economic agents across the space of blocks, one can derive a closed-form distribution of count-valued potential outcomes. \Cref{prop:Pois_Pot_outcome}(b) shows that identification can be credible under assignment mechanisms with verifiable plausibility in given empirical settings.

Finally, the following remark concludes with the existence of a generating latent variable representation of untreated potential outcomes.
\begin{remark}\label{rem:latent_Y_exist}
For the count-valued potential outcomes generated above, let $U_t \sim \mathcal{U}(0,1)$ independently of $Y_t(e)$ and define
\[
    \dot{Y}_t(e;y):=Y_t(e)-y-U_t .
\]
Then, for every integer threshold $y\in\mathbb{N}$,
\[
    \indicator{Y_t(e)\leq y}
    =
    \indicator{\dot{Y}_t(e;y)\leq 0}
    \quad a.s.
\]
Thus, the threshold event for a discrete potential outcome can be represented through a latent continuously distributed variable indexed by $y$.
\end{remark}

\section{Asymptotic Theory}\label{Sect:Asymptotic_Theory}

This section develops the large-sample theory for the distributional DiD estimators introduced above. The goal is to obtain uniform weak convergence for the estimated treated and counterfactual DFs, the associated DTT process, and their convex-weighted counterparts, while allowing the outcome distribution to be discrete, continuous, or mixed. The results are stated under a sampling scheme that treats units as independent across the cross section but allows their observation patterns to vary over time, thereby nesting balanced panels, unbalanced panels, rotating panels, and repeated cross-sections. This formulation is particularly useful for the empirical application, where inference is conducted at the block level and dependence over time within a block must be preserved.

\subsection{Sampling}

Let $S_{jt}\in\{0,1\}$ indicate whether unit $j\in [N]$
is observed at period $t \in [-\T:T] $, where $N$ denotes the number of \emph{unique cross-sectional units} observed at least once over periods
$t\in [-\T:T]$. Let $(Y_{jt}, G_j) $ denote the (latent) full-data vector for unit $j$
at period $t$, and let the observed record be $(S_{jt},\,S_{jt}(Y_{jt}, G_j))$ (equivalently, $(Y_{jt}, G_j)$
is observed only when $S_{jt}=1$). The analysis sample consists of units observed at least once, so that \( \sum_{t\in[-\T:T]}S_{jt}\ge 1 \) a.s. for each $j\in [N]$. Define the unit-level object
$ W_j:= \big(\{(S_{jt},Y_{jt})\}_{t\in [-\T:T]},G_j\big) $.

\begin{assumption}[Random Sampling]\label{ass:Sampling_gen}
The vectors $\{W_j:1\le j\le N\}$ are independent and identically distributed with $S_{jt} \independent (Y_{jt}(0),Y_{jt}(1), G_j) $ for each $t \in [-\T:T] $.
\end{assumption}

Balanced panels obtain when $S_{jt}=1$ a.s. for all $(j,t) \in [N]\times [-\T:T] $; unbalanced panels allow $S_{jt}$ to be stochastic with
\( \displaystyle \sum_{t\in [-\T:T]} S_{jt}\ge 2 \) for some units; and repeated cross-sections obtain when $ \displaystyle \sum_{t\in [-\T:T]} S_{jt} = 1$
a.s. for all $j$. This makes the inference results presented in this paper generally applicable whenever the posited identification conditions, viz. \Cref{ass:NA,ass:PT}, hold. Further, jointly imposing \Cref{ass:PT,ass:NA,ass:Sampling_gen} precludes non-random missingness that may jeopardise inference on the DFs.\footnote{See \citet{wooldridge-2007} and \citet{bellego-benatia-dortet-2025chained} on allowing non-random missingness.} \Cref{ass:Sampling_gen} imposes independence across units while allowing arbitrary dependence over time within a unit through the vector $W_j$. Thus, when the same cross-sectional unit is observed in multiple periods, its within-unit time dependence is preserved in the asymptotic analysis.

The following regularity assumption is imposed on $\displaystyle \pi_t:= \E[S_t], \ t \in [-\T,T] \ \text{and} \ p_g:=\Prob(G=g), \ g\in \mathcal{G} $.
\begin{assumption}[Sampling Proportions]\label{Ass:reg_pi_p}
The constants
\( \pi_t \) and \( p_g\) exist and satisfy \( p_g \in(0,1) \) and \( \pi_t \in (0,1], \quad \text{for all } (g,t)\in \mathcal{G} \times [-\T:T].
\)
\end{assumption}
\noindent\Cref{Ass:reg_pi_p} ensures that the number of observations in each period is not asymptotically negligible relative to the total sample size, and that all treatment groups occur with positive probability in every period. This condition is essential in repeated cross-sections, rotating panels, and unbalanced panels, where the number of observations may vary across periods. The constants $\{\pi_{t'}\}_{t'=-\T}^T$ summarise the sampling design: \(\pi_t\) is the unit-level observation probability in period \(t\). Because observability is independent of group membership under \Cref{ass:Sampling_gen}, $\Prob(S_t=1,G=g)=\pi_t p_g$, which justifies the product normalisation used in the empirical DFs below. Balanced panels satisfy $\pi_t=1$ for all $t \in [-\T:T] $, in which case Assumption~\ref{Ass:reg_pi_p} holds automatically because $\T+T<\infty$.

\begin{remark}[Identifying pre-treatment groups under staggered adoption without balanced panels] 
This paper assumes time-invariant group membership while allowing for a sampling scheme that nests repeated cross-sections and unbalanced panels. Group membership is determined by, or coincides with, observable time-invariant characteristics. Examples include birth cohorts, geographic regions, industries, or eligibility rules that are fixed over time. This allows pre-treatment observations to be recovered as outcomes measured prior to treatment for units sharing the same time-invariant group identifiers.
\end{remark}

\subsection{Estimation}\label{Sect:Estimation}
The parameters of interest are DFs, QFs, $\DTT(y), \ y\in\Y $ and $\QTT(\tau), \ \tau \in (0,1) $. The first terms in \eqref{eqn:DTT} and \eqref{eqn:QTT} are estimable from data using observed outcomes for the treated group in the post-treatment period. It remains to estimate the counterfactual DF and QF, namely $\F_{Y_t(0) \mid G_g=1}(y)$ and $\F_{Y_t(0) \mid G_g=1}^{\leftarrow}(\tau)$, which are identified for every $(y,\tau) \in \Y\times (0,1) $ under the conditions of \Cref{Thm:Identif}. By the analogue and plug-in principles, \( \displaystyle \hat{p}_g:= \frac{1}{N} \sum_{j=1}^{N}\indicator{G_j=g} \) and \( \displaystyle \hat{\pi}_t:= \frac{1}{N} \sum_{j=1}^{N} S_{jt} \) are $\sqrt{N}$-consistent estimators of $ p_g, \ g \in \mathcal{G} $ and $\pi_t, \ t \in [-\T:T] $, respectively.\footnote{Recall that group assignment $G$ is time-invariant and therefore observable for all units.} Similarly, let \( \displaystyle \widehat{\F}_{gt}(y) = \frac{1}{N}\sum_{j=1}^N \frac{S_{jt}\indicator{G_j=g}\indicator{Y_{jt}\leq y}}{\hat{\pi}_t\hat{p}_g} \) be the outcome empirical CDF of group $g \in \mathcal{G} $ at period $t \in [-\T:T] $, which is well-defined thanks to the observability indicator $S_{jt}$. 

For a treated group $g$, post-treatment period $t$, control group $h$, and pre-treatment period $s$, it follows from the proof of \Cref{Thm:Identif} that 
\[
\big(\alpha_s^h(y),\beta_s^{gh}(y),\gamma_{st}^h(y)\big)' = \Big(\Phi^{-1}\big(\F_{hs}(y)\big), \Phi^{-1}\big(\F_{gt}(y)\big) - \alpha_s^h(y), \Phi^{-1}\big(\F_{ht}(y)\big) - \alpha_s^h(y) \Big)'.
\]
Applying the analogue and plug-in principles, the estimators of the DFs are given by 
\begin{align*}
    \widehat{\F}_{Y_t(1) \mid G_g=1}(y) = \widehat{\F}_{gt}(y) \quad \text{and} \quad \widehat{\F}_{Y_t^{hs}(0) \mid G_g=1}(y) = \Phi\Big(\Phi^{-1} \big(\widehat{\F}_{gs}(y)\big) + \Phi^{-1} \big(\widehat{\F}_{ht}(y)\big) - \Phi^{-1} \big(\widehat{\F}_{hs}(y)\big) \Big).
\end{align*}
$\DTT_{gt}$ and $\QTT_{gt}$ can then be estimated using
\[
\widehat{\DTT}_{gt}^{hs}(y) = \widehat{\F}_{Y_t(1) \mid G_g=1}(y) - \widehat{\F}_{Y_t^{hs}(0) \mid G_g=1}(y)
\quad \text{and} \quad 
\widehat{\QTT}_{gt}^{hs}(\tau) = \widehat{\F}_{Y_t(1) \mid G_g=1}^{\leftarrow}(\tau) - \widehat{\F}_{Y_t^{hs}(0) \mid G_g=1}^{\leftarrow}(\tau).
\]
Other causal parameters that can be written as functionals of the treated and counterfactual DFs, such as event probabilities, tail probabilities, means over bounded supports, or inequality indices, can be estimated by the same plug-in principle, replacing the unknown DFs by their estimated counterparts.

\subsection{Weak Convergence}\label{SubSect:Theory_DF}
The next result studies the asymptotic behaviour of $\widehat{\F}_{Y_t(1) \mid G_g=1}(y)$, $\widehat{\F}_{Y_t^{hs}(0) \mid G_g=1}(y)$, and $\widehat{\DTT}_{gt}^{hs}(y)$. The following smoothness and dominance conditions are useful.

\begin{assumption}[Regularity]\label{ass:reg_Phi}
    $\Phi:\mathbb{R}\to[0,1]$ is continuously differentiable, with derivative
    \( \displaystyle
    \phi(b):=\frac{d}{db}\Phi(b).
    \)
    Moreover, there exists a constant $C\in(0,\infty)$ such that
    \( \displaystyle
    \sup_{x\in\mathbb{R}}\phi(x)\leq C
    \)
    and
    \( \displaystyle
    \inf_{(g,t)\in\mathcal{G}\times[-\T:T]}\ \inf_{y\in\mathcal{Y}}
    \phi\big(\Phi^{-1}(\F_{gt}(y))\big)\geq \frac{1}{C}.
    \)
\end{assumption}
\noindent The smoothness condition imposed on $\Phi(\cdot)$ is crucial for the asymptotic results because it makes the counterfactual DF map in \eqref{eqn:DF_Counterfactual} amenable to the functional delta method.

\begin{theorem}[Weak Convergence]\label{Thm:FCLT_DF.DTT}
Suppose \Cref{ass:NA,ass:PT,ass:Sampling_gen,Ass:reg_pi_p,ass:reg_Phi} hold. Then, for every 
$(g,t,h,s)\in \mathcal I $, 
\begin{enumerate}[(a)]
    \item \( \sqrt{N}\big(\widehat{\F}_{Y_t(1) \mid G_g=1} - \F_{Y_t(1) \mid G_g=1}\big) \rightsquigarrow \mathbb{G}_{gt} \) in $ \ell^\infty(\Y) $; 
    \item \( \sqrt{N}\big(\widehat{\F}_{Y_t^{hs}(0) \mid G_g=1} - \F_{Y_t(0) \mid G_g=1}\big) \rightsquigarrow \mathbb{G}_{gt}^{hs} \) in $ \ell^\infty(\Y) $; \quad and
    \item \( \sqrt{N}\big(\widehat{\DTT}_{gt}^{hs} - \DTT_{gt}\big) \rightsquigarrow \mathbb{H}_{gt}^{hs} \)  in $ \ell^\infty(\Y) $.
\end{enumerate}
The limiting processes $\mathbb{G}_{gt}$, $\mathbb{G}_{gt}^{hs}$, and $\mathbb{H}_{gt}^{hs}$ are tight, mean-zero Gaussian processes. Explicit expressions for their covariance functions are provided in the appendix.
\end{theorem}
\noindent A crucial component to the result in \Cref{Thm:FCLT_DF.DTT} is the application of the functional delta method to the counterfactual DF map. The continuous differentiability of $\Phi(\cdot)$ and $\Phi^{-1}(\cdot)$ plays a central role in this step, delivering uniform weak convergence and confidence bands for the treated and counterfactual DFs, and hence for DTT. Subsequent inference on QFs and QTT does not rely on Hadamard differentiability of the quantile map; instead, it uses the band-inversion arguments of \citet{chernozhukov2019generic}, which permits valid inference on QFs and QTT when the outcome distribution is non-continuous. \Cref{Sect:Sim_Est_Inference} in the appendix examines the small sample performance of the proposed distributional DiD estimator via simulations.

\subsection{Weighting Schemes}\label{Subsect:Weighting_DFs} 

Multiple period settings, such as staggered treatment adoption settings, potentially generate heterogeneity in treatment effects that may be interesting from a policy perspective. It may, however, be challenging to interpret heterogeneous treatment effects when there are several groups, post-treatment periods, or both. Therefore, it may be useful to report interesting convex-weighted effects. The following provides a brief treatment; \citet[Sect. 3]{callaway-santanna-2020} provides a detailed treatment.

Convex weighting is most naturally defined at the level of
$(g,t)$-indexed estimands. Let $\big\{\omega(g,t) \ : \ (g,t)\in(\mathcal{G}\setminus\{\infty\})\times[g:T] \big\}$
be non-negative weights satisfying $\displaystyle \sum_{g \in \mathcal{G}\setminus\{\infty\} }\sum_{t=g}^T \omega(g,t)=1$.
Define the convex-weighted treated DF, counterfactual DF, and DTT as
\begin{align*}
\F_{Y_\omega(1)}(y)
&:= \sum_{g\in\mathcal{G}\setminus\{\infty\}}\sum_{t=g}^{T}
\omega(g,t)\,\F_{Y_t(1)\mid G_g=1}(y), \\
\F_{Y_\omega(0)}(y)
&:= \sum_{g\in\mathcal{G}\setminus\{\infty\}}\sum_{t=g}^{T}
\omega(g,t)\,\F_{Y_t(0)\mid G_g=1}(y), \ \text{and} \\
\DTT_\omega(y)
&:= \sum_{g\in\mathcal{G}\setminus\{\infty\}}\sum_{t=g}^{T}
\omega(g,t)\Big(\F_{Y_t(1)\mid G_g=1}(y)-\F_{Y_t(0)\mid G_g=1}(y)\Big).
\end{align*}

Under the identification conditions of \Cref{Thm:Identif}, for each admissible $(g,t)$ the
counterfactual distribution $\F_{Y_t(0)\mid G_g=1}(y)$ is invariant to the
choice of $(h,s)$, in the sense that
\[
\F_{Y_t(0)\mid G_g=1}(y) = \F_{Y_t^{hs}(0)\mid G_g=1}(y)
\quad\text{for all admissible }(h,s),
\] where the superscript $hs$ emphasises the control group and pre-treatment period used for identification. In estimation, one may exploit the multiplicity of valid $(h,s)$ implementations for precision and specification testing. Next, let
$ \big\{\omega(g,t,h,s) \ : \ (g,t,h,s) \in \mathcal{I} \big\} $ collect non-negative weights satisfying
\( \displaystyle \sum_{h\in \mathcal{G}\setminus [1:t]}\sum_{s=-\T}^{g-1}\omega(g,t,h,s)=\omega(g,t) \) for each $(g,t)$. The weighting scheme, which satisfies \( \displaystyle \sum_{(g,t,h,s) \in \mathcal{I}} \omega(g,t,h,s) = 1 \), is known or estimable from data. The quantile treatment on the treated based on weighted DFs can then be estimated as \( \displaystyle \widehat{\QTT}_{\widehat{\omega}}(\tau) = \widehat{\F}_{Y_{\widehat{\omega}}(1)}^{\leftarrow}(\tau) - \widehat{\F}_{Y_{\widehat{\omega}}(0)}^{\leftarrow}(\tau). \) Observe that $\QTT_\omega(\tau)$ is not a convex-weighted QTT but rather QTT obtained from convex-weighted DFs.

For example, \( \displaystyle \omega(g,t,h,s) \propto \indicator{g=g'} \) assigns equal weights across post-treatment periods, pre-treatment periods, and controls for a given group $g' \in \mathcal{G}\setminus \{\infty\} $. Another example applies to $ e \in [0:(T-1)] $ periods after treatment is first received: \( \displaystyle \omega(g,t,h,s) \propto \indicator{t=g+e} \). Similarly, weights can be set proportional to fractions of the population used in identification, namely
\begin{align*}
    \omega(g,t,h,s) =& \frac{p_g\pi_tp_h\pi_s}{ \displaystyle \sum_{(g',t',h',s')\in\mathcal I}\left(p_{g'}\pi_{t'}p_{h'}\pi_{s'}\right)}; \\
    \omega(g,t,h,s) =& \frac{p_gp_h\Prob(S_t+S_s\geq 1)}{  \displaystyle \sum_{(g',t',h',s')\in\mathcal I} \left(p_{g'}p_{h'}\Prob(S_{t'}+S_{s'}\geq 1)\right)}; \\
    \omega_e(g,t,h,s) =& \indicator{t=g+e}\frac{p_g\pi_tp_h\pi_s}{\displaystyle \sum_{(g',t',h',s')\in\mathcal I_e}
\left(p_{g'}\pi_{t'}p_{h'}\pi_{s'}\right)
}; \ \text{and} \\
\omega_e(g,t,h,s) =& \indicator{t=g+e}\frac{p_gp_h\Prob(S_t+S_s\ge 1)}{\displaystyle \sum_{(g',t',h',s')\in\mathcal I_e} 
\left(p_{g'}p_{h'}\Prob(S_{t'}+S_{s'}\ge 1)\right)
}.
\end{align*}
The admissible index set $\mathcal{I}_e$ is defined by \( \displaystyle \mathcal I_e := \Big\{(g,t,h,s): g\in\mathcal G\setminus\{\infty\},\; t=g+e\le T,\; h\in (\mathcal{G}\setminus [1:t]),\; s\in[-\T:g-1] \Big\} \). These population-share weights give more influence to admissible comparisons that are based on larger treated and comparison-group shares and, when relevant, higher observation probabilities in the periods used for identification. The first and third specifications use period-specific observability rates, while the second and fourth use the probability that a unit is observed in at least one of the relevant periods. Thus, the weights are meant to reflect the amount of population information underlying each admissible $(g,t,h,s)$ comparison, rather than to impose any additional identifying restriction. 

In practice, such weights need to be estimated alongside the DFs from data:
\begin{align*}
    \widehat{\F}_{Y_{\widehat{\omega}}(1)}(y) =& \sum_{(g,t,h,s) \in \mathcal{I}} \widehat{\omega}(g,t,h,s) \widehat{\F}_{Y_t(1) \mid G_g=1}(y) \\
    \widehat{\F}_{Y_{\widehat{\omega}}(0)}(y) =& \sum_{(g,t,h,s) \in \mathcal{I}} \widehat{\omega}(g,t,h,s) \widehat{\F}_{Y_t^{hs}(0) \mid G_g=1}(y) \\
    \widehat\DTT_{\widehat{\omega}}(y) =& \sum_{(g,t,h,s) \in \mathcal{I}} \widehat{\omega}(g,t,h,s) \Big( \widehat{\F}_{Y_t(1) \mid G_g=1}(y) - \widehat{\F}_{Y_t^{hs}(0) \mid G_g=1}(y) \Big).
\end{align*}
\noindent To this end, consider the following regularity condition on the estimators of the weights; see, e.g., \citet{callaway-santanna-2020,callaway-tsyawo-2023treatment}.

\begin{assumption}\label{ass:weight_consistent}
For every quadruple $(g,t,h,s) \in \mathcal{I} $, jointly over the finite index set $\mathcal I$,
    \begin{align*}
        \sqrt{N}\big(\widehat{\omega}(g,t,h,s) - \omega(g,t,h,s)\big) = \frac{1}{\sqrt{N}}\sum_{j=1}^{N} \mathcal{W}_j(g,t,h,s) + o_p(1)
    \end{align*}
\end{assumption}
\noindent where $\E[\mathcal{W}_j(g,t,h,s)]=0$ and $\E\big[\mathcal{W}_j(g,t,h,s)^2\big]<\infty$. The following extends \Cref{Thm:FCLT_DF.DTT} to the convex-weighted DF, counterfactual DF and DTT.
\begin{corollary}\label{coro:FCLT_DF.DTT_Weighted}
Suppose \Cref{ass:NA,ass:PT,ass:Sampling_gen,Ass:reg_pi_p,ass:weight_consistent,ass:reg_Phi} hold, then:
\begin{enumerate}[(a)]
    \item \( \sqrt{N}(\widehat{\F}_{Y_{\widehat \omega}(1)} - \F_{Y_\omega(1)}) \rightsquigarrow \mathbb{G}_{\omega} \) in $ \ell^{\infty}(\Y) $;
    \item \( \sqrt{N}(\widehat{\F}_{Y_{\widehat \omega}(0)} - \F_{Y_\omega(0)}) \rightsquigarrow \mathbb{G}_{\omega}^C \) in $ \ell^{\infty}(\Y) $; \quad and
    \item \( \sqrt{N}(\widehat{\DTT}_{\widehat{\omega}} - \DTT_\omega) \rightsquigarrow \mathbb{H}_{\omega} \) in $ \ell^{\infty}(\Y) $.
\end{enumerate}
The limiting processes $\mathbb{G}_{\omega}$, $\mathbb{G}_{\omega}^C$, and $\mathbb{H}_{\omega}$ are tight Gaussian processes with mean $0$ defined on $\ell^\infty(\Y)$. Explicit expressions for their covariance functions are provided in the appendix. 
\end{corollary}

\subsection{Bootstrap Validity}

The bootstrap result is stated in a generic form because the estimators considered above share the same first-order structure. In what follows, $\mathscr F$ denotes any one of the treated DFs, counterfactual DFs, DTT processes, or their convex-weighted counterparts appearing in \Cref{Thm:FCLT_DF.DTT} and \Cref{coro:FCLT_DF.DTT_Weighted}, and $\widehat{\mathscr F}$ denotes the corresponding estimator. The common feature is that each estimator admits an asymptotically linear representation uniformly over the evaluation set $\Y$:
\[
\sqrt{N}\big(\widehat{\mathscr F} - \mathscr F\big)(y)
=
\frac{1}{\sqrt{N}}\sum_{j=1}^N
\Psi_{\mathscr F}(W_j;y)
+o_p(1).
\]
Bootstrap inference is conducted using an exchangeable bootstrap; see, e.g., \citet[Remark 5.1 and Condition EB]{chernozhukov-val-melly-2013}. Let $\{\xi_{Nj}\}_{j=1}^N$ be a non-negative exchangeable random vector, independent of the sample, and define
\( \displaystyle 
\bar\xi_N:=N^{-1}\sum_{j=1}^N\xi_{Nj}.
\)
Assume that, for some $\epsilon>0$,
\[
\sup_N\E\big[\xi_{N1}^{2+\epsilon}\big]<\infty,\qquad
N^{-1}\sum_{j=1}^N(\xi_{Nj}-\bar\xi_N)^2\overset{P}{\to}1,
\quad \text{ and } \quad
\bar\xi_N\overset{P}{\to}1.
\]
The bootstrap analogue of $\widehat{\mathscr F}$ then satisfies
\[
\sqrt{N}\big(\widehat{\mathscr F}^* - \widehat{\mathscr F}\big)(y)
=
\frac{1}{\sqrt{N}}\sum_{j=1}^N
(\xi_{Nj}-\bar\xi_N)\Psi_{\mathscr F}(W_j;y)
+o_p(1),
\quad \text{uniformly in } y\in\Y.
\]
This exchangeable-bootstrap framework covers the non-parametric bootstrap
through multinomial weights and accommodates Bayesian bootstrap-type
auxiliary variables.

The following theorem establishes the validity of the bootstrap for DFs, counterfactual DFs, and DTT. The notation $\mathbb{Z}_N^* \overset{P}{\underset{\xi}{\rightsquigarrow}} \mathbb{Z}$ denotes that, conditional on the data, the bootstrap law of $\mathbb{Z}_N^*$ (with randomness arising from the auxiliary variables $\xi$) converges weakly in probability to the law of the tight random element $\mathbb{Z}$. This generic notation is unrelated to the contrast process $\widehat{\bm Z}_{gt}$ used below for the over-identifying restrictions tests.

\begin{theorem}[Bootstrap Validity]\label{Thm:BootValid} 
    Suppose \Cref{Ass:reg_pi_p,ass:Sampling_gen,ass:PT,ass:NA,ass:weight_consistent,ass:reg_Phi} hold. Then, for every 
$(g,t,h,s)\in \mathcal I $,
    \begin{enumerate}[(a)]
        \item  $\sqrt{N}\big(\widehat{\F}_{Y_t(1) \mid G_g=1}^* - \widehat \F_{Y_t(1) \mid G_g=1}\big) \overset{P}{\underset{\xi}{\rightsquigarrow}} \mathbb{G}_{gt}$;
        \item $\sqrt{N}\big(\widehat{\F}_{Y_t^{hs}(0) \mid G_g=1}^* - \widehat \F_{Y_t^{hs}(0) \mid G_g=1}\big) \overset{P}{\underset{\xi}{\rightsquigarrow}} \mathbb{G}_{gt}^{hs}$;
        \item $\sqrt{N}\big(\widehat{\DTT}_{gt}^{hs*} - \widehat \DTT_{gt}^{hs}\big) \overset{P}{\underset{\xi}{\rightsquigarrow}} \mathbb{H}_{gt}^{hs}$; 
        \item $\sqrt{N}\big(\widehat{\F}_{Y_{\widehat{\omega}}(1)}^* - \widehat \F_{Y_{\widehat{\omega}}(1)}\big) \overset{P}{\underset{\xi}{\rightsquigarrow}} \mathbb{G}_{\omega}$;
        \item $\sqrt{N}\big(\widehat{\F}_{Y_{\widehat{\omega}}(0)}^* - \widehat \F_{Y_{\widehat{\omega}}(0)}\big) \overset{P}{\underset{\xi}{\rightsquigarrow}} \mathbb{G}_{\omega}^C$; \quad and
        \item  $\sqrt{N}\big(\widehat{\DTT}_{\widehat{\omega}}^* - \widehat \DTT_{\widehat{\omega}}\big) \overset{P}{\underset{\xi}{\rightsquigarrow}} \mathbb{H}_{\omega}$.
    \end{enumerate}
\end{theorem}

\noindent The proof of \Cref{Thm:BootValid} is provided in
\Cref{App:Sect_Proofs}. An immediate implication is that valid uniform confidence bands for distribution functions and DTTs can be constructed using bootstrap-based procedures, such as Algorithm~3 of \citet{chernozhukov-val-melly-2013}. In addition, uniformly valid confidence bands on QFs and QTTs can be constructed using Algorithm~1 of \citet{chernozhukov2019generic}. For completeness, the inference procedure on QFs and QTT is presented in \Cref{Sect:Uniform_Infer}.

\section{Testing Functional Over-identifying Restrictions}\label{Subsect:spec_test}

The identification result in this paper relies jointly on a distributional parallel trends assumption, a distributional no-anticipation assumption, and the specification of $\Phi(\cdot)$. This interplay between identification and functional specification naturally calls for a specification test. Moreover, when $\Phi$ is taken to be a proper cumulative distribution function—rather than the identity map—the testable implication exploited in \citet{kim-wooldridge-2024-difference} is no longer available. Nevertheless, staggered treatment assignment designs with multiple control groups, pre-treatment periods, or both, provide an alternative source of testable implications. In particular, they imply the existence of multiple valid representations of the \emph{same} counterfactual distribution, thereby enabling an identification test based on functional over-identifying restrictions; see \Cref{Thm:Identif}.

To fix ideas, consider a three-group two-period setting with two control groups. This provides two possible estimands of $ \F_{Y_t(0)\mid G_g=1}(y), y\in\Y $. Under the assumption that both control groups are valid, i.e., satisfy \Cref{ass:PT} for a given treated group $g$, both estimands ought to equal $ \F_{Y_t(0)\mid G_g=1}(y)$ for every $ y\in\Y $. An analogous argument applies to multiple pre-treatment periods. For the general case, fix $(g,t) \in \mathcal{G}\setminus \{\infty\} \times [g:T] $ and define the index set \( \displaystyle \mathcal I_{gt} := \Big\{(h,s): h\in \big(\mathcal{G}\setminus [1:t]\big) ,\; s\in[-\T:(g-1)] \Big\} \). For clarity, $ \F_{Y_t^{hs}(0)\mid G_g=1}(y) $ denotes the representation of the counterfactual distribution obtained using control group $h$ and pre-treatment period $s$. The test is available only when $K_{gt}:=|\mathcal I_{gt}|\ge 2$; if $K_{gt}<2$, there is no over-identifying restriction for the pair $(g,t)$. Under the maintained identifying assumptions, \Cref{Thm:Identif} implies that all admissible representations equal the same counterfactual distribution, and hence equal each other, over $\Y$.

The operational test hypotheses can then be stated as follows:
\begin{align*}
\mathbb H_0 &: \ \F_{Y_t^{hs}(0)\mid G_g=1}(y)=
\F_{Y_t^{h's'}(0)\mid G_g=1}(y)
\quad \forall \ (h,s),(h',s')\in \mathcal{I}_{gt}^2,\ y\in\Y,\\
\mathbb H_a &: \ \exists\, (h^\dagger,s^\dagger),(h',s')\in \mathcal{I}_{gt},\ y^\dagger\in\Y
\text{ s.t. } \F_{Y_t^{h^\dagger s^\dagger}(0)\mid G_g=1}(y^\dagger) \neq
\F_{Y_t^{h's'}(0)\mid G_g=1}(y^\dagger).
\end{align*}
\noindent Define
\( \displaystyle \widehat{\mathbf F}_{Y_t(0)\mid G_g=1}(y) := \big(
\widehat \F_{Y_t^{hs}(0)\mid G_g=1}(y) \big)_{(h,s)\in \mathcal{I}_{gt}}' \in\mathbb R^{K_{gt}} \). The Cram\'er--von Mises and Kolmogorov--Smirnov type test statistics are defined as
\begin{align*}
\widehat{\mathrm{CvM}}_{gt}
=&
N\sum_{\substack{(h,s,h',s') \in \mathcal{I}_{gt}^2 \\ (h,s)<(h',s')}}
\int_{\Y} \Big( \widehat \F_{Y_t^{hs}(0)\mid G_g=1}(y)
- \widehat \F_{Y_t^{h's'}(0)\mid G_g=1}(y) \Big)^2
\,\widehat H(dy) = \int_{\Y} \big\|\widehat{\bm Z}_{gt}(y)\big\|_2^2\,
\widehat H(dy),\\[1ex]
\widehat{\mathrm{KS}}_{gt}
=& 
\sqrt{N}\max_{\substack{(h,s,h',s') \in \mathcal{I}_{gt}^2 \\ (h,s)<(h',s')}}
\sup_{y\in\Y}
\Big|
\widehat \F_{Y_t^{hs}(0)\mid G_g=1}(y)
-
\widehat \F_{Y_t^{h's'}(0)\mid G_g=1}(y)
\Big| =
\sup_{y\in\Y}
\big\|\widehat{\bm Z}_{gt}(y)\big\|_{\infty},
\end{align*}
where ``$ < $" denotes any fixed total ordering on $\mathcal{I}_{gt}$,
\( \widehat{\bm Z}_{gt}(y) := \sqrt{N}\,\bm B_{gt}'\,\widehat{\mathbf F}_{Y_t(0)\mid G_g=1}(y),\) $\widehat H$ is a weighting measure over $\Y$, taken here to be the empirical CDF of all observations used in the construction of the test statistics, and $\bm B_{gt}\in\mathbb R^{ K_{gt} \times L_{gt} }, \ L_{gt}:= K_{gt}(K_{gt}-1)/2 $ is the oriented incidence matrix of the complete graph on $[K_{gt}]$.\footnote{Here, an oriented incidence matrix is simply a signed contrast matrix: each column is associated with one ordered pair of admissible representations and has one entry equal to $1$, one entry equal to $-1$, and all remaining entries equal to zero.} The fixed ordering identifies the $r$th element of $\mathcal I_{gt}$ with vertex $r\in[K_{gt}]$. Each column of $\bm B_{gt}$ corresponds to a pair of distinct elements $(h,s)$ and $(h',s'), \ (h,s)<(h',s')$, and encodes the oriented contrast $e_{hs}-e_{h's'}$ where $e_{hs} \in \mathbb{R}^{K_{gt}} $ is the standard basis vector whose entry corresponding to the $(h,s)$'th element equals $1$ with zeroes elsewhere. As special cases, the above tests nest (1) a \emph{functional} pre-trends test (using only pre-treatment periods and a single control group) and (2) a \emph{functional} DiD over-identifying restrictions test (using a single pre-treatment period and multiple control groups). 

Define the bootstrapped versions of the test statistics as
{\footnotesize
\begin{align*}
\widehat{\mathrm{CvM}}_{gt}^* =&
N\sum_{\substack{(h,s,h',s') \in \mathcal{I}_{gt}^2 \\ (h,s)<(h',s')}}
\int_{\Y}
\Big(
\big(\widehat \F_{Y_t^{hs}(0)\mid G_g=1}^*(y)-\widehat \F_{Y_t^{hs}(0)\mid G_g=1}(y)\big) 
-
\big(\widehat \F_{Y_t^{h's'}(0)\mid G_g=1}^*(y) - \widehat \F_{Y_t^{h's'}(0)\mid G_g=1}(y)\big)\Big)^2
\,\widehat H(dy) \\
=&:\int_{\Y} \big\|\widehat{\bm Z}_{gt}^*(y)\big\|_2^2\,
\widehat H(dy), \ \text{and} \\[1ex]
\widehat{\mathrm{KS}}_{gt}^*
=&
\sqrt{N}\max_{\substack{(h,s,h',s') \in \mathcal{I}_{gt}^2 \\ (h,s)<(h',s')}}
\sup_{y\in\Y}
\Big|
\big(\widehat \F_{Y_t^{hs}(0)\mid G_g=1}^*(y) - \widehat \F_{Y_t^{hs}(0)\mid G_g=1}(y)\big)
-
\big(\widehat \F_{Y_t^{h's'}(0)\mid G_g=1}^*(y) - \widehat \F_{Y_t^{h's'}(0)\mid G_g=1}(y) \big)
\Big|\\
=&: \sup_{y\in\Y}
\big\|\widehat{\bm Z}_{gt}^*(y)\big\|_{\infty}.
\end{align*}
}

\noindent In vector notation, the bootstrap contrast process is
\[
\widehat{\bm Z}_{gt}^*(y):=
\sqrt N\,\bm B_{gt}'\big(\widehat{\mathbf F}_{Y_t(0)\mid G_g=1}^*(y)
-\widehat{\mathbf F}_{Y_t(0)\mid G_g=1}(y)\big).
\]
Observe that the bootstrapped DFs in the bootstrapped statistics, unlike in the test statistics, are centred at their estimates. With $B$ bootstrap replications, let 
$\big\{\widehat{\mathrm{CvM}}_{gt,b}^* \big\}_{b=1}^B$ and 
$\big\{\widehat{\mathrm{KS}}_{gt,b}^* \big\}_{b=1}^B$ denote the corresponding bootstrap test statistics. 
The bootstrap p-values are computed as
\[
\widehat{pv}_{\mathrm{CvM}}
=
\frac{1+\sum_{b=1}^B \indicator{\widehat{\mathrm{CvM}}_{gt,b}^* \ge \widehat{\mathrm{CvM}}_{gt}}}{B+1} \quad \text{and} \quad
\widehat{pv}_{\mathrm{KS}}
=
\frac{1+\sum_{b=1}^B \indicator{\widehat{\mathrm{KS}}_{gt,b}^* \ge \widehat{\mathrm{KS}}_{gt}}}{B+1}.
\]
For a pre-specified statistic, the null is rejected at level $\alpha\in(0,1)$ whenever the corresponding p-value is no larger than $\alpha$. To examine the local power of the test, consider the following sequence of alternatives converging to the null at the $N^{-1/2}$-rate, where $F_{gt}^0$ denotes a common baseline distribution over $\Y$.
\begin{equation}\label{eqn:local_alt}
\begin{split}
\mathbb{H}_{an}:\quad 
F_{Y_t^{hs}(0)\mid G_g=1}(y)
&=
F_{gt}^0(y)
+ N^{-1/2}\,\varrho_{gt}^{hs}(y)
\quad \forall \, (h,s)\in \mathcal{I}_{gt},\; y\in\Y, \\
&\text{with } \sup_{(h,s)\in\mathcal{I}_{gt},\,y\in\Y}|\varrho_{gt}^{hs}(y)| < \infty, \\
&\text{and } \exists\,(h^\dagger,s^\dagger),(h',s')\in\mathcal{I}_{gt},\ y^\dagger\in\Y
\text{ such that } \varrho_{gt}^{h^\dagger s^\dagger}(y^\dagger)-\varrho_{gt}^{h's'}(y^\dagger) \neq 0.
\end{split}
\end{equation}

Like over-identifying restrictions tests more generally, the proposed test has no power against violations of the maintained identifying assumptions under which the over-identifying restrictions remain satisfied. In the present context, this occurs when
\begin{equation}\label{eqn:Ha_triv}
\begin{split}
  F_{Y_t^{hs}(0)\mid G_g=1}(y)
  =
  F_{Y_t^{h's'}(0)\mid G_g=1}(y)
  \quad \forall \, (h,s,h',s',y)\in \mathcal{I}_{gt}^2\times \Y,
\end{split}
\end{equation}
that is, either \Cref{ass:PT} or \Cref{ass:NA} is violated \emph{at at least} one pre-treatment period or for at least one control group, yet all admissible combinations $(h,s)$ yield the same counterfactual distribution. In such cases, the identifying restrictions are violated but observationally indistinguishable from the operational null, so that no test based solely on these restrictions can detect the violation; see \citet{guggenberger2012note}. This scenario requires misspecification to offset exactly across all admissible comparisons and is therefore a special case rather than the generic implication of a violation of the maintained assumptions. Consequently, whenever multiple pre-treatment periods or control groups are available, the test is expected to have non-trivial power against empirically relevant alternatives, making it a practically useful diagnostic for both identification and specification.

The following establishes the limiting distributions of the test statistics and the validity of the test. Define $\bm R_{gt}(y):= \big(\varrho_{gt}^{hs}(y)\big)_{(h,s)\in \mathcal{I}_{gt}}' $.

\begin{theorem}[Asymptotic Validity of the Functional Over-identifying Restrictions Tests]
\label{Thm:AsyBoot_CvM_KS}
Suppose $K_{gt}\ge 2$, \Cref{Ass:reg_pi_p,ass:Sampling_gen,ass:reg_Phi} hold, and the joint uniform asymptotic linear representations for the admissible counterfactual estimators hold. Then the following results obtain.

\begin{enumerate}[(a)]
\item Under $\mathbb H_0$, (i) \(\widehat{\bm Z}_{gt} \rightsquigarrow \bm Z_{gt}
\quad \text{in } \ell^\infty(\Y)^{L_{gt}},\) where $\bm Z_{gt}$ is a tight, mean-zero Gaussian process with covariance kernel characterised in the proof; (ii)
\(\displaystyle
\widehat{\mathrm{CvM}}_{gt} \xrightarrow{d} 
\int_{\Y}
\|\bm Z_{gt}(y)\|_2^2\, H(dy)\); and (iii) \( \displaystyle \widehat{\mathrm{KS}}_{gt} \xrightarrow{d} \sup_{y\in\Y}
\|\bm Z_{gt}(y)\|_\infty
\).

\item Under the sequence of local alternatives $\mathbb H_{an}$, \(
\widehat{\bm Z}_{gt} \rightsquigarrow \bm Z_{gt} + \bm \Delta_{gt}
\quad \text{in } \ell^\infty(\Y)^{L_{gt}},
\)
where $\bm \Delta_{gt}:\Y\to\mathbb R^{L_{gt}}$ is a deterministic drift
function defined by \(
\bm \Delta_{gt}(y) := \bm B_{gt}'\bm R_{gt}(y)
\) with \( \lVert \bm \Delta_{gt}(y^\dagger) \rVert_2 > 0 \) by the detectable-local-alternative condition in \eqref{eqn:local_alt}.

\item Under $\mathbb H_a$,
\(
\widehat{\mathrm{KS}}_{gt} \xrightarrow{p} \infty 
\). If, in addition, there exists a pair $(h,s)<(h',s')$ such that
\(\int_{\Y}\big(\F_{Y_t^{hs}(0)\mid G_g=1}(y)-\F_{Y_t^{h's'}(0)\mid G_g=1}(y)\big)^2H(dy)>0\),
then \( \widehat{\mathrm{CvM}}_{gt} \xrightarrow{p} \infty \).

\item Conditional on the data,
\( \widehat{\bm Z}_{gt}^* \overset{P}{\underset{\xi}{\rightsquigarrow}}
\bm Z_{gt} \quad \text{in } \ell^\infty(\Y)^{L_{gt}}, \)
and, consequently,
\( \displaystyle
\widehat{\mathrm{CvM}}_{gt}^*
\overset{P}{\underset{\xi}{\rightsquigarrow}}
\int_{\Y} \|\bm Z_{gt}(y)\|_2^2\, H(dy)\), and \( \displaystyle
\widehat{\mathrm{KS}}_{gt}^*
\overset{P}{\underset{\xi}{\rightsquigarrow}}
\sup_{y\in\Y} \|\bm Z_{gt}(y)\|_\infty.
\)
\end{enumerate}
\end{theorem}

\noindent The above results show that the tests are valid under the null, have non-trivial power under the sequence of local alternatives, and have power against fixed alternatives that generate non-zero admissible contrasts. Simulation studies in \Cref{Sect:Sim_Spec_Test} show that both variants of the test control size meaningfully and have non-trivial power against fixed alternatives.

\section{Empirical Application}\label{Sect:Empirical_Appl}

This section uses the proposed distributional DiD to estimate the DTT and QTT of increased police presence on crime, using data from \citet{diTella-2004}. The data cover car thefts over nine months, from April 1, 1994, to December 31, 1994. Car thefts are reported monthly in 876 blocks in three neighbourhoods in Buenos Aires, Argentina. This non-staggered setting, featuring multiple pre- and post-treatment periods, is a special case of the general staggered treatment adoption framework used in this paper.

On July 25, 1994, a ``treatment" was implemented, which involved providing 24-hour police protection to more than 270 Jewish and Muslim institutions in Buenos Aires following a terrorist attack on the main Jewish Center on July 18, 1994. Since treatment occurred in July and there is a seven-day time lag between the terrorist attack and treatment, the month of July is not used in the analyses partly to avoid anticipation bias. Thus, April through June constitute the pre-treatment months and August through December constitute the post-treatment months. This intervention affected 37 blocks, where a treated block contains a Jewish institution. The outcome variable is  the weekly average number of car thefts per block; it is discrete. 

\begin{figure}[!htbp]
\centering 
\caption{QTT/DTT, $\Phi$- Standard Normal}
\begin{subfigure}{0.49\textwidth}
\centering
\caption{DTT}
\includegraphics[width=1\textwidth]{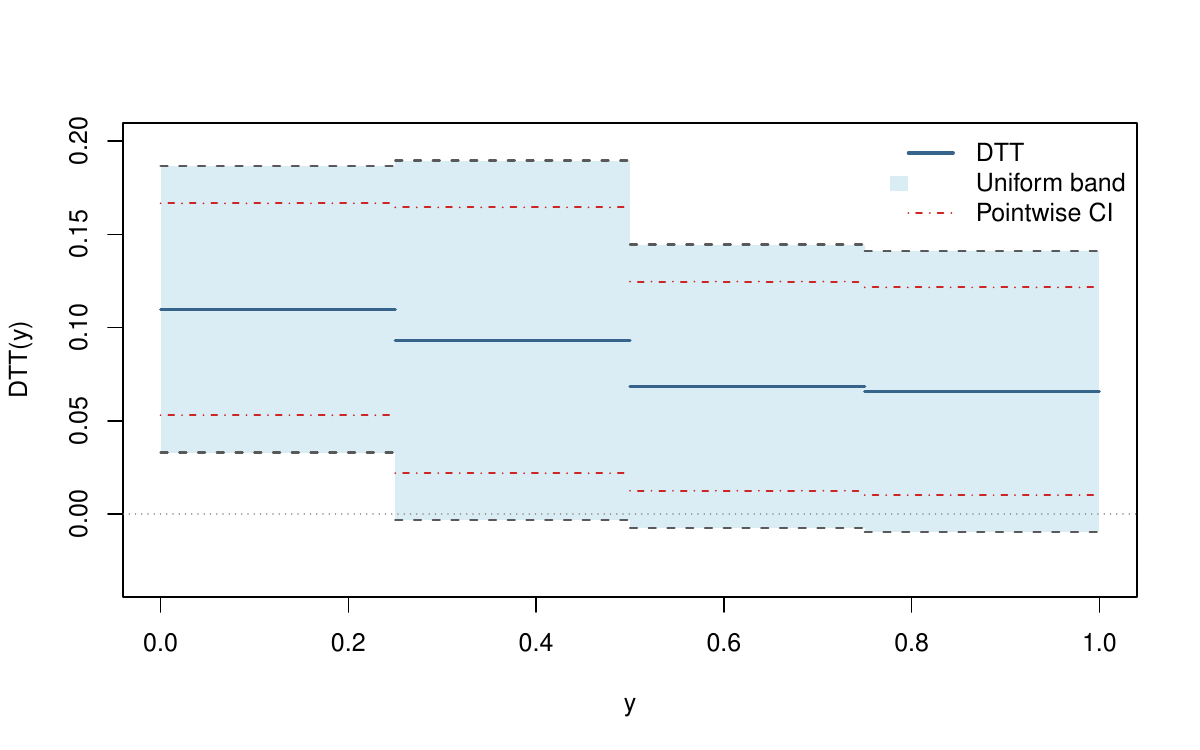}
\end{subfigure}
\begin{subfigure}{0.49\textwidth}
\centering
\caption{QTT}
\includegraphics[width=1\textwidth]{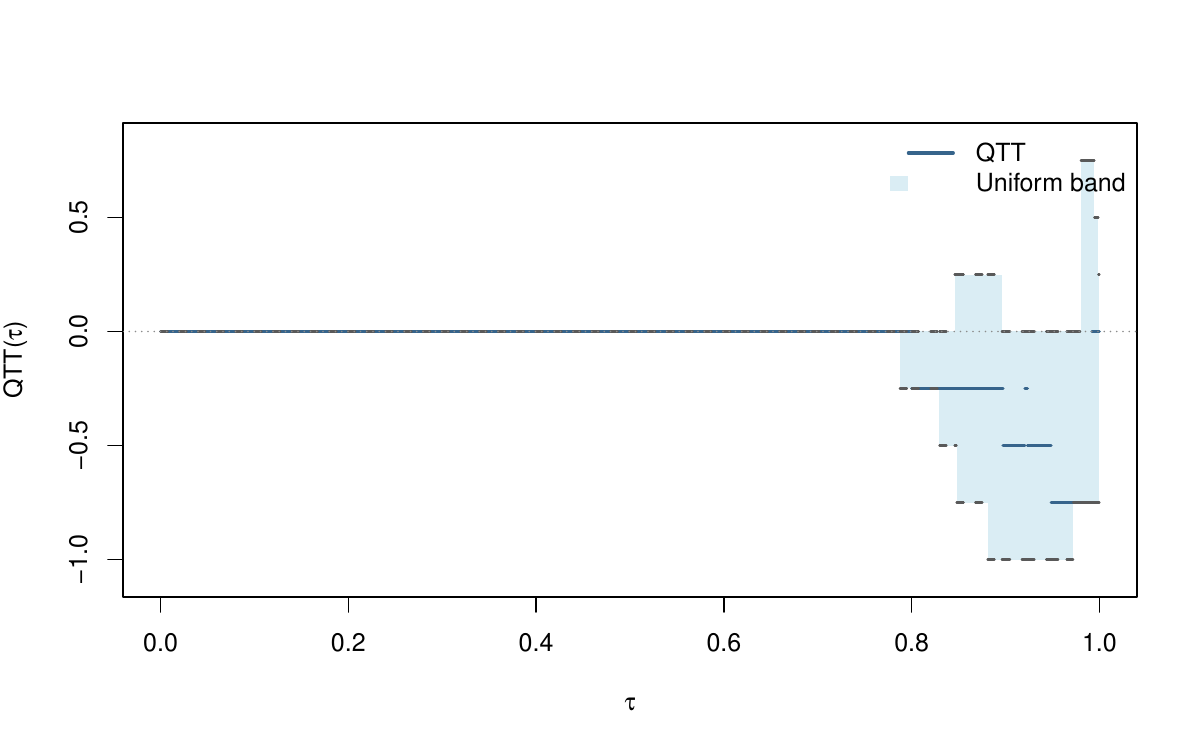}
\end{subfigure}

{\footnotesize
	\begin{justify}
		\textit{Notes:} DTT and QTT are computed from equally weighted DFs using all 15 pre-post period combinations in the data. 999 empirical bootstrap samples (at the block level) are used. For DTT, the shaded region is the $90\%$ uniform confidence band and the dot-dashed lines are $90\%$ pointwise confidence intervals based on the IQR-normalised bootstrap scale. QTT confidence bands are constructed from DF bands with joint $90\%$ coverage following \citet{chernozhukov2019generic}. The standard normal is the working CDF. The p-values of the CvM and KS over-identifying restrictions tests are $0.469$ and $0.559$, respectively.
	\end{justify}
}\label{Fig:DTT_QTT_normal}
\end{figure}

\begin{figure}[!htbp]
\centering 
\caption{QTT/DTT, $\Phi$- Standard Logistic}
\begin{subfigure}{0.49\textwidth}
\centering
\caption{DTT}
\includegraphics[width=1\textwidth]{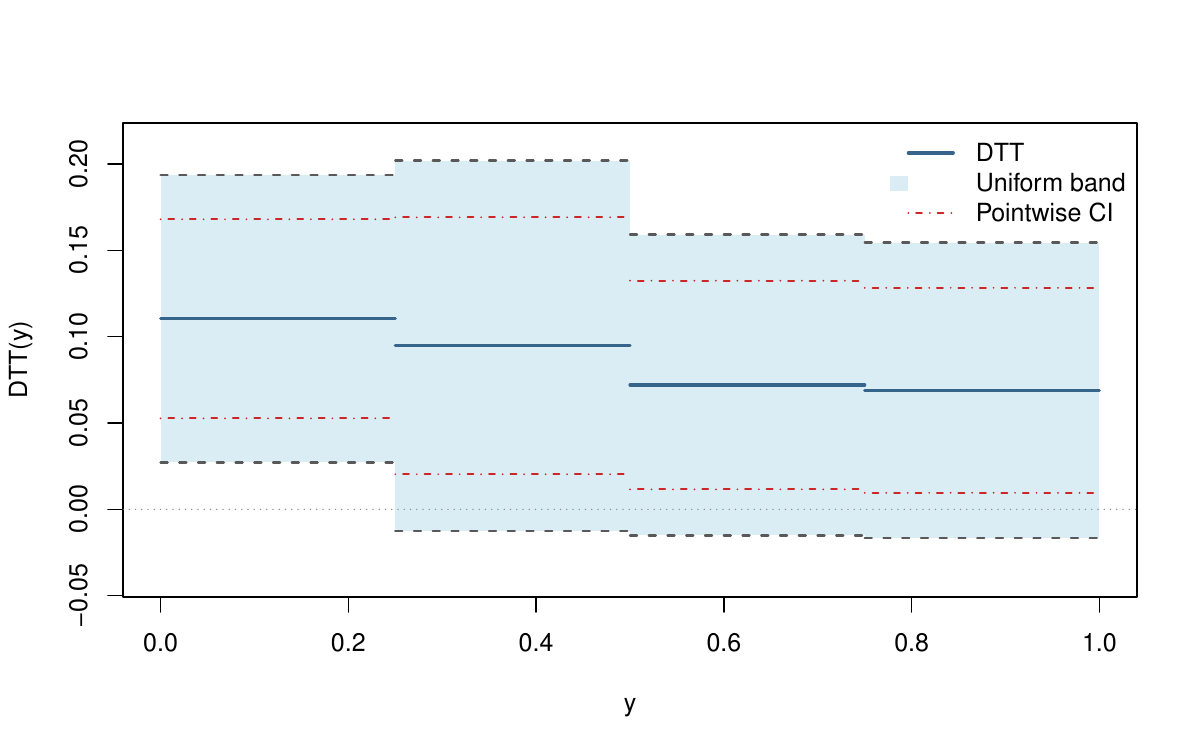}
\end{subfigure}
\begin{subfigure}{0.49\textwidth}
\centering
\caption{QTT}
\includegraphics[width=1\textwidth]{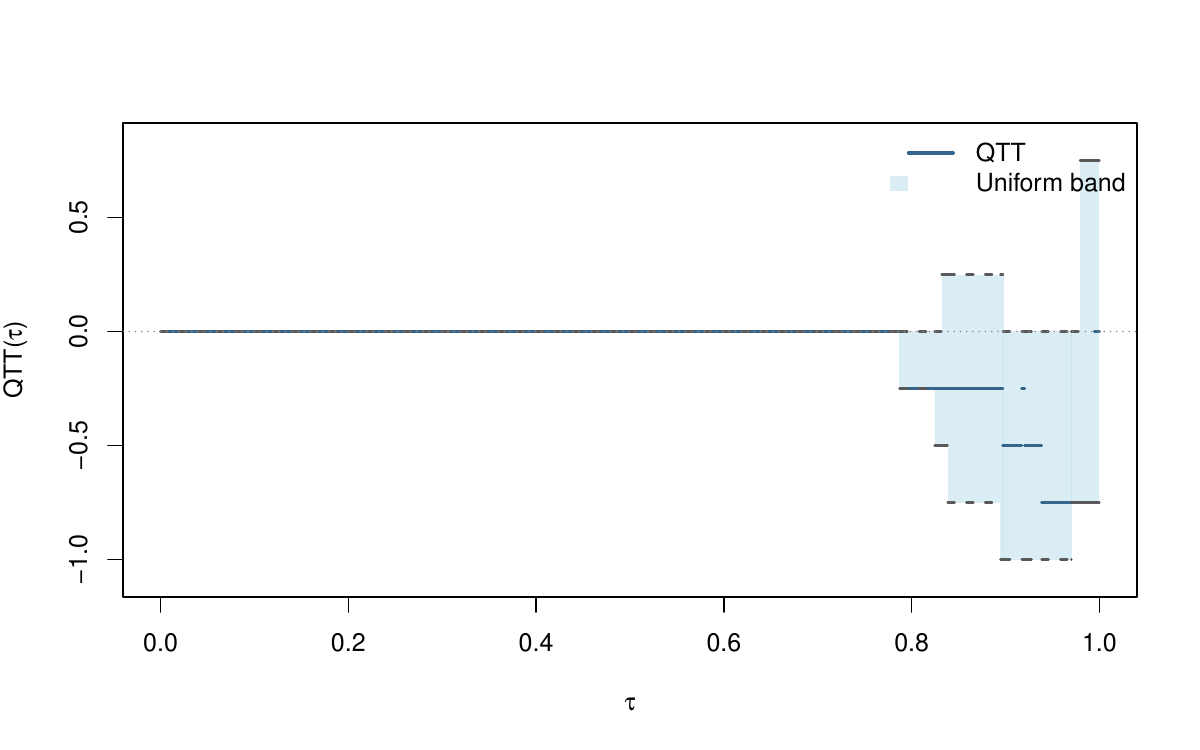}
\end{subfigure}

{\footnotesize
	\begin{justify}
		\textit{Notes:} DTT and QTT are computed from equally weighted DFs using all 15 pre-post period combinations in the data. 999 empirical bootstrap samples (at the block level) are used. For DTT, the shaded region is the $90\%$ uniform confidence band and the dot-dashed lines are $90\%$ pointwise confidence intervals based on the IQR-normalised bootstrap scale. QTT confidence bands are constructed from DF bands with joint $90\%$ coverage following \citet{chernozhukov2019generic}. The standard logistic is the working CDF. The p-values of the CvM and KS over-identifying restrictions tests are $0.465$ and $0.590$, respectively.
	\end{justify}
}\label{Fig:DTT_QTT_logis}
\end{figure}

\citet{diTella-2004} employed a standard linear Difference-in-Differences (DiD) model with time and block fixed effects -- see the paper for more details. The analysis below estimates the DTT and QTT of police presence on crime. There are 15 unique pre- and post-treatment period pairs. Following \Cref{Subsect:Weighting_DFs}, pair-specific DFs and DTTs are equally averaged to obtain the DTT and QTT shown in \Cref{Fig:DTT_QTT_normal,Fig:DTT_QTT_logis}. The preferred empirical specification uses the standard normal working CDF, given its stable finite-sample performance in the simulations in \Cref{App:Sect_Sim}. The standard logistic working CDF is reported as a robustness check, since it is also a smooth symmetric proper CDF but allows heavier tails. 999 empirical bootstrap samples at the block level, which preserve temporal dependence for each block, are used on the evaluation grid $\Y:=\{0.00,0.25,0.50,0.75,1.0\}$, a subset of the empirical support $\overline{\Y}$ of the outcome, to compute p-values of the over-identifying restrictions tests and the DF and DTT confidence bands.\footnote{About $1\%$ of outcomes in the sample exceed $1.0$.} The band-inversion technique of \citet{chernozhukov2019generic} is used to compute the confidence bands on QTT. 

\Cref{Fig:DTT_QTT_normal} presents the DTT and QTT estimates under the preferred standard normal working CDF, and \Cref{Fig:DTT_QTT_logis} reports the corresponding standard logistic robustness check. For both working CDF specifications, the CvM and KS over-identifying restrictions tests do not reject at conventional levels, with p-values around $0.46$--$0.59$; this provides no evidence against the maintained distributional restrictions used to construct the reported DTT and QTT estimates. The dotted horizontal line corresponds to the null effect. As can be observed, there is heterogeneity in both DTT and QTT. The uniform confidence bands indicate stronger evidence for effects on the distribution function than for effects on the quantile treatment effect. Under the usual two-sided alternative, the DTT estimates are statistically different from zero at the bottom of the evaluation grid, but not uniformly over all $y\in\Y$. With the standard normal and standard logistic working CDFs, the $90\%$ uniform band excludes zero at $y=0$, while it includes zero at $y\in \{0.25,0.5,0.75,1.0\}$. At $y=0$, this estimate has a direct event-probability interpretation: increased police presence raises the probability that a treated block is crime-free by about 11 percentage points, or equivalently lowers the probability of any theft by about 11 percentage points. This conclusion is based on the reported uniform bands and is therefore conservative relative to pointwise inference for the scalar parameter $\DTT(0)$. The pointwise confidence intervals, by contrast, exclude zero at all reported thresholds, suggesting threshold-by-threshold evidence of positive DTT effects throughout $\Y$. This pointwise evidence is informative, but it is interpreted as secondary to the simultaneous bands, which control coverage uniformly over $\Y$. In both cases, the point estimates of $\DTT$ are positive. Thus, one concludes that increased police presence does reduce the likelihood that car thefts in a given month exceed certain thresholds, with the strongest simultaneous evidence at the bottom of the evaluation grid. By contrast, for the QTT the simultaneous bands include zero for both working CDF specifications at all reported quantiles, so under the conventional two-sided criterion, one does not obtain statistically significant quantile effects.\footnote{Since the outcome is discrete, the estimated QF and the confidence bands obtained by inverting DF bands need not resemble the smooth bands familiar from continuous-outcome settings. In particular, flat regions and ties can make a QTT estimate coincide with a band endpoint at some quantiles; this is a feature of the inversion procedure with discrete support, not a numerical error.}

Because the economically relevant alternative is directional, the DTT evidence can also be read through the lens of stochastic dominance: $\DTT(y)\geq 0$ for all $y\in\Y$, with strict inequality for some $y\in\Y$. Under both the normal and logistic working CDFs, the DTT point estimates are positive at all reported grid points, and the lower uniform band is positive at $y=0$. This pattern is consistent with a shift toward lower crime counts over the reported evaluation grid. At the same time, because the bands include zero at higher thresholds and $\Y$ is a subset of $\overline{\Y}$, the evidence does not justify a claim of uniform strict dominance over the full outcome support.

For $\QTT$, the estimates are predominantly non-positive under both working CDFs, which is consistent with the directional hypothesis that the treatment weakly lowers quantiles of the untreated potential outcome distribution for the treated. However, the simultaneous bands still contain zero throughout, so the data do not permit a sharp rejection of the null of no quantile effect.\footnote{This is perhaps due to the marginal conservativeness of the QTT bands derived from the Minkowski difference of the QF bands --- see \citet[124 \& 129]{chernozhukov2019generic}.} Thus, the quantile evidence is best described as sign-consistent but not statistically decisive. Overall, the empirical results provide clearer support for a positive distribution effect in the DTT representation than for a strictly negative effect in the QTT representation, and this conclusion is robust across the normal and logistic CDF specifications.

\section{Conclusion}\label{Sect:Conclusion}

This paper proposes a distributional DiD strategy to identify, estimate, and conduct valid uniform inference on DTT and QTT in the presence of possibly non-continuous outcomes—namely, discrete, mixed, or continuous outcomes—in staggered treatment adoption settings. Identification of DTT and QTT is based on distributional extensions of the parallel trends and no-anticipation assumptions of \citet{wooldridge-2023-simple}. 

By modelling the counterfactual DF, rather than the counterfactual QF, the paper provides a pathway to valid inference under possibly non-continuously distributed outcomes. Specifically, it delivers (i) valid joint uniform confidence bands for DFs and (ii) confidence bands for QFs and QTT constructed via the band-inverting arguments of \citet{chernozhukov2019generic}. This approach avoids relying on Hadamard differentiability of the quantile map for non-continuously distributed outcomes, using instead valid DF bands and band inversion to conduct uniform inference on QFs and QTT.

The general sampling framework considered here encompasses balanced panels, unbalanced panels, rotating panels, and repeated cross-sections, thereby allowing the inference results to apply across a broad class of empirical designs. Further, the paper proposes and analyses a functional over-identifying restrictions test for DiD settings, providing a useful diagnostic for the over-identifying implications of the identifying assumptions and the specification of the working CDF.

The proposed framework is further complemented by a micro-founded random utility model of crime, which illustrates how discrete, count-valued potential outcomes can arise naturally from economic primitives. An empirical application to the distributional effects of police on crime reveals the strongest evidence of distributional effects at the bottom of the crime distribution. In contrast, the corresponding QTT estimates are sign-consistent but not statistically decisive.

\vspace{0.5cm}
\noindent \texttt{Supplementary appendix and replication files:} The supplementary appendix is available at \url{https://estsyawo.github.io/files/appendices/2026_Djuazon_Tsyawo_online_appendix.pdf}. The replication files are available on author E.S.T.'s website.

\vspace{0.5cm}
\noindent \texttt{Funding and competing interests:} The authors have no funding sources or competing interests to declare.

\vspace{0.5cm}
\noindent \texttt{Declaration of AI use:} The authors used OpenAI's ChatGPT/Codex to assist with manuscript editing, LaTeX formatting, and wording suggestions. The authors reviewed, revised, and take full responsibility for all content, analyses, and conclusions.

\printbibliography
\end{refsection}

\newpage
\setcounter{page}{1}
\begin{refsection}
\appendix
\renewcommand{\thetable}{S.\arabic{table}}
\renewcommand{\thefigure}{S.\arabic{figure}}
\renewcommand{\thesection}{S.\arabic{section}}
\setcounter{equation}{0}
\renewcommand{\theequation}{S.\arabic{equation}}

\begin{center}
    \Large{\bf Supplementary Material}
\end{center}
\begin{center}
    Nelly K. Djuazon  ~~~~~~~~Emmanuel Selorm Tsyawo
\end{center}

\section{Proofs of results in the main text}\label{App:Sect_Proofs}
\subsection{Proof of \Cref{Thm:Identif}}

First, by the strict monotonicity (invertibility) of $\Phi(\cdot)$ assumed in \Cref{ass:PT},
\[
\F_{Y_s(0)\mid G_g=0,G_g+G_h=1}(y)
=
\Phi\big(\alpha_s^h(y)\big)
\quad \Longleftrightarrow \quad
\alpha_s^h(y)=\Phi^{-1}\big(\F_{hs}(y)\big).
\]
Here and below, within the event $G_g+G_h=1$, the condition $G_g=0$ selects comparison group $h$. Hence $\F_{hs}(y)=\F_{Y_s(0)\mid G_g=0,G_g+G_h=1}(y)$ is identified from the data sampling  process, and $\Phi(\cdot)$ is a known strictly increasing function. $\alpha_s^h(y)$ is therefore identified for every $y\in\Y$.

Second, by \Cref{ass:PT}, $\F_{Y_s(0) \mid G_g=1,G_g+G_h=1}(y) = \Phi\big(\alpha_s^h(y) + \beta_s^{gh}(y)\big)$. Since $G_g=1$ selects the treated group in the pairwise comparison sample, $\F_{Y_s(0) \mid G_g=1,G_g+G_h=1}(y)=\F_{Y_s(0) \mid G_g=1}(y)$. $\F_{Y_s(0) \mid G_g=1}(y)$ is not identified from the data sampling process, but thanks to the no-anticipation (\Cref{ass:NA}), namely, $ \F_{Y_s(0) \mid G_g=1}(y) = \F_{Y_s(1) \mid G_g=1}(y) = \F_{gs}(y)$ for all $y\in\Y$. Thus, by the invertibility of $\Phi(\cdot)$ assumed in \Cref{ass:PT}, $\F_{gs}(y) = \Phi\big(\alpha_s^h(y) + \beta_s^{gh}(y)\big) \Longleftrightarrow \beta_s^{gh}(y) = \Phi^{-1}(\F_{gs}(y)) - \alpha_s^h(y)$. As $\alpha_s^h(y)$ is identified from the preceding step for every $y\in\Y$ and $\F_{gs}(y)$ is identified from the sampling process for every $y\in\Y$, $\beta_s^{gh}(y)$ is identified for every $y\in\Y$.

Third, it remains to identify $\gamma_{st}^h(y)$ for every $y\in\Y$. By the invertibility assumption in \Cref{ass:PT} for comparison group $h$,
\[
\F_{Y_t(0) \mid G_g=0,G_g+G_h=1}(y)
= \F_{ht}(y)
= \Phi(\alpha_s^h(y) + \gamma_{st}^h(y))
\quad \Longleftrightarrow \quad
\gamma_{st}^h(y) = \Phi^{-1}(\F_{ht}(y)) -\alpha_s^h(y).
\]
$\F_{ht}(y)$ is identified from the data sampling process for every $y\in\Y$, and $\alpha_s^h(y)$ is identified from the first step for every $y\in\Y$; $\gamma_{st}^h(y)$ is thus identified for every $y\in\Y$.

Putting the above three steps together, $ \big(\alpha_s^h(y),\beta_s^{gh}(y),\gamma_{st}^h(y)\big)', \ y\in\Y $ is identified for $\Phi(\cdot)$ assumed known in \Cref{ass:PT}. Moreover, $\F_{Y_t(0) \mid G_g=1}(y)=\F_{Y_t(0) \mid G_g=1,G_g+G_h=1}(y)$ because $G_g=1$ selects the treated group in the pairwise comparison sample. Therefore, $\F_{Y_t(0) \mid G_g=1}(y)$ is identified:
\begin{align*}
    \F_{Y_t(0) \mid G_g=1}(y) =& \Phi\big(\alpha_s^h(y) + \beta_s^{gh}(y) + \gamma_{st}^h(y)\big) \\
    =& \Phi\Big(\Phi^{-1}\big(\F_{gs}(y)\big) + \Phi^{-1}\big(\F_{ht}(y)\big) - \Phi^{-1}\big(\F_{hs}(y)\big) \Big).
\end{align*}

\qed

\subsection{An empirical process building block}

For any $ (g,t) \in \mathcal{G}\times [-\T:T] $, define the empirical process
\[
\widetilde{H}_{gt}(y)
:=
\sqrt{N}\big(\widehat{\F}_{gt} - \F_{gt} \big)(y), \, y\in\Y.
\]
Empirical processes of the form $\widetilde{H}_{gt}(y)$ constitute the fundamental building blocks of the asymptotic theory in this paper. The following lemma states a standard weak convergence result for such processes.

\begin{lemma}\label{Lem:Donsk_Hdt}
Under \Cref{ass:Sampling_gen,Ass:reg_pi_p}, the process $\widetilde{H}_{gt}$ converges weakly to a tight Gaussian process,
\[
\widetilde{H}_{gt} \;\rightsquigarrow\; \mathbb{G}_{gt}
\quad \text{in } \ell^\infty(\Y),
\]
for every $(g,t) \in \mathcal{G} \times [-\T:T] $. The limiting process $\mathbb{G}_{gt}$ has mean zero and covariance function $\Omega_{gt}(y_1,y_2)$ defined on $\Y\times\Y$. An explicit expression for the covariance function is provided in the proof.
\end{lemma}

\textbf{Proof:}
First, obtain the asymptotically linear expression of $\widetilde{H}_{gt}(y)$. 

Recall
\[
\F_{gt}(y):=\Prob(Y_t\le y\mid G=g).
\]
Since \(S_t\independent (Y_t,G)\) under \Cref{ass:Sampling_gen}, one has
\[
\Prob(Y_t\le y\mid G=g)
=
\Prob(Y_t\le y\mid G=g,S_t=1),
\] which is directly identified from the data sampling process. Thus, under the independence condition in \Cref{ass:Sampling_gen} and the regularity conditions of \Cref{Ass:reg_pi_p},
\[
\F_{gt}(y)
=
\frac{\E\big[S_t\indicator{G=g}\indicator{Y_t\le y}\big]}
{\E\big[S_t\indicator{G=g}\big]}
=
\frac{1}{\pi_tp_g}\E\big[S_t\indicator{G=g}\indicator{Y_t\le y}\big].
\]

\noindent The natural sample analogue of $\F_{gt}$ has the convenient expression
\[ 
\widehat{\F}_{gt}(y):= \frac{1}{\hat{\pi}_t\hat{p}_g}\frac{1}{N}\sum_{j=1}^{N}S_{jt} \indicator{G_j=g} \indicator{Y_{jt}\leq y}
\] using observations in period $t \in [-\T:T] $. Recall \( \displaystyle \hat{\pi}_t = N^{-1} \sum_{j=1}^N S_{jt} \) and \( \hat{p}_g = N^{-1} \sum_{j=1}^N \indicator{G_j = g} \). $ \pi_t = \E[\hat{\pi}_t] $ and $p_g=\E[\hat{p}_g]$ under the $i.i.d.$ sampling condition of \Cref{ass:Sampling_gen}.

Then,
\begin{align*}
    \widehat{\F}_{gt}(y)- \F_{gt}(y)
    &= \frac{1}{N}\sum_{j=1}^N \Big\{ \frac{1}{\hat{\pi}_t\hat{p}_g}S_{jt}\indicator{G_j=g} \indicator{Y_{jt}\leq y} - \frac{1}{\pi_tp_g}\E\big[S_t\indicator{G=g}\indicator{Y_t\leq y}\big] \Big\}.
\end{align*}The following decomposition holds:
\begin{align*}
\sqrt{N}\big(\widehat{\F}_{gt} - \F_{gt}\big)(y) 
=& \frac{1}{\pi_t p_g \sqrt{N}}\sum_{j=1}^N 
\Big(
S_{jt}\indicator{G_j=g}\indicator{Y_{jt}\le y}
-
\E\big[S_t\indicator{G=g}\indicator{Y_t\le y}\big]
\Big) \\
&\quad
- \frac{\F_{gt}(y)}{\pi_t\sqrt{N}}\sum_{j=1}^N \big(S_{jt}-\pi_t\big)
- \frac{\F_{gt}(y)}{p_g\sqrt{N}}\sum_{j=1}^N \big(\indicator{G_j=g}-p_g\big)
+ \widehat{\mathcal R}_{gt}(y)
\end{align*}where the remainder term is given by
\begin{align*}
\widehat{\mathcal R}_{gt}(y)
:=&\;
-\frac{\sqrt N(\hat \pi_t\hat p_g - \pi_tp_g)}{\hat\pi_t\hat p_g}
\cdot \frac{1}{\pi_tp_g}\frac{1}{N}\sum_{j=1}^N
\Big(
S_{jt}\indicator{G_j=g}\indicator{Y_{jt}\le y}
-
\E\big[S_t\indicator{G=g}\indicator{Y_t\le y}\big]
\Big) \\
&\quad
+
\frac{\F_{gt}(y)}{\pi_tp_g}\sqrt N\frac{(\hat\pi_t\hat p_g - \pi_tp_g)^2}{\hat\pi_t\hat p_g} - \frac{\F_{gt}(y)}{\pi_tp_g}\sqrt N(\hat\pi_t-\pi_t)(\hat p_g - p_g)
=:\widehat{\mathcal R}_{gt}^{(1)}(y) + \widehat{\mathcal R}_{gt}^{(2)}(y) + \widehat{\mathcal R}_{gt}^{(3)}(y).
\end{align*}

It is shown that \( \displaystyle \sup_{y\in\Y}|\widehat{\mathcal R}_{gt}(y)|=o_p(1)\). First, by \Cref{ass:Sampling_gen},
\(
\hat p_g=\frac{1}{N}\sum_{j=1}^N \indicator{G_j=g}
\)
is the sample mean of \(i.i.d.\) Bernoulli random variables with mean \(p_g\). Hence
\(
\hat p_g\xrightarrow{p} p_g,
\text{ and }
\sqrt N(\hat p_g-p_g)=\mathcal{O}_p(1).
\)
Since \(p_g\in(0,1)\) by \Cref{Ass:reg_pi_p}, it follows that \(1/\hat p_g=\mathcal{O}_p(1)\). Likewise, since
\(
\hat\pi_t=\frac{1}{N}\sum_{j=1}^N S_{jt}
\)
and \(\pi_t\in(0,1]\), one has
\(
\hat \pi_t \xrightarrow{p} \pi_t, \sqrt N(\hat\pi_t-\pi_t)=\mathcal{O}_p(1), \text{ and } 1/\hat\pi_t=\mathcal{O}_p(1).
\)
Therefore, 
\(
\hat\pi_t\hat p_g-\pi_t p_g
=
\pi_t(\hat p_g-p_g)+p_g(\hat\pi_t-\pi_t)+(\hat\pi_t-\pi_t)(\hat p_g-p_g)
=
\mathcal{O}_p(N^{-1/2}),
\)
and so
\begin{equation}\label{eqn:pi_hat_phat}
    \frac{\sqrt N(\hat\pi_t\hat p_g-\pi_t p_g)}{\hat\pi_t\hat p_g} = \mathcal{O}_p(1).
\end{equation}

Next, define
\(
f_y(W_j):=
S_{jt}\indicator{G_j=g}\indicator{Y_{jt}\le y},
\quad y\in\Y.
\)
The class \(\{f_y:y\in\Y\}\) is uniformly bounded and VC-subgraph. Hence, by the Glivenko--Cantelli theorem,
\[
\sup_{y\in\Y}
\left|
\frac{1}{N}\sum_{j=1}^N
\Big(
f_y(W_j)-\E[f_y(W_j)]
\Big)
\right|
=o_p(1).
\]
Therefore,
\begin{align*}
\sup_{y\in\Y}|\widehat{\mathcal R}_{gt}^{(1)}(y)|
&\le
\left|
\frac{\sqrt N(\hat\pi_t\hat p_g-\pi_t p_g)}{\hat\pi_t\hat p_g}
\right|
\frac{1}{\pi_t p_g}
\sup_{y\in\Y}
\left|
\frac{1}{N}\sum_{j=1}^N
\Big(
f_y(W_j)-\E[f_y(W_j)]
\Big)
\right| \\
&=
\mathcal{O}_p(1)\cdot o_p(1)
=
o_p(1).
\end{align*}

Next, since \(0\le \F_{gt}(y)\le 1\),
\( \displaystyle 
\sup_{y\in\Y}|\widehat{\mathcal R}_{gt}^{(2)}(y)|
\le
\frac{1}{\pi_t p_g}
\left|
\frac{1}{\hat\pi_t\hat p_g}
\right|
\sqrt N\big(\hat\pi_t\hat p_g-\pi_t p_g\big)^2.
\)
Because
\(
\hat\pi_t\hat p_g-\pi_t p_g=\mathcal{O}_p(N^{-1/2})
\)
and \(1/(\hat\pi_t\hat p_g)=\mathcal{O}_p(1)\), it follows that
\( \displaystyle
\sup_{y\in\Y}|\widehat{\mathcal R}_{gt}^{(2)}(y)|
=
\mathcal{O}_p(N^{-1/2})
=
o_p(1).
\)

Finally, again using \(0\le \F_{gt}(y)\le 1\),
\( \displaystyle
\sup_{y\in\Y}|\widehat{\mathcal R}_{gt}^{(3)}(y)|
\le
\frac{1}{\pi_t p_g}\sqrt N\,|\hat\pi_t-\pi_t|\,|\hat p_g-p_g|.
\)
Since
\(
\hat\pi_t-\pi_t=\mathcal{O}_p(N^{-1/2})
\text{ and } \hat p_g-p_g=\mathcal{O}_p(N^{-1/2}),
\)
one obtains
\( \displaystyle
\sup_{y\in\Y}|\widehat{\mathcal R}_{gt}^{(3)}(y)|
=
\mathcal{O}_p(N^{-1/2})
=
o_p(1).
\)

Combining the three bounds via the triangle inequality yields
\[
\sup_{y\in\Y}|\widehat{\mathcal R}_{gt}(y)|
\le
\sup_{y\in\Y}|\widehat{\mathcal R}_{gt}^{(1)}(y)|
+
\sup_{y\in\Y}|\widehat{\mathcal R}_{gt}^{(2)}(y)|
+
\sup_{y\in\Y}|\widehat{\mathcal R}_{gt}^{(3)}(y)|
=
o_p(1).
\]

From the foregoing,
\begin{align}\label{eqn:exp_Hdt_n}
\sqrt{N}\big(\widehat{\F}_{gt} - \F_{gt}\big)(y) 
=& \frac{1}{\sqrt{N}}\sum_{j=1}^N \psi_{gt}(W_j;y) + o_p(1),
\end{align}
where
\begin{align*}
   \psi_{gt}(W_j;y)
   =&\Big(\frac{S_{jt}\indicator{G_j=g}\indicator{Y_{jt}\le y}}{\pi_t p_g}
   - \F_{gt}(y) \Big) 
   - \frac{\F_{gt}(y)}{\pi_t}\big(S_{jt}-\pi_t\big)
   - \frac{\F_{gt}(y)}{p_g}\big(\indicator{G_j=g}-p_g\big),
\end{align*}
since
\(
\E\big[S_t\indicator{G=g}\indicator{Y_t\le y}\big]
=
\pi_t p_g \F_{gt}(y)
\)
under \Cref{ass:Sampling_gen}.

Consider the function class
\(
\mathcal{Z}_{gt}
:=
\big\{
\psi_{gt}(\cdot;y):y\in\Y
\big\}.
\)
Recall \(Y_t\) denotes the generic observed outcome at period \(t\). The class \(
\{\indicator{Y_t\le y}: y\in\Y\}
\)
is VC-subgraph \citep[Example~2.5.4]{vaart-wellner-1996}. Since \(S_t\indicator{G=g}\) is a bounded measurable multiplier, it follows that
\(
\big\{
S_t\indicator{G=g}\indicator{Y_t\le y}:y\in\Y
\big\}
\)
is again VC-subgraph, hence \(P\)-Donsker. Moreover, \(\F_{gt}(y)\) is a bounded deterministic function of \(y\), and \(S_t-\pi_t\) and \(\indicator{G=g}-p_g\) are bounded random variables not indexed by \(y\). Therefore, by the preservation of Donsker classes under centring, bounded multipliers, and finite linear combinations, \(\mathcal{Z}_{gt}\) is \(P\)-Donsker; see, e.g., \citet[Sections~2.6.5 and ~2.10]{vaart-wellner-1996}.

Hence, the \(P\)-Donsker property of \(\mathcal Z_{gt}\) implies that the empirical process
\[
\left\{
\frac{1}{\sqrt N}\sum_{j=1}^N \psi_{gt}(W_j;y): y\in\Y
\right\}
\]
is asymptotically equicontinuous and converges weakly in \(\ell^\infty(\Y)\); see \citet[Theorem~2.1]{kosorok-2007}. In particular, for any finite collection
\(y_1,\ldots,y_L\in\Y\), the vector
\[
\left(
\frac{1}{\sqrt N}\sum_{j=1}^N \psi_{gt}(W_j;y_\ell)
\right)_{\ell=1}^L
\]
satisfies a multivariate central limit theorem by \Cref{ass:Sampling_gen} and the boundedness of \(\psi_{gt}(W_j;y)\), and the Donsker property supplies the required asymptotic equicontinuity. Therefore,
\( \displaystyle 
\sqrt N\big(\widehat \F_{gt}-\F_{gt}\big)
\rightsquigarrow
\mathbb{G}_{gt} \quad \text{ in }\ell^\infty(\Y),
\)
for a tight mean-zero Gaussian process \(\mathbb{G}_{gt}\). Its covariance function is
\[
\Omega_{gt}(y_1,y_2)
:=
\E\big[\psi_{gt}(W;y_1)\psi_{gt}(W;y_2)\big],
\qquad (y_1,y_2)\in\Y\times\Y.
\]

\qed

\subsection{Proof of \Cref{Thm:FCLT_DF.DTT}}

\paragraph{Part (a):} This part follows directly from \Cref{Lem:Donsk_Hdt} as $\F_{Y_t(1) \mid G_g=1}(y)$ is identified from the data sampling process: $\F_{Y_t(1) \mid G_g=1}(y)=\F_{gt}(y)$.

\paragraph{Part (b):} Recall the estimator of the counterfactual distribution $\widehat{\F}_{Y_t(0) \mid G_g=1}(y) = \Phi\big(\widehat{\alpha}_s^h(y) + \widehat{\beta}_s^{gh}(y) + \widehat{\gamma}_{st}^h(y)\big)$. From the identification result, namely \Cref{Thm:Identif},
\[
\F_{Y_t(0) \mid G_g=1}(y) = \Phi\Big(\Phi^{-1}(\F_{gs}(y)) + \Phi^{-1}(\F_{ht}(y)) - \Phi^{-1}(\F_{hs}(y)) \Big). 
\] Define the function
\[
    h_{CF}(a) := \Phi\big(\Phi^{-1}(a_1) + \Phi^{-1}(a_2) - \Phi^{-1}(a_3) \big)
\]
\noindent where $a:=(a_1,a_2,a_3)'$. By the continuous differentiability and dominance conditions on $\Phi(\cdot)$ and $\Phi^{-1}(\cdot)$ in \Cref{ass:reg_Phi}, the induced composition map from $\ell^\infty(\Y)^3$ to $\ell^\infty(\Y)$ is Hadamard differentiable and the functional delta method (\citet[Examples 3.9.2, Lemma 3.9.3, and Theorem 3.9.4]{vaart-wellner-1996}) applies uniformly over $y\in\Y$:
 \begin{align*}
     \sqrt{N}\big(\widehat{\F}_{Y_t(0) \mid G_g=1} - \F_{Y_t(0) \mid G_g=1}\big)(y)
     =&\nabla_{h_{CF}}(\F_{gs}(y),\F_{ht}(y),\F_{hs}(y))\\ 
     &\times \sqrt{N}\big((\widehat{\F}_{gs} - \F_{gs})(y), (\widehat{\F}_{ht} - \F_{ht})(y), (\widehat{\F}_{hs} - \F_{hs})(y)\big) + o_p(1) \\ 
     =& \nabla_{h_{CF}}\big(\F_{gs}(y), \F_{ht}(y), \F_{hs}(y)\big) \times \big(\widetilde{H}_{gs}(y), \widetilde{H}_{ht}(y), \widetilde{H}_{hs}(y)\big)' +o_p(1)\\
     =& \nabla_{gs}(y)\widetilde{H}_{gs}(y) + \nabla_{ht}(y)\widetilde{H}_{ht}(y) - \nabla_{hs}(y)\widetilde{H}_{hs}(y) + o_p(1)
 \end{align*}

\noindent uniformly in $y\in\Y$ where 
 \begin{align*}
     \nabla_{h_{CF}}\big(\F_{gs}(y),\F_{ht}(y),\F_{hs}(y)\big) =& \phi\Big( \Phi^{-1}\big(\F_{gs}(y)\big) + \Phi^{-1}\big(\F_{ht}(y)\big) - \Phi^{-1}\big(\F_{hs}(y)\big) \Big)\\
     &\times
 \bigg(\frac{1}{\phi\big(\Phi^{-1}(\F_{gs}(y))\big)}, \ \frac{1}{\phi\big(\Phi^{-1}(\F_{ht}(y))\big)},\ -\frac{1}{\phi\big(\Phi^{-1}(\F_{hs}(y))\big)}\bigg)\\
 =&:\big(\nabla_{gs}(y),\nabla_{ht}(y),-\nabla_{hs}(y)\big)
 \end{align*}
 \noindent is the gradient row vector of $h_{CF}$ and $\phi(b):=d\Phi(b)/db$. 
 
 Under the regularity conditions of \Cref{ass:reg_Phi}, \( \displaystyle \sup_{y\in\Y} \max \big\{\nabla_{gs}(y),\nabla_{ht}(y),\nabla_{hs}(y)\big\} \leq C^2 < \infty  \).  In addition to \eqref{eqn:exp_Hdt_n}, one has
\begin{align}\label{eqn:exp_CF}
    \sqrt{N}&\big(\widehat{\F}_{Y_t^{hs}(0) \mid G_g=1} - \F_{Y_t(0) \mid G_g=1}\big)(y) \nonumber \\
    =& \frac{1}{\sqrt{N}}\sum_{j=1}^{N}\Big\{ \nabla_{gs}(y)\psi_{gs}(W_j;y) + \nabla_{ht}(y)\psi_{ht}(W_j;y) - \nabla_{hs}(y)\psi_{hs}(W_j;y) \Big\} + o_p(1) \nonumber \\
    =&: \frac{1}{\sqrt{N}}\sum_{j=1}^{N} \Psi_{gt}^{hs}(W_j;y) + o_p(1)
\end{align} with \( \displaystyle \E\big[\Psi_{gt}^{hs}(W;y)\big] = 0 \ \forall y \in\Y \) and $ \displaystyle \sup_{y\in\Y} \E |\Psi_{gt}^{hs}(W;y)|^2 < \infty $ under the conditions of \Cref{Lem:Donsk_Hdt} and \Cref{ass:reg_Phi}. This implies that under the sampling conditions of \Cref{ass:Sampling_gen}, the Multivariate Lindeberg--L\'evy Central Limit Theorem applies:  
\[
\Big(\frac{1}{\sqrt{N}}\sum_{j=1}^N \Psi_{gt}^{hs}(W_j;y_1), \ldots, \frac{1}{\sqrt{N}}\sum_{j=1}^N \Psi_{gt}^{hs}(W_j;y_L)\Big)'
\]
converges in distribution to the multivariate normal with the $(l,l')$'th entry of the covariance matrix: $\E\big[\Psi_{gt}^{hs}(W;y_l)\Psi_{gt}^{hs}(W;y_{l'})\big]$ for every finite collection $ \big\{y_l \ : \ 1 \leq l \leq L \big\} \subset \Y$. 

Consider the indexing class
\[
\mathcal{Z}_{gt}^{hs}
:=
\big\{\Psi_{gt}^{hs}(\cdot;y):y\in\Y\big\}.
\]
By Lemma~\ref{Lem:Donsk_Hdt}, for each $(g,t) \in \mathcal{G}\times [-\T:T] $, the class
$\{\psi_{gt}(\cdot;y):y\in\Y\}$ is $P$-Donsker. Under the boundedness
conditions in \Cref{Ass:reg_pi_p} and \Cref{ass:reg_Phi} (in particular, $p_g, \pi_t$ bounded away from zero)
and $\sup_{y\in\Y}|\nabla_{gs}(y)|+\sup_{y\in\Y}|\nabla_{ht}(y)|+\sup_{y\in\Y}|\nabla_{hs}(y)|<\infty$,
it follows that $\mathcal{Z}_{gt}^{hs}$ is obtained from a finite linear combination of $P$-Donsker classes with bounded multipliers. Hence $\mathcal{Z}_{gt}^{hs}$ is $P$-Donsker and the corresponding
empirical process
\[
\left\{\frac{1}{\sqrt{N}}\sum_{j=1}^{N}\Psi_{gt}^{hs}(W_j;y):y\in\mathcal{Y}\right\}
\]
is asymptotically equicontinuous and converges weakly in $\ell^\infty(\Y)$.

Combining the foregoing yields the weak convergence result in $\ell^\infty(\Y)$ \citep[Theorem 2.1]{kosorok-2007}. Therefore, the empirical process $\sqrt{N}\big( \widehat{\F}_{Y_t^{hs}(0) \mid G_g=1} - \F_{Y_t(0) \mid G_g=1} \big)$ converges to a tight mean-zero Gaussian process, namely $\mathbb{G}_{gt}^{hs}$. Under \Cref{ass:Sampling_gen}, the covariance function of $\mathbb{G}_{gt}^{hs}$ is given by $\Omega_{gt}^{hs}(y_1,y_2):=\E\big[\Psi_{gt}^{hs}(W;y_1)\Psi_{gt}^{hs}(W;y_2)\big]$.

\paragraph{Part (c):} From parts (a) and (b) above, in addition to
 \eqref{eqn:exp_Hdt_n} and \eqref{eqn:exp_CF},
\begin{align*}
    \sqrt{N}\big( \widehat{\DTT}_{gt}^{hs} - \DTT_{gt} \big)(y) =& \sqrt{N}\big(\widehat{\F}_{Y_t(1) \mid G_g=1} - \F_{Y_t(1)\mid G_g=1} \big)(y) -  \sqrt{N}\big(\widehat{\F}_{Y_t^{hs}(0) \mid G_g=1} - \F_{Y_t(0) \mid G_g=1} \big)(y) \\
    =& \frac{1}{\sqrt{N}} \sum_{j=1}^{N} \Big\{ \psi_{gt}(W_j;y) - \Psi_{gt}^{hs}(W_j;y) \Big\} + o_p(1)\\
    =&: \frac{1}{\sqrt{N}} \sum_{j=1}^{N} \Upsilon_{gt}^{hs}\big( W_j;y\big) + o_p(1)\\
    &\rightsquigarrow \mathbb{H}_{gt}^{hs}
\end{align*}
using the sequence of arguments akin to part (b) above and \Cref{Lem:Donsk_Hdt}. The associated function class is $P$-Donsker by closure under finite sums of $P$-Donsker classes with bounded coefficients, so the corresponding empirical process is asymptotically equicontinuous and converges weakly in $\ell^\infty(\Y)$. Thanks to \Cref{ass:Sampling_gen}, the covariance function of $\mathbb{H}_{gt}^{hs}$ is given by $ \Sigma_{gt}^{hs}(y_1,y_2):=\E\big[\Upsilon_{gt}^{hs}\big( W;y_1\big)\Upsilon_{gt}^{hs}\big( W;y_2\big) \big] $.

\qed

\subsection{Proof of \Cref{coro:FCLT_DF.DTT_Weighted} }

Because $\mathcal I$ is finite, the joint expansion in \Cref{ass:weight_consistent} implies that the estimated-weight contribution is finite-dimensional and tight. Products of the form $\sqrt N(\widehat\omega-\omega)(\widehat{\mathscr F}-\mathscr F)$ are $o_p(1)$ uniformly over $\Y$, since $\sqrt N(\widehat\omega-\omega)=O_p(1)$ jointly over $\mathcal I$ and the relevant first-stage DF or DTT estimation error is $o_p(1)$ uniformly over $\Y$.

\paragraph{Part (a):} 

The following asymptotically linear representation holds uniformly in $y\in\Y$ thanks to \Cref{ass:weight_consistent} and \Cref{Lem:Donsk_Hdt}:
\begin{align*}
    \sqrt{N}(\widehat{\F}_{Y_{\widehat{\omega}}(1)} - \F_{Y_{\omega}(1)})(y) =& \sum_{(g,t,h,s) \in \mathcal{I}} \Big\{\F_{gt}(y) \sqrt{N} \big(\widehat{\omega}(g,t,h,s) - \omega(g,t,h,s)\big) +  \omega(g,t,h,s)\times \sqrt{N}\big(\widehat{\F}_{gt} - \F_{gt}\big)(y) \\ 
    & \qquad + \sqrt{N}\big(\widehat{\omega}(g,t,h,s) - \omega(g,t,h,s)\big)\big(\widehat{\F}_{gt} - \F_{gt}\big)(y)\Big\} \\
    =& \frac{1}{\sqrt{N}} \sum_{j=1}^{N} \Bigg\{ 
	    \sum_{(g,t,h,s) \in \mathcal{I}} \Big\{ \F_{gt}(y)\mathcal{W}_j(g,t,h,s) + \omega(g,t,h,s)\psi_{gt}(W_j;y) \Big\} \Bigg\}  + o_p(1)\\
	    =&: \frac{1}{\sqrt{N}}\sum_{j=1}^{N} \Psi_{\omega}(W_j;y) + o_p(1).
\end{align*} Following the sequence of arguments in \Cref{Lem:Donsk_Hdt}, together with finite sums over $\mathcal I$, the associated class is $P$-Donsker and the corresponding empirical process is asymptotically equicontinuous. Thus, $\sqrt{N}(\widehat{\F}_{Y_{\widehat{\omega}}(1)} - \F_{Y_{\omega}(1)})$ converges weakly to a tight Gaussian process with mean $0$, namely $\mathbb{G}_{\omega}$. Under \Cref{ass:Sampling_gen}, the covariance function is given by $\Omega_{\omega}(y_1,y_2):=\E\big[\Psi_{\omega}(W;y_1)\Psi_{\omega}(W;y_2)\big]$.

\paragraph{Part (b):}
The following asymptotically linear representation holds thanks to \Cref{ass:weight_consistent} and part (b) of the proof of \Cref{Thm:FCLT_DF.DTT}:
\begin{align*}
    \sqrt{N}(\widehat{\F}_{Y_{\widehat{\omega}}(0)} - \F_{Y_{\omega}(0)})(y) =& \sum_{(g,t,h,s) \in \mathcal{I}} \Big\{\F_{Y_t(0) \mid G_g=1}(y) \sqrt{N} \big(\widehat{\omega}(g,t,h,s) - \omega(g,t,h,s)\big) \\ 
    & \qquad +  \omega(g,t,h,s)\times \sqrt{N}\big(\widehat{\F}_{Y_t^{hs}(0) \mid G_g=1} - \F_{Y_t(0) \mid G_g=1}\big)(y)\Big\} +  o_p(1) \\
    =& \frac{1}{\sqrt{N}} \sum_{j=1}^{N} \Bigg\{ 
	    \sum_{(g,t,h,s) \in \mathcal{I}} \Big\{ \F_{Y_t(0) \mid G_g=1}(y)\mathcal{W}_j(g,t,h,s) + \omega(g,t,h,s)\Psi_{gt}^{hs}(W_j;y) \Big\} \Bigg\}  + o_p(1)\\
	    =&: \frac{1}{\sqrt{N}}\sum_{j=1}^{N} \Psi_{\omega}^C(W_j;y) + o_p(1).
\end{align*} Following the sequence of arguments in part (b) of the proof of \Cref{Thm:FCLT_DF.DTT}, together with finite sums over $\mathcal I$, the associated class is $P$-Donsker and the corresponding empirical process is asymptotically equicontinuous. Thus, $\sqrt{N}\big( \widehat{\F}_{Y_{\widehat{\omega}}(0)} - \F_{Y_{\omega}(0)} \big)$ converges weakly to a tight Gaussian process with mean $0$, namely $\mathbb{G}_{\omega}^C$. Under \Cref{ass:Sampling_gen}, the covariance function is given by $\Omega_{\omega}^C(y_1,y_2):=\E\big[\Psi_{\omega}^C(W;y_1)\Psi_{\omega}^C(W;y_2)\big]$. 

\paragraph{Part (c):}
The following asymptotically linear representation holds thanks to \Cref{ass:weight_consistent} and part (c) of the proof of \Cref{Thm:FCLT_DF.DTT}:
\begin{align*}
    \sqrt{N}\big(\widehat{\DTT}_{\widehat{\omega}} - \DTT_{\omega}\big)(y) =& \sum_{(g,t,h,s) \in \mathcal{I}} \Big\{\DTT_{gt}(y) \sqrt{N} \big(\widehat{\omega}(g,t,h,s) - \omega(g,t,h,s)\big) \\ 
    &+  \omega(g,t,h,s)\times \sqrt{N}\big( \widehat{\DTT}_{gt}^{hs} - \DTT_{gt} \big)(y)\Big\} +  o_p(1) \\
    =& \frac{1}{\sqrt{N}} \sum_{j=1}^{N} \Bigg\{ 
    \sum_{(g,t,h,s) \in \mathcal{I}} \Big\{ \DTT_{gt}(y)\mathcal{W}_j(g,t,h,s) + \omega(g,t,h,s)\Upsilon_{gt}^{hs}(W_j;y) \Big\} \Bigg\} + o_p(1)\\
    =&: \frac{1}{\sqrt{N}}\sum_{j=1}^{N} \Upsilon_{\omega}(W_j;y) + o_p(1).
\end{align*}

\noindent Thus, following the Donsker and asymptotic-equicontinuity arguments in the proof of parts (b) and (c) of \Cref{Thm:FCLT_DF.DTT}, in addition to part (b) above:
\[
\sqrt{N}\big( \widehat{\DTT}_{\widehat{\omega}} - \DTT_{\omega} \big) \rightsquigarrow \mathbb{H}_{\omega},
\] where $ \mathbb{H}_{\omega} $ is a tight mean-zero Gaussian process. Thanks to \Cref{ass:Sampling_gen}, the covariance function is given by $ \Sigma_{\omega}(y_1,y_2):= \E\Big[\Upsilon_{\omega}\big( W;y_1\big)\Upsilon_{\omega}\big( W;y_2\big) \Big] $.

\qed

\subsection{Proof of \Cref{Thm:BootValid}}

Let $\widehat{\mathscr F}$ denote an estimator of a functional
$\mathscr F$ defined on $\Y$, e.g., $\F_{Y_t(0)\mid G_g=1}$, $\mathrm{DTT}_{gt}$, $\F_{Y_\omega(0)}$, or $\DTT_\omega$. From the uniform asymptotic linear representations established in \Cref{Lem:Donsk_Hdt}, \Cref{Thm:FCLT_DF.DTT}, and \Cref{coro:FCLT_DF.DTT_Weighted}, there exists a measurable influence function
$\Psi_{\mathscr F}(W;y)$ such that
\[
\sqrt{N}\big(\widehat{\mathscr F} - \mathscr F\big)(y)
=
\frac{1}{\sqrt{N}}\sum_{j=1}^N
\Psi_{\mathscr F}(W_j;y)
+o_p(1)
\quad\text{in }\ell^\infty(\Y),
\]
where the associated function class
$\{\Psi_{\mathscr F}(\cdot;y):y\in\Y\}$ is
P--Donsker, has mean zero, and is square integrable.

Consider the exchangeable bootstrap analogue satisfying the uniform expansion
\begin{equation}\label{eqn:boot_infl_fun}
    \sqrt{N}\big( \widehat{\mathscr F}^* - \widehat{\mathscr F} \big)(y) 
= \frac{1}{\sqrt{N}}\sum_{j=1}^N
(\xi_{Nj}-\bar\xi_N)\Psi_{\mathscr F}(W_j;y)
+o_p(1)
\quad\text{in }\ell^\infty(\Y).
\end{equation}
This representation follows from the uniform asymptotic linearity of $\widehat{\mathscr F}$ and the exchangeably weighted bootstrap theorem of
\citet[Theorem~3.6.13]{vaart-wellner-1996}, applied under the exchangeability, moment, sample-variance, and sample-mean requirements stated in the bootstrap subsection. Conditional on the data,
the bootstrap empirical process
\[
\frac{1}{\sqrt{N}}\sum_{j=1}^N(\xi_{Nj}-\bar\xi_N)\Psi_{\mathscr F}(W_j;\cdot)
\]
converges weakly in probability in $\ell^\infty(\Y)$ to the \emph{same} tight Gaussian limit as the original empirical process. Hence, the exchangeable bootstrap consistently estimates the law of
$\sqrt{N}(\widehat{\mathscr F}-\mathscr F)$ uniformly over $\Y$.

\qed 

\subsection{Proof of \Cref{Thm:AsyBoot_CvM_KS}}

\paragraph{Part (a):} 
\noindent \textbf{(i)} Since the admissible representations satisfy a joint uniform asymptotic linear representation, the stacked process $\sqrt{N}\big(\widehat{\mathbf F}_{Y_t(0)\mid G_g=1} - \mathbf{F}_{Y_t(0)\mid G_g=1} \big)$ is tight in
$\ell^\infty(\Y)^{K_{gt}}$, where \( \displaystyle \mathbf{F}_{Y_t(0)\mid G_g=1}(y) := \big(
\F_{Y_t^{hs}(0)\mid G_g=1}(y) \big)_{(h,s)\in \mathcal{I}_{gt}}' \in\mathbb R^{K_{gt}}\). The map induced by $\bm B_{gt}'$ is continuous and linear, so under $\mathbb{H}_0$ and for every $y\in\Y$,
\begin{align*}
    \widehat{\bm Z}_{gt}(y) 
    &=\sqrt{N}\,\bm B_{gt}'\,\widehat{\mathbf F}_{Y_t(0)\mid G_g=1}(y) \\
    &= \sqrt{N}\,\bm B_{gt}'\,\big(\widehat{\mathbf F}_{Y_t(0)\mid G_g=1} - \mathbf F_{Y_t(0)\mid G_g=1} \big)(y) + \sqrt{N}\,\bm B_{gt}'\,\mathbf F_{Y_t(0)\mid G_g=1}(y) \\
    &\overset{\mathbb{H}_0}{=} \sqrt{N}\,\bm B_{gt}'\,\big(\widehat{\mathbf F}_{Y_t(0)\mid G_g=1} - \mathbf F_{Y_t(0)\mid G_g=1} \big)(y) + \sqrt{N}\,\bm B_{gt}'\,\mathbf 1_{K_{gt}}F_{gt}^0(y) \\
&= \sqrt{N}\,\bm B_{gt}'\,\big(\widehat{\mathbf F}_{Y_t(0)\mid G_g=1} - \mathbf F_{Y_t(0)\mid G_g=1} \big)(y),
\end{align*}
where $ \displaystyle \mathbf 1_{K_{gt}} $ is a $ K_{gt} \times 1 $ vector of ones and $F_{gt}^0(y)$ denotes the common value of the admissible representations under $\mathbb H_0$. The third equality holds under $\mathbb{H}_0$ since \( \displaystyle \mathbf F_{Y_t(0)\mid G_g=1}(y) = \mathbf 1_{K_{gt}}F_{gt}^0(y), \ \forall y \in \Y \), and the fourth uses \( \bm B_{gt}'\,\mathbf 1_{K_{gt}} = \mathbf 0_{L_{gt}} \), since each column of $\bm B_{gt}$ contains one entry equal to $1$, one entry equal to $-1$, and zeroes elsewhere, so $\bm B_{gt}'$ annihilates vectors that are constant across admissible representations. It then follows that $ \widehat{\bm Z}_{gt} $  converges weakly in $\ell^\infty(\Y)^{L_{gt}}$ to
\(
\bm Z_{gt} := \bm B_{gt}'\mathbb Z_{gt},
\)
where $\mathbb Z_{gt}(y):=\big(\mathbb{G}_{gt}^{hs}(y)\big)_{(h,s)\in\mathcal{I}_{gt}}'$ (see \Cref{Thm:FCLT_DF.DTT}(b)) denotes the mean-zero tight Gaussian limit of the stacked process, with covariance kernel
$\Cov\big(\mathbb Z_{gt}(y_1),\mathbb Z_{gt}(y_2)\big) \in\mathbb R^{K_{gt}\times K_{gt}}$. Since $\bm Z_{gt}(y)=\bm B_{gt}'\,\mathbb Z_{gt}(y)$, the process $\bm Z_{gt}$ is mean-zero tight Gaussian in
$\ell^\infty(\Y)^{L_{gt}}$ with covariance kernel
\[
\Cov(\bm Z_{gt}(y_1),\bm Z_{gt}(y_2))=\bm B_{gt}'\,\Cov\big(\mathbb Z_{gt}(y_1),\mathbb Z_{gt}(y_2)\big)\,\bm B_{gt}
\in\mathbb R^{L_{gt} \times L_{gt}}.
\]
Equivalently, the column of $\bm B_{gt}$ associated with distinct elements $(h,s)$ and $(h',s')$ yields the pairwise contrast
\[
\widehat{\bm Z}_{gt}(y)_{(h,s),(h',s')}
=
\frac{1}{\sqrt{N}}\sum_{j=1}^{N}
\Big( \Psi_{gt}^{hs}(W_j;y)
-
\Psi_{gt}^{h's'}(W_j;y)
\Big) + o_p(1)
\]
under $\mathbb H_0$.

\noindent \textbf{(ii)} Consider the following decomposition:
\begin{align*}
    \widehat{\mathrm{CvM}}_{gt}
    =& \int_{\Y} \| \widehat{\bm Z}_{gt}(y)\|_2^2\, \widehat H(dy) \\
      =& \int_{\Y} \|\widehat{\bm Z}_{gt}(y)\|_2^2\, H(dy) + \int_{\Y} \| \widehat{\bm Z}_{gt}(y)\|_2^2\, (\widehat H - H)(dy)
\end{align*}

\noindent For fixed $H$, the map $ \displaystyle z\mapsto \int_{\Y}\|z(y)\|_2^2\,H(dy)$ is continuous on $\ell^\infty(\Y)^{L_{gt}}$ under the supremum norm, since
\[
\big|\|a\|_2^2-\|b\|_2^2\big|
\leq \|a-b\|_2\big(\|a\|_2+\|b\|_2\big).
\]
Thus, the weak convergence result in part (i) and the continuous mapping theorem imply
\[
\int_{\Y}\|\widehat{\bm Z}_{gt}(y)\|_2^2\,H(dy)
\xrightarrow{d}
\int_{\Y}\|\bm Z_{gt}(y)\|_2^2\,H(dy).
\]
For the empirical weighting measure used here, $\widehat{H}$ is an empirical CDF; hence \( \displaystyle \sup_{y\in \Y} \big| \big(\widehat{H} - H\big)(y) \big | = o_p(1) \) by the sampling condition in \Cref{ass:Sampling_gen} and the Glivenko-Cantelli Theorem \citep[Theorem 19.1]{van-der-vaart-2000}. The condition \( \displaystyle \int_{\Y} \| \widehat{\bm Z}_{gt}(y)\|_2^2\, \big(\widehat H - H\big)(dy) = o_p(1) \) follows from the asymptotic equicontinuity and boundedness of the process, using the argument in the concluding part of the proof of Corollary 1 in \citet{santana-song-2019specification}. Therefore, \( \displaystyle \widehat{\mathrm{CvM}}_{gt} \xrightarrow{d} \;\; \int_{\Y} \| \bm Z_{gt}(y)\|_2^2\, H(dy) \).

\noindent \textbf{(iii)} By the continuous mapping theorem together with part (i) above, the conclusion follows from \citet[Theorem 1.3.6]{vaart-wellner-1996}.

\paragraph{Part (b):} Under $ \mathbb{H}_{an} $, for each pair $(h,s)<(h',s')$,

\begin{equation*}
    \begin{split}
\widehat{\bm Z}_{gt}&(y)_{(h,s),(h',s')} = \sqrt{N}\big(\widehat \F_{Y_t^{hs}(0)\mid G_g=1}
-
\widehat \F_{Y_t^{h's'}(0)\mid G_g=1}
\big)(y) \\
=& \sqrt{N}\big(
\widehat \F_{Y_t^{hs}(0)\mid G_g=1} - F_{gt}^0 \big)(y)
-
\sqrt{N}\big( \widehat \F_{Y_t^{h's'}(0)\mid G_g=1} - F_{gt}^0 \big)(y) \\
=& \sqrt{N}\big(
\widehat \F_{Y_t^{hs}(0)\mid G_g=1} - \F_{Y_t^{hs}(0)\mid G_g=1} \big)(y)
-
\sqrt{N}\big(
\widehat \F_{Y_t^{h's'}(0)\mid G_g=1} - \F_{Y_t^{h's'}(0)\mid G_g=1}
\big)(y) \\
&+ \varrho_{gt}^{hs}(y) - \varrho_{gt}^{h's'}(y) \\
=& \frac{1}{\sqrt{N}}\sum_{j=1}^{N}
\Big( \Psi_{gt}^{hs}(W_j;y)
-
\Psi_{gt}^{h's'}(W_j;y)
\Big) + \big( \varrho_{gt}^{hs}(y) - \varrho_{gt}^{h's'}(y) \big) + o_p(1).
\end{split}
\end{equation*}
\noindent Stacking these pairwise drifts over all columns of $\bm B_{gt}$ gives $\bm \Delta_{gt}(y)=\bm B_{gt}'\bm R_{gt}(y)$. The conclusion then follows from the above decomposition and part (a)(i) above.

\paragraph{Part (c):} The KS conclusion follows from a direct lower-bound argument. Under a fixed alternative, there exists at least one pair $(h,s)<(h',s') \in \mathcal{I}_{gt} $ and $y^\dagger \in \Y $ such that 
\begin{align*}
\Big|\big(\F_{Y_t^{hs}(0)\mid G_g=1}
- \F_{Y_t^{h's'}(0)\mid G_g=1} \big)(y^\dagger)\Big|>0.
\end{align*}
The estimation error in the corresponding pairwise contrast is $\mathcal{O}_p(1)$ after multiplication by $\sqrt N$, whereas the fixed discrepancy contributes
\[
\sqrt{N}\Big|\big(\F_{Y_t^{hs}(0)\mid G_g=1}
- \F_{Y_t^{h's'}(0)\mid G_g=1} \big)(y^\dagger)\Big|
\rightarrow \infty
\]
as $N\rightarrow \infty$. Therefore, \( \widehat{\mathrm{KS}}_{gt}\xrightarrow{p}\infty\).

For the CvM statistic, let $d_{gt}^{hs,h's'}(y):=\F_{Y_t^{hs}(0)\mid G_g=1}(y)-\F_{Y_t^{h's'}(0)\mid G_g=1}(y)$. If $\displaystyle \int_{\Y}\big(d_{gt}^{hs,h's'}(y)\big)^2H(dy)>0$ for some admissible pair, then empirical-measure consistency gives
\[
\int_{\Y}\big(d_{gt}^{hs,h's'}(y)\big)^2\widehat H(dy)
=
\int_{\Y}\big(d_{gt}^{hs,h's'}(y)\big)^2H(dy)+o_p(1).
\]
The corresponding component of the statistic contains
\[
N\int_{\Y}\big(d_{gt}^{hs,h's'}(y)\big)^2H(dy)+O_p(\sqrt{N})+O_p(1),
\]
up to an additional $o_p(N)$ term from replacing $H$ by $\widehat H$. Since the leading deterministic term is of order $N$ and strictly positive, while the remaining terms are of smaller stochastic order, \( \widehat{\mathrm{CvM}}_{gt}\xrightarrow{p}\infty \).

\paragraph{Part (d):} Using \eqref{eqn:boot_infl_fun}, the $(h,s)<(h',s')$-th entry of the bootstrapped process has the following representation:
\begin{equation*}
    \begin{split}
\widehat{\bm Z}_{gt}^*(y)_{(h,s),(h',s')} =& \sqrt{N}\big(
\widehat \F_{Y_t^{hs}(0)\mid G_g=1}^* - \widehat{\F}_{Y_t^{hs}(0)\mid G_g=1} \big)(y)
-
\sqrt{N}\big(
\widehat \F_{Y_t^{h's'}(0)\mid G_g=1}^* - \widehat{\F}_{Y_t^{h's'}(0)\mid G_g=1}
\big)(y) \\
=& \frac{1}{\sqrt{N}}\sum_{j=1}^{N}
(\xi_{Nj}-\bar\xi_N)\Big( \Psi_{gt}^{hs}(W_j;y)
-
\Psi_{gt}^{h's'}(W_j;y)
\Big) + o_p(1).
\end{split}
\end{equation*}
Stacking the pairwise contrasts over all columns of $\bm B_{gt}$ gives
\[
\widehat{\bm Z}_{gt}^*(y)
=
\frac{1}{\sqrt{N}}\sum_{j=1}^{N}
(\xi_{Nj}-\bar\xi_N)\bm B_{gt}'\bm\Psi_{gt}(W_j;y)
+o_p(1)
\quad\text{in }\ell^\infty(\Y)^{L_{gt}},
\]
where \( \bm\Psi_{gt}(W_j;y):=\big(\Psi_{gt}^{hs}(W_j;y)\big)_{(h,s)\in\mathcal I_{gt}} \). By \eqref{eqn:boot_infl_fun} and the exchangeably weighted bootstrap theorem applied to the joint uniform asymptotic linear representation of the admissible counterfactual estimators,
\[
\widehat{\bm Z}_{gt}^*
\overset{P}{\underset{\xi}{\rightsquigarrow}}
\bm Z_{gt}
\quad\text{in }\ell^\infty(\Y)^{L_{gt}}.
\]
The bootstrap KS conclusion follows by the continuous mapping theorem applied to \( \displaystyle z\mapsto\sup_{y\in\Y}\lVert z(y)\rVert_\infty\). For the CvM statistic, the map \( \displaystyle z\mapsto\int_{\Y}\lVert z(y)\rVert_2^2\,H(dy)\) is continuous along bounded uniformly continuous sample paths. Replacing \(H\) by \(\widehat H\) is asymptotically negligible by the same asymptotic equicontinuity and boundedness argument used in part (a), now applied conditionally to \(\widehat{\bm Z}_{gt}^*\). Hence
\[
\widehat{\mathrm{CvM}}_{gt}^*
\overset{P}{\underset{\xi}{\rightsquigarrow}}
\int_{\Y}\lVert \bm Z_{gt}(y)\rVert_2^2\,H(dy)
\quad \text{and} \quad
\widehat{\mathrm{KS}}_{gt}^*
\overset{P}{\underset{\xi}{\rightsquigarrow}}
\sup_{y\in\Y}\lVert \bm Z_{gt}(y)\rVert_\infty.
\]

\qed
\section{Extension - Covariates}\label{Sect:Covariates}
\citet{kim-wooldridge-2024-difference} employs an Inverse Probability Weighted (IPW) approach to introducing covariates, where the link function $\Phi(\cdot)$ is the identity function. With possibly non-linear link functions in the current paper, an IPW approach to introducing covariates is less direct. This paper complements the \citet{kim-wooldridge-2024-difference} approach by using an Outcome Regression approach via Distribution Regression (DR) --- see also \citet{fernandez-meier-vuuren-vella-2024distribution}. Let $p(X) \in \mathbb{R}^{p_x}$ denote a dictionary of transformations of a set of some elementary time-invariant characteristics $X$, including the constant term $1$, and $p_x < \infty $ fixed. 

\subsection{Identification}
The following states versions of \Cref{ass:NA,ass:PT} that are conditional on pre-treatment covariates $X$ for a fixed group $ g $ at a post-treatment period $t\geq g$ using a valid control group $h> t$. 
\begin{namedassumption}{\ref{ass:PT}-X}[Conditional Distributional Parallel Trends]\label{ass:PT_X}
For every $y\in\Y$, every $(g,t,h,s)\in \mathcal I $, and $(d,\nu)\in(\{0,1\}\times \{s,t\}) $, the following restriction and representation hold on the treated covariate support, i.e., for $\Prob_{X\mid G_g=1}$-almost every $x$:
\begin{align*}
    \F_{Y_\nu(0) \mid X=x, G_g=d,G_g+G_h=1}(y)
    = \Phi\Big(p(x)\big(\bm{\alpha}_s^{h}(y) + d\bm{\beta}_s^{gh}(y) + \indicator{\nu=t}\bm{\gamma}_{st}^h(y)\big)\Big).
\end{align*}
\end{namedassumption}
\noindent Observe in this case that $ \big(\bm{\alpha}_s^h(y)', \bm{\beta}_s^{gh}(y)', \bm{\gamma}_{st}^h(y)'\big)' \in \mathbb{R}^{3p_x} $. Similarly, the conditional distributional no-anticipation condition is stated as follows.

\begin{namedassumption}{\ref{ass:NA}-X}[Conditional Distributional No-Anticipation]\label{ass:NA_X} For every $y\in\Y$, every $ g\in \mathcal{G} \setminus \{\infty\} $, and any $ s \in [-\T:(g-1)] $, 
\begin{align*}
    \F_{Y_s(1) \mid X=x,G_g=1}(y) = \F_{Y_s(0) \mid X=x,G_g=1}(y) \ \text{for} \ \Prob_{X\mid G_g=1}-a.e. \ x.
\end{align*}
\end{namedassumption}

\Cref{ass:PT_X,ass:NA_X} constitute weaker forms of \Cref{ass:PT,ass:NA} as they only require that the distributional parallel trends and no-anticipation conditions hold in sub-populations characterised by $X=x$ but not necessarily unconditionally. The overlap condition below ensures that the comparison-group conditional distributions are defined on the treated covariate support. Further, by using dictionaries of transformations of $X$, namely $p(X)$, \Cref{ass:PT_X,ass:NA_X} are more likely to hold in given applications. 

The following provides a suitable overlap condition that applies to the current setting.
\begin{assumption}[Overlap]\label{ass:overlap}
    For every $ (g,t) \in \mathcal{G}\setminus \{\infty\} \times [g:T] $ and any $ h \in \big(\mathcal{G}\setminus [1:t]\big) $, \( \displaystyle \Prob(G_g=1 \mid X, \, G_g + G_h = 1 ) < (1- c) \ a.s. \) for some $ c \in (0,1) $.
\end{assumption}

\noindent The following regularity condition concerns $p(X)$.

\begin{assumption}[No Perfect Collinearity]\label{ass:Full_Rank_X}
    $p(X) \mid G=g $ has full column rank almost surely on its support for each group $ g \in \mathcal{G} $.
\end{assumption}

Let $F_X(\cdot)$ and $F_{X \mid G_g=1}(\cdot)$ denote, respectively, the joint unconditional and conditional CDFs of $X$. The following extends \Cref{Thm:Identif} to the covariate setting.
\begin{theorem}[Identification]\label{Thm:Identif_X}
Suppose Assumptions \ref{ass:PT_X}, \ref{ass:NA_X}, \ref{ass:overlap}, and \ref{ass:Full_Rank_X} hold. Then, for every $y\in\Y$ and every $(g,t,h,s)\in \mathcal I $, 
\begin{enumerate}[(a)]
    \item the parameter vector $\big(\bm{\alpha}_s^h(y)', \bm{\beta}_s^{gh}(y)', \bm{\gamma}_{st}^h(y)'\big)'$ is identified;
    \item $\F_{Y_t(0) \mid X=x, G_g=1}(y)$ is identified: \( \displaystyle \F_{Y_t(0) \mid X=x, G_g=1}(y) = \Phi\Big(p(x)\big(\bm{\alpha}_s^h(y) + \bm{\beta}_s^{gh}(y) + \bm{\gamma}_{st}^h(y)\big)\Big) \); and
    \item $\F_{Y_t(0) \mid G_g=1}(y)$ is identified: \( \displaystyle \F_{Y_t(0) \mid G_g=1}(y) = \E\Big[\Phi\Big(p(X)\big(\bm{\alpha}_s^h(y) + \bm{\beta}_s^{gh}(y) + \bm{\gamma}_{st}^h(y)\big)\Big) \Big| G_g=1\Big]. \)
\end{enumerate}

\end{theorem} 

\subsection{Estimation}
Unlike the estimators introduced in the main text, which have closed-form expressions, the introduction of covariates $X$ does not lead to closed-form expressions. Estimation is therefore required, for example by quasi-maximum-likelihood Distribution Regression (DR); see \citet{wooldridge-2023-simple,chernozhukov-val-melly-2013}. One may estimate a binary response model with working CDF $\Phi(\cdot)$. This yields the distribution regression approach of \citet{foresi-peracchi-1995,chernozhukov-val-melly-2013}. Under \Cref{ass:Full_Rank_X}, and because the retained cells are $(g,s)$, $(h,s)$, and $(h,t)$, the regressors $p(X)$, $d\,p(X)$, and $\indicator{\nu=t} p(X)$ are not perfectly collinear.

The following condition relies on the sampling condition in \Cref{ass:Sampling_gen} with covariates included in observed data, namely with the modification $ W_j:= \big((S_{jt},Y_{jt})_{t\in [-\T:T]},G_j,X_j\big) $.

\begin{namedassumption}{\ref{ass:Sampling_gen}-X}[Random Sampling with $X$]\label{ass:Sampling_gen_X} The vectors $ \big\{W_j:= \big((S_{jt},Y_{jt})_{t\in [-\T:T]},G_j,X_j\big) \ : \ 1\le j\le N \big\}$ are independent and identically distributed with $S_{jt} \independent (Y_{jt}(0),Y_{jt}(1),G_j,X_j) $ for each $t \in [-\T:T] $. In addition, $\E[\|p(X)\|^4]<\infty$.
\end{namedassumption}
 
By the Law of Iterated Expectations (LIE), the conditions of \Cref{Thm:Identif_X}, and \Cref{ass:Sampling_gen_X},
\begin{align}\label{eqn:F_X_estimand}
    \F_{Y_t(0) \mid G_g=1}(y) =& \E[\indicator{Y_t(0) \leq y} \mid G_g=1] \nonumber \\
    =& \E[\indicator{Y_t(0) \leq y} \mid G_g=1,S_t=1] \nonumber \\ 
    =& \E\Big[\E\big[\indicator{Y_t(0) \leq y} \mid G_g=1,S_t=1,X\big]\big | G_g=1,S_t=1\Big] \nonumber \\ 
    =& \E\Big[\E\big[\indicator{Y_t(0) \leq y} \mid G_g=1,X\big]\big | G_g=1,S_t=1\Big] \nonumber \\ 
    =& \int \F_{Y_t(0) \mid X=x, G_g=1}(y)dF_{X\mid G_g=1,S_t=1}(x) \nonumber \\
    =& \frac{1}{\pi_tp_g}\E\Big[ S_t\indicator{G=g} \Phi\Big(p(X)\big(\bm{\alpha}_s^h(y) + \bm{\beta}_s^{gh}(y) + \bm{\gamma}_{st}^h(y)\big)\Big) \Big]
\end{align} where the second and fourth equalities follow by the independence of observability $S_t$ from $ (Y_t(0),G_g,X) $ (\Cref{ass:Sampling_gen_X}). The last equality shows that in practice, $\F_{Y_t(0) \mid G_g=1}(y)$ is computed by averaging $\F_{Y_t(0) \mid X=x, G_g=1}(y)$ over $x$ of the treated group $g$. The estimator of $\F_{Y_t(0) \mid G_g=1}(y)$ with covariates using control group $h$ and pre-treatment period $s$ is given by
\begin{equation}\label{eqn:F_X_estimator}
    \widehat{\F}_{Y_t^{X,hs}(0) \mid G_g=1}(y) = \frac{1}{\hat{\pi}_t\hat{p}_gN} \sum_{j=1}^N S_{jt}\indicator{G_j=g} \Phi\Big(p(X_j)\big(\widehat{\bm{\alpha}}_s^h(y) + \widehat{\bm{\beta}}_s^{gh}(y) + \widehat{\bm{\gamma}}_{st}^h(y)\big)\Big),
\end{equation}where the coefficients are obtained from fitting
\[ 
\F_{Y_\nu(0) \mid X=x, G_g=d,G_g+G_h=1}(y) = \Phi\Big(p(x)\bm{\alpha}_s^h(y) + dp(x)\bm{\beta}_s^{gh}(y) + \indicator{\nu=t}p(x)\bm{\gamma}_{st}^h(y)\Big) 
\] via DR on the pooled sample $\{g,h\} \times \{s,t\} \setminus \{(g,t)\}$. $\DTT_{gt}(y)$ can subsequently be estimated as $\widehat{\DTT}_{gt}^{X,hs}(y) = \widehat{\F}_{gt}(y) - \widehat{\F}_{Y_t^{X,hs}(0) \mid G_g=1}(y) $.

\subsection{Asymptotic Theory}
Consider the following high-level uniform (in $\Y$) asymptotic linearity condition on the estimators: $ \big(\widehat{\bm{\alpha}}_s^h(y)', \widehat{\bm{\beta}}_s^{gh}(y)', \widehat{\bm{\gamma}}_{st}^h(y)' \big)' $. This is a useful step in extending \Cref{Thm:FCLT_DF.DTT} to the covariate setting.

\begin{assumption}[Asymptotic linear representation]\label{ass:eta_expand}
For every $y\in\Y$ and every $(g,t,h,s)\in \mathcal I $, the estimators 
\[
\big(\widehat{\bm{\alpha}}_{s}^h(y)',\, \widehat{\bm{\beta}}_s^{gh}(y)',\, \widehat{\bm{\gamma}}_{st}^h(y)'\big)'
\]
admit the asymptotically linear representation
\begin{align*}
\sqrt{N}\begin{pmatrix}
\widehat{\bm{\alpha}}_s^h(y)-\bm{\alpha}_s^h(y) \\
\widehat{\bm{\beta}}_s^{gh}(y)-\bm{\beta}_s^{gh}(y) \\
\widehat{\bm{\gamma}}_{st}^h(y)-\bm{\gamma}_{st}^h(y)
\end{pmatrix}
=
\frac{1}{\sqrt{N}}\sum_{j=1}^{N}
\begin{pmatrix}
\mathcal S_s^h(W_j;y) \\
\mathcal S_s^{gh}(W_j;y) \\
\mathcal S_{st}^h(W_j;y)
\end{pmatrix}
+ o_p(1),
\end{align*}
where uniformly in $y\in\Y$, the influence functions are mean zero, have bounded second moments, and the collection \(\displaystyle \big\{ \mathcal S_s^h(\cdot;y), \ \mathcal S_s^{gh}(\cdot;y), \ \mathcal S_{st}^h(\cdot;y): y \in \Y \big\} \) forms a P-Donsker class. In the covariate extension, $\phi$ is continuously differentiable with derivative $\dot\phi$ that is bounded and Lipschitz.
\end{assumption}

\noindent \Cref{ass:eta_expand} is a high-level regularity condition on the DR coefficient processes. Its validity is established under primitive conditions in \citet[Corollary 5.3 and Lemma E.3]{chernozhukov-val-melly-2013}.

The following result extends \Cref{Thm:FCLT_DF.DTT} to the covariate setting.
\begin{theorem}\label{Thm:FCLT_DF.DTT_X}
Suppose Assumptions \ref{ass:PT_X}, \ref{ass:NA_X}, \ref{ass:Sampling_gen_X}, \ref{Ass:reg_pi_p}, \ref{ass:reg_Phi}, \ref{ass:Full_Rank_X}, \ref{ass:overlap}, and \ref{ass:eta_expand} hold, then for every $y\in \Y $ and every $(g,t,h,s)\in \mathcal I $,
\begin{enumerate}[(a)]
    \item \( \displaystyle \sqrt{N}\big(\widehat{\F}_{Y_t^{X,hs}(0) \mid G_g=1} - \F_{Y_t(0) \mid G_g=1}\big) \rightsquigarrow \mathbb{G}_{gt}^{X,hs} \) in $ \ell^{\infty}(\Y) $ and
    \item \( \displaystyle \sqrt{N}\big(\widehat{\DTT}_{gt}^{X,hs} - \DTT_{gt}\big) \rightsquigarrow \mathbb{H}_{gt}^{X,hs} \) in $ \ell^{\infty}(\Y) $
\end{enumerate}
where $\mathbb{G}_{gt}^{X,hs}$ and $\mathbb{H}_{gt}^{X,hs}$ are tight Gaussian processes with mean $0$ and respective covariance kernels $\Omega_{gt}^{X,hs}(y_1,y_2)$ and $\Sigma_{gt}^{X,hs}(y_1,y_2)$ defined on $\Y\times\Y$. The expressions of the covariance kernels are provided in the proof.
\end{theorem}

\begin{center}
    \textbf{ Proofs of results in \Cref{Sect:Covariates} }
\end{center}

\subsubsection*{Proof of \Cref{Thm:Identif_X}}
\paragraph{Part (a):} The proof of identification in the covariate setting proceeds along the same lines as that of \Cref{Thm:Identif}. 
Recall that $\bm{\alpha}_s^h(y)$, $\bm{\beta}_s^{gh}(y)$, and 
$\bm{\gamma}_{st}^h(y)$ are each vectors in $\mathbb{R}^{p_x}$. 
The argument follows the same three-step structure as in the baseline $gt$-case, with the additional use of covariate variation. In each step, strict monotonicity of $\Phi(\cdot)$ and \Cref{ass:Full_Rank_X} ensure that the corresponding coefficient vector is identified; see also the classical identification result of \citet{manski-1988-identification} for binary response models.

Under \Cref{ass:PT_X} and the overlap condition \Cref{ass:overlap}, 
\[
\F_{Y_s(0) \mid X=x,G_g=0,G_g+G_h=1}(y) = \Phi\big(p(x)\bm{\alpha}_s^h(y) \big),\ \Prob_{X\mid G_g=1} -a.e. \ x
\]
for control group $h$ and pre-treatment period $s$. Since $\Phi(\cdot)$ is strictly increasing and $p(X)$ has full column rank, $\bm{\alpha}_s^h(y)$ is identified for every $y\in\Y$.

Under \Cref{ass:PT_X} and the overlap condition \Cref{ass:overlap},
\begin{align*}
    \F_{Y_t(0) \mid X=x, G_g=0, G_g+G_h=1}(y)
    &= \Phi\big(p(x)(\bm{\alpha}_s^h(y) + \bm{\gamma}_{st}^h(y))\big) =: \Phi\big(p(x)\bm{\eta}_{ht}(y)\big),\ \Prob_{X\mid G_g=1} -a.e. \ x
\end{align*}
\noindent for all $y\in\Y$ with control group $h$ and post-treatment period $t$. The identification of $\bm{\eta}_{ht}(y)$ follows from strict monotonicity of $\Phi(\cdot)$ and \Cref{ass:Full_Rank_X}. In particular, $\Phi\big(p(x)\bm{\eta}_{ht}(y)\big)=\Phi\big(p(x)(\bm{\alpha}_s^h(y) + \bm{\gamma}_{st}^h(y))\big)$ is equivalent to $p(x)\bm{\eta}_{ht}(y)= p(x)\big(\bm{\alpha}_s^h(y) + \bm{\gamma}_{st}^h(y)\big)$ for $\Prob_{X\mid G_g=1}$-a.e. $x$ and all $y\in\Y$. By the full rank condition on $p(X)$ (\Cref{ass:Full_Rank_X}) and because $\bm{\alpha}_s^h(y)$ is identified in the first step, $\bm{\gamma}_{st}^h(y)$ is identified for every $y\in\Y$: $\bm{\gamma}_{st}^h(y) = \bm{\eta}_{ht}(y) - \bm{\alpha}_s^h(y)$.

Under \Cref{ass:NA_X,ass:PT_X}, 
\begin{align*}
    \F_{Y_s(0) \mid X=x,G_g=1}(y) =&  \F_{Y_s(1) \mid X=x,G_g=1}(y) = \F_{Y_s(1) \mid X=x,G_g=1,G_g+G_h=1}(y) \\ 
    =& \Phi\big(p(x)(\bm{\alpha}_s^h(y) + \bm{\beta}_s^{gh}(y))\big)=: \Phi\big(p(x)\bm{\eta}_{gs}(y)\big), \ \Prob_{X\mid G_g=1}-a.e. \ x.
\end{align*}
\noindent $\bm{\eta}_{gs}(y)$ is identified by arguments akin to those of $\bm{\eta}_{ht}(y)$. Thus, $\bm{\beta}_s^{gh}(y)$ is identified for every $y\in\Y$: $\bm{\beta}_s^{gh}(y) = \bm{\eta}_{gs}(y) - \bm{\alpha}_s^h(y)$.

The conclusion follows by combining the above three parts.

\paragraph{Part (b):} This follows directly from the identification of $ \big( \bm{\alpha}_s^h(y)',\bm{\beta}_s^{gh}(y)',\bm{\gamma}_{st}^h(y)' \big)' $ in part (a) and \Cref{ass:PT_X}.

\paragraph{Part (c):} In addition to part (b), part (c) follows by integrating the identified conditional counterfactual distribution over the covariate distribution of the treated group. Under \Cref{ass:Sampling_gen_X}, observability is independent of $(Y_t(0),G_g,X)$, so the distribution of $X$ among treated units is identified from observed treated units. Hence, $\F_{Y_t(0)\mid G_g=1}(y)$ is identified for every $y\in\Y$.

\qed

\subsubsection*{Proof of \Cref{Thm:FCLT_DF.DTT_X}}
\paragraph{Part (a):} The proof of \Cref{Thm:FCLT_DF.DTT_X}(a) is structured into several steps to facilitate the exposition of the asymptotic results.

\noindent\textbf{Influence function decomposition:} Adopting the following short-hand notation: \( \displaystyle \widehat{\theta}_{st}^{gh}(y):= \widehat{\bm{\alpha}}_s^h(y) + \widehat{\bm{\beta}}_s^{gh}(y) + \widehat{\bm{\gamma}}_{st}^h(y) \) and \( \displaystyle \theta_{st}^{gh}(y):= \bm{\alpha}_s^h(y) + \bm{\beta}_s^{gh}(y) + \bm{\gamma}_{st}^h(y)\), consider the following decomposition using \eqref{eqn:F_X_estimand} and \eqref{eqn:F_X_estimator}: 

\begin{align*}
    \sqrt{N}&\big(\widehat{\F}_{Y_t^{X,hs}(0) \mid G_g=1} - \F_{Y_t(0) \mid G_g=1} \big)(y)\\
    =&   \frac{1}{\sqrt{N}} \sum_{j=1}^N \Big\{ \frac{S_{jt}\indicator{G_j=g}}{\hat{\pi}_t\hat{p}_g} \Phi\Big(p(X_j)\widehat{\theta}_{st}^{gh}(y)\Big) - \E\Big[ \frac{S_t\indicator{G=g}}{\pi_tp_g} \Phi\Big(p(X)\theta_{st}^{gh}(y)\Big) \Big]\Big\} \\
    =&   \frac{1}{\sqrt{N}} \sum_{j=1}^N \Big\{ \frac{S_{jt}\indicator{G_j=g}}{\pi_tp_g} \Big(\Phi\big(p(X_j)\widehat{\theta}_{st}^{gh}(y)\big)- \Phi\big(p(X_j)\theta_{st}^{gh}(y)\big)\Big)\Big\} \\
    &- \frac{1}{N}\sum_{j=1}^N \Big\{ S_{jt}\indicator{G_j=g} \Phi\big(p(X_j)\theta_{st}^{gh}(y)\big)\Big\} \times \frac{\sqrt{N}\big( \hat{\pi}_t\hat{p}_g - \pi_tp_g \big)}{\pi_tp_g\hat{\pi}_t\hat{p}_g} \\
    &+ \frac{1}{\sqrt{N}} \sum_{j=1}^N \Big\{\frac{S_{jt}\indicator{G_j=g}\Phi\big(p(X_j)\theta_{st}^{gh}(y)\big) - \E\big[S_t\indicator{G=g}\Phi\big(p(X)\theta_{st}^{gh}(y)\big) \big]}{\pi_tp_g}\Big\} \\
    & \underbrace{- \frac{\sqrt{N}( \hat{\pi}_t\hat{p}_g - \pi_tp_g )}{\hat{\pi}_t\hat{p}_g} \frac{1}{N} \sum_{j=1}^N \Big\{ \frac{S_{jt}\indicator{G_j=g}}{\pi_tp_g} \Big(\Phi\big(p(X_j)\widehat{\theta}_{st}^{gh}(y)\big)- \Phi\big(p(X_j)\theta_{st}^{gh}(y)\big)\Big)\Big\}}_{\widehat{\mathcal R}_{gt,X}^{(1)}(y)}.
\end{align*}

Observe the following decomposition of the term
\begin{align*}
    \frac{\sqrt{N}\big(\hat{\pi}_t\hat{p}_g - \pi_tp_g \big)}{\pi_tp_g\hat{\pi}_t\hat{p}_g} = \frac{1}{\pi_tp_g\hat{\pi}_t}\sqrt{N}(\hat{\pi}_t - \pi_t ) + \frac{1}{p_g\hat{\pi}_t\hat{p}_g}\sqrt{N}(\hat{p}_g - p_g ).
\end{align*} In addition to the differentiability of $\Phi(\cdot)$ in \Cref{ass:reg_Phi}, obtain the following decomposition:

{\footnotesize
\begin{equation}\label{eqn:decomp_Fhat_X}
    \begin{split}
        \sqrt{N}&\big(\widehat{\F}_{Y_t^{X,hs}(0) \mid G_g=1} - \F_{Y_t(0) \mid G_g=1} \big)(y)\\
    =& \frac{1}{N} \sum_{j=1}^N \Big\{ \frac{S_{jt}\indicator{G_j=g}}{\pi_tp_g} \phi\big(p(X_j)\theta_{st}^{gh}(y)\big)p(X_j)\Big\}\sqrt{N}(\widehat{\theta}_{st}^{gh} - \theta_{st}^{gh})(y) \\
    &- \frac{1}{N}\sum_{j=1}^N \Big\{ S_{jt}\indicator{G_j=g} \Phi\big(p(X_j)\theta_{st}^{gh}(y)\big)\Big\} \Big( \frac{1}{\pi_tp_g\hat{\pi}_t}\sqrt{N}(\hat{\pi}_t - \pi_t ) + \frac{1}{p_g\hat{\pi}_t\hat{p}_g}\sqrt{N}(\hat{p}_g - p_g ) \Big) \\
    &+ \frac{1}{\sqrt{N}} \sum_{j=1}^N \Big\{\frac{S_{jt}\indicator{G_j=g}\Phi\big(p(X_j)\theta_{st}^{gh}(y)\big) - \E\big[S_t\indicator{G=g}\Phi\big(p(X)\theta_{st}^{gh}(y)\big) \big]}{\pi_tp_g}\Big\} \\
    & \underbrace{- \frac{\sqrt{N}\big( \hat{\pi}_t\hat{p}_g - \pi_tp_g \big)}{\hat{\pi}_t\hat{p}_g} \frac{1}{N} \sum_{j=1}^N \Big\{ \frac{S_{jt}\indicator{G_j=g}}{\pi_tp_g} \Big(\Phi\big(p(X_j)\widehat{\theta}_{st}^{gh}(y)\big)- \Phi\big(p(X_j)\theta_{st}^{gh}(y)\big)\Big)\Big\}}_{\widehat{\mathcal R}_{gt,X}^{(1)}(y)} \\
    &+ \underbrace{\frac{1}{\sqrt{N}} \sum_{j=1}^N \Big\{ \frac{S_{jt}\indicator{G_j=g}}{\pi_tp_g} \Big(\Phi\big(p(X_j)\widehat{\theta}_{st}^{gh}(y)\big)- \Phi\big(p(X_j)\theta_{st}^{gh}(y)\big) - \phi\big(p(X_j)\theta_{st}^{gh}(y)\big)p(X_j)(\widehat{\theta}_{st}^{gh} - \theta_{st}^{gh})(y)\Big)\Big\}}_{\widehat{\mathcal R}_{gt,X}^{(2)}(y)}.
    \end{split}
\end{equation}
}

\noindent\textbf{Asymptotic negligibility of remainder terms:} Let
\(
\Delta_{st}^{gh}(y):=\big(\widehat{\theta}_{st}^{gh}-\theta_{st}^{gh}\big)(y).
\) Observe that 
\( \displaystyle 
\sup_{y\in\Y}\|\Delta_{st}^{gh}(y)\|=o_p(1) \text{ and } \sup_{y\in\Y}\sqrt N\,\|\Delta_{st}^{gh}(y)\|=\mathcal{O}_p(1),
\) under \Cref{ass:eta_expand}. It is first shown that
\(\displaystyle 
\sup_{y\in\Y}\big|\widehat{\mathcal R}_{gt,X}^{(1)}(y)\big|=o_p(1).
\)
By the mean-value theorem and the boundedness of \(\phi\) in \Cref{ass:reg_Phi}, 
\[
\Big|
\Phi\big(p(X_j)\widehat\theta_{st}^{gh}(y)\big)
-
\Phi\big(p(X_j)\theta_{st}^{gh}(y)\big)
\Big|
\le
C\,\|p(X_j)\|\,\|\Delta_{st}^{gh}(y)\|.
\]
In addition to the result \( \displaystyle \frac{\sqrt{N}\big(\hat{\pi}_t\hat{p}_g - \pi_tp_g \big)}{\pi_tp_g\hat{\pi}_t\hat{p}_g} = \mathcal{O}_p(1) \), which holds under \Cref{ass:Sampling_gen,Ass:reg_pi_p} (see \eqref{eqn:pi_hat_phat}),
\begin{align*}
\sup_{y\in\Y}\big|\widehat{\mathcal R}_{gt,X}^{(1)}(y)\big|
&\le
\left|
\frac{\sqrt N(\hat\pi_t\hat p_g-\pi_t p_g)}{\hat\pi_t\hat p_g}
\right|
\frac{1}{\pi_t p_g}
\frac1N\sum_{j=1}^N
S_{jt}\indicator{G_j=g}
\sup_{y\in\Y}
\Big|
\Phi\big(p(X_j)\widehat\theta_{st}^{gh}(y)\big)
-
\Phi\big(p(X_j)\theta_{st}^{gh}(y)\big)
\Big| \\
&\le
\left|
\frac{\sqrt N(\hat\pi_t\hat p_g-\pi_t p_g)}{\hat\pi_t\hat p_g}
\right|
\frac{C}{\pi_t p_g}
\left(\frac1N\sum_{j=1}^N \|p(X_j)\|\right)
\sup_{y\in\Y}\|\Delta_{st}^{gh}(y)\|.
\end{align*}
Therefore, in addition to the moment condition on $p(X)$ in \Cref{ass:Sampling_gen_X}, 
\(\displaystyle 
\sup_{y\in\Y}\big|\widehat{\mathcal R}_{gt,X}^{(1)}(y)\big|
=
\mathcal{O}_p(1)\cdot \mathcal{O}_p(1)\cdot o_p(1)
=
o_p(1).
\)

To handle \(\widehat{\mathcal R}_{gt,X}^{(2)}\), apply Taylor's theorem to
\(\vartheta \mapsto \Phi(p(X_j)\vartheta)\). Since $\dot\phi$ is bounded under
\Cref{ass:eta_expand}, there exists a constant $C<\infty$ such that, uniformly
in $y\in\Y$,
\[
\Big|
\Phi\big(p(X_j)\widehat\theta_{st}^{gh}(y)\big)
-
\Phi\big(p(X_j)\theta_{st}^{gh}(y)\big)
-
\phi\big(p(X_j)\theta_{st}^{gh}(y)\big)
p(X_j)\Delta_{st}^{gh}(y)
\Big|
\le
C\|p(X_j)\|^2\|\Delta_{st}^{gh}(y)\|^2.
\]
Therefore, using \Cref{ass:Sampling_gen_X,ass:eta_expand},
\begin{align*}
\sup_{y\in\Y}\big|\widehat{\mathcal R}_{gt,X}^{(2)}(y)\big|
&\le
\frac{C}{\sqrt N}
\sum_{j=1}^N
\frac{S_{jt}\indicator{G_j=g}}{\pi_t p_g}
\|p(X_j)\|^2
\sup_{y\in\Y}\|\Delta_{st}^{gh}(y)\|^2 \\
&=
C\sqrt N
\left(
\frac{1}{N}\sum_{j=1}^N
\frac{S_{jt}\indicator{G_j=g}}{\pi_t p_g}
\|p(X_j)\|^2
\right)
O_p(N^{-1})
=o_p(1).
\end{align*}

Since the remainder terms are asymptotically negligible uniformly in $y\in\Y$, i.e., \( \displaystyle 
\sup_{y\in\Y}\big|\widehat{\mathcal R}_{gt,X}^{(1)}(y)\big|
=
o_p(1)
\) and 
\( \displaystyle 
\sup_{y\in\Y}\big|\widehat{\mathcal R}_{gt,X}^{(2)}(y)\big|
=
o_p(1)
\), it follows from the decomposition in \eqref{eqn:decomp_Fhat_X} that 
\begin{align*}
    \sqrt{N}&\big(\widehat{\F}_{Y_t^{X,hs}(0) \mid G_g=1} - \F_{Y_t(0) \mid G_g=1} \big)(y)\\
    =& \frac{1}{N} \sum_{j=1}^N \Big\{ \frac{S_{jt}\indicator{G_j=g}}{\pi_tp_g} \phi\big(p(X_j)\theta_{st}^{gh}(y)\big)p(X_j)\Big\}\sqrt{N}(\widehat{\theta}_{st}^{gh} - \theta_{st}^{gh})(y) \\
    &- \frac{1}{N}\sum_{j=1}^N \Big\{ S_{jt}\indicator{G_j=g} \Phi\big(p(X_j)\theta_{st}^{gh}(y)\big)\Big\} \Big( \frac{1}{\pi_tp_g\hat{\pi}_t}\sqrt{N}(\hat{\pi}_t - \pi_t ) + \frac{1}{p_g\hat{\pi}_t\hat{p}_g}\sqrt{N}(\hat{p}_g - p_g ) \Big) \\
    &+ \frac{1}{\sqrt{N}} \sum_{j=1}^N \Big\{\frac{S_{jt}\indicator{G_j=g}\Phi\big(p(X_j)\theta_{st}^{gh}(y)\big) - \E\big[S_t\indicator{G=g}\Phi\big(p(X)\theta_{st}^{gh}(y)\big) \big]}{\pi_tp_g}\Big\} \\
    & + o_p(1).
\end{align*}

\noindent\textbf{Asymptotic linearity:} The next part of the proof studies the non-negligible summands of the above in turn.

First, thanks to \Cref{ass:Sampling_gen_X,ass:reg_Phi}, it follows from the Uniform Weak Law of Large Numbers that
\[
\sup_{y\in\Y}
\left\|
\frac{1}{N} \sum_{j=1}^N \Big\{ \frac{S_{jt}\indicator{G_j=g}}{\pi_tp_g} \phi\big(p(X_j)\theta_{st}^{gh}(y)\big)p(X_j)\Big\}
-
    \E\big[\phi\big(p(X)\theta_{st}^{gh}(y)\big)p(X) \mid G_g=1 \big]
\right\|
\xrightarrow{p} 0
\] where \( \displaystyle \E\big[\phi\big(p(X)\theta_{st}^{gh}(y)\big)p(X) \mid G_g=1 \big] = \E\Big[
\frac{S_t\indicator{G=g}}{\pi_t p_g}
\phi\big(p(X)\theta_{st}^{gh}(y)\big)p(X)
\Big] \) under \Cref{ass:Sampling_gen_X}. Also, under \Cref{ass:eta_expand}, 
\[ 
\sqrt{N}(\widehat{\theta}_{st}^{gh} - \theta_{st}^{gh})(y) = \frac{1}{\sqrt{N}} \sum_{j=1}^N  \big( \mathcal S_s^h(W_j;y) + \mathcal S_s^{gh}(W_j;y) + \mathcal S_{st}^h(W_j;y) \big) + o_p(1).
\] Combining both terms gives uniformly in $y\in\Y$:
\begin{align*}
    \frac{1}{N} &\sum_{j=1}^N \Big\{ \frac{S_{jt}\indicator{G_j=g}}{\pi_tp_g} \phi\big(p(X_j)\theta_{st}^{gh}(y)\big)p(X_j)\Big\}\sqrt{N}(\widehat{\theta}_{st}^{gh} - \theta_{st}^{gh})(y) \\
=& \E\big[\phi\big(p(X)\theta_{st}^{gh}(y)\big)p(X) \mid G_g=1 \big]\frac{1}{\sqrt{N}} \sum_{j=1}^N  \big( \mathcal S_s^h(W_j;y) + \mathcal S_s^{gh}(W_j;y) + \mathcal S_{st}^h(W_j;y) \big) + o_p(1).
\end{align*}

Second, by \Cref{ass:Sampling_gen_X,Ass:reg_pi_p}, it follows from the UWLLN and the continuous mapping theorem that
\begin{align*}
    \frac{1}{N}&\sum_{j=1}^N \Big\{ S_{jt}\indicator{G_j=g} \Phi\big(p(X_j)\theta_{st}^{gh}(y)\big)\Big\} \Big( \frac{1}{\pi_tp_g\hat{\pi}_t}\sqrt{N}(\hat{\pi}_t - \pi_t ) + \frac{1}{p_g\hat{\pi}_t\hat{p}_g}\sqrt{N}(\hat{p}_g - p_g ) \Big) \\
    =& \E\big[S_t\indicator{G=g} \Phi\big(p(X)\theta_{st}^{gh}(y)\big)\big] \Big( \frac{1}{\pi_t^2p_g}\sqrt{N}(\hat{\pi}_t - \pi_t ) + \frac{1}{\pi_tp_g^2}\sqrt{N}(\hat{p}_g - p_g ) \Big) + o_p(1) \\
    =& \frac{\F_{Y_t(0)\mid G_g=1}(y)}{\pi_t} \sqrt{N}(\hat{\pi}_t - \pi_t ) + \frac{\F_{Y_t(0)\mid G_g=1}(y)}{p_g} \sqrt{N}(\hat{p}_g - p_g ) + o_p(1)
\end{align*}

Third, since \( \displaystyle \frac{\E\big[S_t\indicator{G=g}\Phi\big(p(X)\theta_{st}^{gh}(y)\big) \big]}{\pi_tp_g} = \int
\F_{Y_t(0)\mid X=x,\,G_g=1}(y)\, dF_{X\mid G_g=1,\,S_t=1}(x) \) under \Cref{ass:PT_X} and the independence of observability $S_t$ from $ (Y_t(0),G_g,X) $ (\Cref{ass:Sampling_gen_X}),

\[
\frac{1}{\sqrt N}\sum_{j=1}^N \left\{
\frac{S_{jt}\indicator{G_j=g}\Phi\big(p(X_j)\theta_{st}^{gh}(y)\big)
-
\E\big[S_t\indicator{G=g}\Phi\big(p(X)\theta_{st}^{gh}(y)\big)\big]}
{\pi_t p_g}
\right\}
\]
\[
=
\int
\F_{Y_t(0)\mid X=x,\,G_g=1}(y)\,
\sqrt N\big(\widehat F_{X\mid G_g=1,S_t=1}-F_{X\mid G_g=1,S_t=1}\big)(dx)
\] where \( \displaystyle \widehat F_{X\mid G_g=1,S_t=1} \) denotes the empirical conditional CDF.

It therefore follows that
\begin{align*}
    \sqrt{N}&\big(\widehat{\F}_{Y_t^{X,hs}(0) \mid G_g=1} - \F_{Y_t(0) \mid G_g=1} \big)(y)\\
    =& \underbrace{\E\big[\phi\big(p(X)\theta_{st}^{gh}(y)\big)p(X) \mid G_g=1 \big]\frac{1}{\sqrt{N}} \sum_{j=1}^N \big( \mathcal S_s^h(W_j;y) + \mathcal S_s^{gh}(W_j;y) + \mathcal S_{st}^h(W_j;y) \big)}_{\frac{1}{\sqrt{N}}\sum_{j=1}^N \Psi_{F,X}^A(W_j;y)} \\
    &- \underbrace{\frac{\F_{Y_t(0)\mid G_g=1}(y)}{\pi_t} \frac{1}{\sqrt{N}} \sum_{j=1}^N(S_{jt}-\pi_t ) - \frac{\F_{Y_t(0)\mid G_g=1}(y)}{p_g}\frac{1}{\sqrt{N}} \sum_{j=1}^N(\indicator{G_j=g} - p_g )}_{\frac{1}{\sqrt{N}}\sum_{j=1}^N \Psi_{F,X}^B(W_j;y)} \\
    &+ \underbrace{\int \F_{Y_t(0)\mid X=x,\,G_g=1}(y)\,
\sqrt N\big(\widehat F_{X\mid G_g=1,S_t=1}-F_{X\mid G_g=1,S_t=1}\big)(dx)}_{\frac{1}{\sqrt{N}}\sum_{j=1}^N \Psi_{F,X}^C(W_j;y)} + o_p(1) \\
    =:& \frac{1}{\sqrt{N}}\sum_{j=1}^N \Psi_{gt}^{X,hs}(W_j;y) + o_p(1).
\end{align*}

\noindent\textbf{Donsker property of the associated function class:} 
The next step is to investigate the Donsker property of the function class associated with $ \Psi_{gt}^{X,hs}(W_j;y) $. The map
\(
c(y):=\E\big[\phi\big(p(X)\theta_{st}^{gh}(y)\big)p(X) \mid G_g=1 \big]
\)
satisfies $ \displaystyle \sup_{y\in\Y}\lVert c(y)\rVert < \infty$ under \Cref{ass:reg_Phi} and the dominance condition on $p(X)$ in \Cref{ass:eta_expand}. $y\mapsto c(y)$ is bounded, hence multiplication by $c(y)$ preserves Donskerness. In addition to the Donskerness of the function class associated with the DR coefficient process (\Cref{ass:eta_expand}), the functional class associated with $ \Psi_{F,X}^A(W;y) $ is $P$-Donsker.

Let 
\(
Z(W) := \Big(\frac{S_t}{\pi_t} -1\Big)
+ \Big(\frac{\indicator{G=g}}{p_g} - 1\Big).
\)
Then
\(
\Psi_{F,X}^B(W;y)= -Z(W)\,\F_{Y_t(0)\mid G_g=1}(y).
\)
The index $y$ enters only through the deterministic scalar $\F_{Y_t(0)\mid G_g=1}(y)\in[0,1]$. Thus, 
\(
\{\,w\mapsto -Z(w)\,a : a\in[0,1]\,\}
\)
is a uniformly bounded one-dimensional linear class, hence $P$-Donsker.

The third summand $ \Psi_{F,X}^C(W_j;y) $ is cast in the same form as the $ \widehat{G}_k(f) $ empirical process on page 2224 with the limit in equation 4.2 of \citet{chernozhukov-val-melly-2013} for distribution regression. It follows from \citet[Theorem 4.1]{chernozhukov-val-melly-2013} that the associated function class of $ \Psi_{F,X}^C(W_j;y) $ is $P$-Donsker.

Each of the function classes corresponding to $\Psi_{F,X}^A(W_j;y), \ \Psi_{F,X}^B(W_j;y), \ \Psi_{F,X}^C(W_j;y) $ is $P$-Donsker, and $P$-Donsker classes are closed under finite sums. Therefore the associated class of $\Psi_{gt}^{X,hs}(W_j;y)$ is $P$-Donsker.

\( \displaystyle \sup_{y\in\Y} \E \Big| \Psi_{gt}^{X,hs}(W;y) \Big|^2 < \infty \) under \Cref{ass:reg_Phi,ass:Sampling_gen_X}. Thus, for any arbitrary finite subset $\{y_1,\ldots,y_L\}\subset \Y$, it follows from the Multivariate Lindeberg--L\'evy Central Limit Theorem that 
$$
\Big(\frac{1}{\sqrt{N}}\sum_{j=1}^N \Psi_{gt}^{X,hs}(W_j;y_1), \ldots, \frac{1}{\sqrt{N}}\sum_{j=1}^N \Psi_{gt}^{X,hs}(W_j;y_L)\Big)
$$ converges in distribution to the multivariate normal with the $(l,l')$'th entry of the covariance matrix: $\E\big[\Psi_{gt}^{X,hs}(W;y_l)\Psi_{gt}^{X,hs}(W;y_{l'})\big]$. The $P$-Donsker property supplies asymptotic equicontinuity, so the empirical process converges weakly in $\ell^\infty(\Y)$ to a tight Gaussian process, namely $\mathbb{G}_{gt}^{X,hs}$, with covariance kernel \( \Omega_{gt}^{X,hs}(y_1,y_2) := \E\big[\Psi_{gt}^{X,hs}(W;y_1)\Psi_{gt}^{X,hs}(W;y_2)\big]\); see \citet[Theorem 2.1]{kosorok-2007}.

\paragraph{Part (b):}
From \eqref{eqn:exp_Hdt_n} and part (a) above, the following decomposition holds:
\begin{align*}
    \sqrt{N}\big(\widehat{\DTT}_{gt}^{X,hs} &- \DTT_{gt}\big)(y) \\
    =& \sqrt{N}\big(\widehat{\F}_{Y_t(1) \mid G_g=1} - \F_{Y_t(1)\mid G_g=1}\big)(y) -  \sqrt{N}\big(\widehat{\F}_{Y_t^{X,hs}(0) \mid G_g=1} - \F_{Y_t(0) \mid G_g=1} \big)(y) \\
    =& \sqrt{N}\big(\widehat{\F}_{gt} - \F_{gt}\big)(y) - \sqrt{N}\big(\widehat{\F}_{Y_t^{X,hs}(0) \mid G_g=1} - \F_{Y_t(0) \mid G_g=1} \big)(y) \\
    =& \frac{1}{\sqrt{N}} \sum_{j=1}^{N} \Big\{ \psi_{gt}(W_j;y) - \Psi_{gt}^{X,hs}(W_j;y) \Big\} + o_p(1)\\
    =&: \frac{1}{\sqrt{N}} \sum_{j=1}^{N} \Upsilon_{gt}^{X,hs}\big( W_j;y\big) + o_p(1).
\end{align*} From part (a) above and \Cref{Lem:Donsk_Hdt}, conclude using the Donsker and asymptotic-equicontinuity arguments in the proof of part (c) of \Cref{Thm:FCLT_DF.DTT}. The covariance kernel of the limiting process is \( \Sigma_{gt}^{X,hs}(y_1,y_2):=\E\big[\Upsilon_{gt}^{X,hs}(W;y_1)\Upsilon_{gt}^{X,hs}(W;y_2)\big] \).

\qed

\section{Uniform Inference}\label{Sect:Uniform_Infer} 

The results in this section are specialised to $ (g,t) \in \mathcal{G}\setminus \{\infty\} \times [g:T] $. Results on convex-weighted DFs, convex-weighted DTTs, and QTTs constructed from convex-weighted DFs follow straightforwardly. Let
$\widehat I_{gt}^{(1)}(y) = \big[\widehat L_{gt}^{(1)}(y),\ \widehat U_{gt}^{(1)}(y)\big]$ and
$\widehat I_{gt}^{(0)}(y) = \big[\widehat L_{gt}^{(0)}(y),\ \widehat U_{gt}^{(0)}(y)\big]$ denote joint
$p$-level confidence bands for $\F_{Y_t(1)\mid G_g=1}(y)$ and
$\F_{Y_t(0)\mid G_g=1}(y)$, respectively, and let
$\widehat I_{gt}^{\DTT}(y) = \big[\widehat L_{gt}^{\DTT}(y),\ \widehat U_{gt}^{\DTT}(y)\big]$ be the $p$-level confidence band for $\DTT_{gt}$. The confidence bands can be constructed using steps 1-5 of Algorithm 1 in \citet{chernozhukov2019generic} --- see also \citet[Algorithm 3]{chernozhukov-val-melly-2013}. For the band-inversion step used below, the estimated DFs and DF-band endpoints are clipped to $[0,1]$ and monotonised over $\Y$ before taking left inverses. This rearrangement enforces the nondecreasing property required of CDFs for the QF and QTT construction and is first-order equivalent for the estimated DFs and associated endpoints under standard rearrangement arguments; see \citet{chernozhukov-val-galichon-2010} and \citet[2252]{chernozhukov-val-melly-2013}. The following states the validity of confidence bands on DFs and DTT.
\begin{result}[Coverage of Uniform Confidence Bands on DFs and DTT]\label{Res:DF_Bands_valid}
Under the conditions of \Cref{Thm:BootValid},
\begin{align*}
    &\liminf_{N\rightarrow\infty} \Prob\Big( \big\{\F_{Y_t(1) \mid G_g=1}(y) \in \widehat{I}_{gt}^{(1)}(y) \big\} \bigcap \big\{\F_{Y_t(0) \mid G_g=1}(y) \in \widehat{I}_{gt}^{(0)}(y) \big\} \ \forall y \in \Y \Big) \geq p, \text{ and}\\
    &\liminf_{N\rightarrow\infty} \Prob\Big( \DTT_{gt}(y) \in \widehat{I}_{gt}^{\DTT}(y) \ \forall y \in \Y \Big) \geq p
\end{align*}by Lemma SA.1(b) in the Supplemental material of \citet{chernozhukov-val-melly-2013}.
\end{result}

This paper adopts the band-inversion approach of \citet{chernozhukov2019generic} to conduct uniform inference for QFs and QTT. A crucial ingredient for using the authors' results is valid confidence bands on DFs (\Cref{Res:DF_Bands_valid}). For completeness, theoretical results from \citet{chernozhukov2019generic} are adapted to QFs and QTT. The confidence bands on the QFs are given by $\widehat{I}_{gt}^{(1)\leftarrow}(\tau) = \big[\widehat{U}_{gt}^{(1)\leftarrow}(\tau), \ \widehat{L}_{gt}^{(1)\leftarrow}(\tau)\big]$ and $\widehat{I}_{gt}^{(0)\leftarrow}(\tau) = \big[\widehat{U}_{gt}^{(0)\leftarrow}(\tau), \ \widehat{L}_{gt}^{(0)\leftarrow}(\tau)\big],\ \tau \in (0,1)$, respectively, where $\widehat{L}^{\leftarrow}$ and $\widehat{U}^{\leftarrow}$ denote the left inverses of $\widehat{L}$ and $\widehat{U}$ --- see \Cref{Def:QF}. The following definition is essential in deriving the confidence bands of $\QTT_{gt}$.

\begin{definition}[Minkowski Difference] \label{Def:MinkowDiff}
The Minkowski difference between two subsets $I$
and $J$ of a vector space is $I\ominus J := \{i - j : i \in I, j\in J\}$. If $I$ and $J$ are
intervals $[i_1, i_2]$ and $[j_1, j_2]$, then
$I\ominus J = [i_1, i_2] \ominus [j_1, j_2] = [i_1 - j_2, i_2 - j_1]$.
\end{definition}

Following \citet{chernozhukov2019generic}, the confidence bands on $\QTT_{gt}$ are given by $\widehat{I}_{gt}^{\QTT\leftarrow}(\tau) = \widehat{I}_{gt}^{(1)\leftarrow}(\tau) \ominus \widehat{I}_{gt}^{(0)\leftarrow}(\tau)=: \big[ \widehat{L}_{gt}^{\QTT}(\tau), \ \widehat{U}_{gt}^{\QTT}(\tau) \big] $. The following states the validity of the band-inverted confidence bands on QFs and QTT \`a la \citet{chernozhukov2019generic}.
\begin{result}[Coverage of Uniform Confidence Bands on QFs and QTT]\label{Res:Qf_Qtt_Bands}
Under the conditions of \Cref{Thm:BootValid},
\begin{align*}
    \liminf_{N\rightarrow\infty} \Prob\Bigg(
    \Big\{&
    \F_{Y_t(1) \mid G_g=1}^{\leftarrow}(\tau)
    \in \widehat{I}_{gt}^{(1)\leftarrow}(\tau), \\
    &\F_{Y_t(0) \mid G_g=1}^{\leftarrow}(\tau)
    \in \widehat{I}_{gt}^{(0)\leftarrow}(\tau), \\
    &\QTT_{gt}(\tau)
    \in \widehat{I}_{gt}^{\QTT\leftarrow}(\tau),
    \ \forall \tau \in (0,1)
    \Big\}
    \Bigg) \geq p.
\end{align*}
This follows by \Cref{Res:DF_Bands_valid} and Theorem 2 of \citet{chernozhukov2019generic}.
\end{result}

\section{Simulations}\label{App:Sect_Sim}
This section conducts two sets of simulation studies to examine the small sample performance of the distributional DiD estimator and the functional over-identifying restrictions tests.
\subsection{Estimation and Inference}\label{Sect:Sim_Est_Inference}

This simulation study is designed to assess the finite-sample performance of the proposed distributional DiD procedure. Three data-generating processes are considered. The first is an empirical design, labelled DGP0, which is calibrated directly to the empirical application. The remaining two designs are stylised benchmark models, one discrete and one continuous.

Under the benchmark designs, the latent outcome is generated as
\[
\dot{Y}_i=\alpha + D_i\beta + t_i\gamma + D_it_i\delta + U_i,
\]
where $D_i \overset{i.i.d.}{\sim} \mathrm{Ber}(0.5)$, $t_i=\indicator{i>N/2}$, $U_i$ is an i.i.d.\ disturbance, $\alpha=0.1$, $\beta=0.2$, $\gamma=-0.1$, and $\delta=0$. The disturbance $U_i$ is taken to follow either the standard normal distribution or an asymmetric Laplace distribution. The observed outcome is then defined by
\[
Y_i=
\begin{cases}
\max\{\lceil \dot{Y}_i+1\rceil,0\}, & \text{DGP1},\\[0.3em]
\dot{Y}_i, & \text{DGP2}.
\end{cases}
\]
Hence, DGP1 yields a discrete and censored outcome, whereas DGP2 yields a continuous outcome.

DGP0 is calibrated to the empirical distribution of the outcome variable in \Cref{Sect:Empirical_Appl}. Let $\widehat{\F}_{00}$, $\widehat{\F}_{01}$, and $\widehat{\F}_{10}$ denote the empirical CDFs of the outcome for untreated blocks in the pre-treatment and post-treatment periods, and for treated blocks in the pre-treatment period, respectively. Under the null of no treatment effect, the treated post-treatment counterfactual distribution is constructed using the additive DiD relation under the identity link: 
\[
\widehat{\F}_{11}^{\,0}(y)
=
\widehat{\F}_{10}(y)+\widehat{\F}_{01}(y)-\widehat{\F}_{00}(y).
\]
$\widehat{\F}_{11}^{\,0}(y)$ is truncated to lie in $[0,1]$, monotonised over the empirical outcome grid, and normalised so that its upper endpoint equals one. Simulation draws are then generated from the empirical quantile functions associated with $\widehat{\F}_{00}$, $\widehat{\F}_{01}$, $\widehat{\F}_{10}$, and the projected version of $\widehat{\F}_{11}^{\,0}$. In this way, DGP0 preserves the truncated empirical support and the principal distributional features of the application data while imposing the null hypothesis $\DTT(y)=0$ for all $y\in\Y$.

The designs are calibrated so that there is no treatment effect in the underlying data-generating process, and hence provide a finite-sample assessment of estimation error, confidence-band coverage, and sensitivity to the working CDF. DGP0 is useful for assessing performance under an empirically calibrated discrete design, while DGP1 and DGP2 provide complementary evidence under stylised discrete and continuous settings. In the latter designs, the comparison across normal, uniform, and identity links should be interpreted as a robustness exercise for the working-CDF specification rather than as imposing exact size validity for every link by construction.

\Cref{Tab:DGP_Empirical,Tab:DGP_Discrete,Tab:DGP_Continuous} report simulation results for DGP0, DGP1, and DGP2, respectively. For DGP1 and DGP2, each panel is characterised by the distribution of $U$, namely the standard normal or the Asymmetric Laplace Distribution $ALD(0,1,\kappa)$ with $\kappa \in \{0.1,0.25,0.5\}$. The inclusion of the ALD cases is useful in examining the performance of the procedure when $U$ is asymmetric. For each table, sample sizes $N \in \{200,400,600,800,1000\}$ are considered. For DGP2, the evaluation grid is formed from the sample quantiles indexed by $\{\tau_\ell\}_{\ell=1}^{101}$, whereas for the discrete designs it is formed from the realised support after trimming the upper tail as implemented in the simulation code. Results are summarised by $\mathcal{L}_2(\widehat{\DTT})$ where $ \displaystyle \mathcal{L}_p(\widehat{\F}):= \Big(\frac{1}{K}\sum_{k=1}^K |\widehat{\F}(y_k) - F(y_k)|^p\Big)^{1/p} $, where $K$ denotes the number of grid-points used in estimating the function $F(\cdot)$, and the empirical rejection rate of the 10\% uniform confidence band for $\DTT$. For each design, $499$ non-parametric bootstrap samples are used within each of the $500$ Monte Carlo replications.

\begin{table}[!htbp]
\caption{\label{Tab:DGP_Empirical} DGP0: empirically calibrated discrete design}
\centering
\begin{tabular}{@{}lcccccccc@{}}
\toprule
& \multicolumn{2}{c}{ DiD, $\Phi$ -  Normal} & \multicolumn{2}{c}{DiD, $\Phi$ -  Unif.} & \multicolumn{2}{c}{DiD, $\Phi$ -  Identity} & \multicolumn{2}{c}{ CiC } \\ \cmidrule(lr){2-3} \cmidrule(lr){4-5} \cmidrule(lr){6-7} \cmidrule(lr){8-9}
$N$ & $\mathcal{L}_2(\widehat{\DTT})$ & 10\% Rej. & $\mathcal{L}_2(\widehat{\DTT})$ & 10\% Rej. & $\mathcal{L}_2(\widehat{\DTT})$ & 10\% Rej. & $\mathcal{L}_2(\widehat{\DTT})$ & 10\% Rej. \\ \midrule
200 & 0.108 & 0.144 & 0.082 & 0.156 & 0.084 & 0.120 & 0.082 & 0.168 \\
400 & 0.066 & 0.076 & 0.059 & 0.108 & 0.059 & 0.086 & 0.069 & 0.190 \\
600 & 0.053 & 0.094 & 0.049 & 0.110 & 0.049 & 0.094 & 0.064 & 0.222 \\
800 & 0.044 & 0.080 & 0.042 & 0.108 & 0.042 & 0.100 & 0.060 & 0.230 \\
1000 & 0.037 & 0.088 & 0.036 & 0.088 & 0.036 & 0.086 & 0.061 & 0.234 \\
\bottomrule
\end{tabular}
\end{table}

\begin{table}[!htbp]
\caption{\label{Tab:DGP_Discrete} DGP1: stylised discrete design}
\centering
\begin{tabular}{@{}lcccccccc@{}}
\toprule
& \multicolumn{2}{c}{ DiD, $\Phi$ -  Normal} & \multicolumn{2}{c}{DiD, $\Phi$ -  Unif.} & \multicolumn{2}{c}{DiD, $\Phi$ -  Identity} & \multicolumn{2}{c}{ CiC } \\ \cmidrule(lr){2-3} \cmidrule(lr){4-5} \cmidrule(lr){6-7} \cmidrule(lr){8-9}
$N$ & $\mathcal{L}_2(\widehat{\DTT})$ & 10\% Rej. & $\mathcal{L}_2(\widehat{\DTT})$ & 10\% Rej. & $\mathcal{L}_2(\widehat{\DTT})$ & 10\% Rej. & $\mathcal{L}_2(\widehat{\DTT})$ & 10\% Rej. \\ \midrule
& \multicolumn{8}{l}{$U\sim \mathrm{ALD}(0,1,0.10)$} \\ \cmidrule{2-9}
200 & 0.124 & 0.100 & 0.105 & 0.132 & 0.106 & 0.094 & 0.107 & 0.056 \\
400 & 0.078 & 0.066 & 0.075 & 0.156 & 0.075 & 0.112 & 0.077 & 0.072 \\
600 & 0.064 & 0.082 & 0.063 & 0.158 & 0.063 & 0.116 & 0.067 & 0.094 \\
800 & 0.054 & 0.090 & 0.054 & 0.138 & 0.054 & 0.102 & 0.058 & 0.086 \\
1000 & 0.048 & 0.078 & 0.047 & 0.114 & 0.047 & 0.088 & 0.053 & 0.106 \\
\midrule
& \multicolumn{8}{l}{$U\sim \mathrm{ALD}(0,1,0.25)$} \\ \cmidrule{2-9}
200 & 0.111 & 0.088 & 0.104 & 0.162 & 0.105 & 0.094 & 0.113 & 0.086 \\
400 & 0.075 & 0.090 & 0.075 & 0.140 & 0.075 & 0.080 & 0.087 & 0.112 \\
600 & 0.063 & 0.108 & 0.063 & 0.152 & 0.063 & 0.118 & 0.080 & 0.180 \\
800 & 0.053 & 0.096 & 0.053 & 0.130 & 0.053 & 0.104 & 0.076 & 0.224 \\
1000 & 0.047 & 0.094 & 0.047 & 0.102 & 0.047 & 0.088 & 0.074 & 0.260 \\
\midrule
& \multicolumn{8}{l}{$U\sim \mathrm{ALD}(0,1,0.50)$} \\ \cmidrule{2-9}
200 & 0.112 & 0.100 & 0.104 & 0.132 & 0.105 & 0.084 & 0.124 & 0.132 \\
400 & 0.076 & 0.090 & 0.075 & 0.138 & 0.075 & 0.102 & 0.104 & 0.186 \\
600 & 0.063 & 0.100 & 0.062 & 0.130 & 0.062 & 0.106 & 0.101 & 0.226 \\
800 & 0.054 & 0.094 & 0.053 & 0.108 & 0.053 & 0.094 & 0.100 & 0.324 \\
1000 & 0.048 & 0.084 & 0.047 & 0.106 & 0.047 & 0.092 & 0.099 & 0.366 \\
\midrule
& \multicolumn{8}{l}{$U\sim \mathcal{N}(0,1)$} \\ \cmidrule{2-9}
200 & 0.118 & 0.106 & 0.100 & 0.158 & 0.102 & 0.124 & 0.210 & 0.372 \\
400 & 0.086 & 0.098 & 0.071 & 0.106 & 0.071 & 0.098 & 0.202 & 0.452 \\
600 & 0.057 & 0.094 & 0.056 & 0.136 & 0.056 & 0.110 & 0.200 & 0.480 \\
800 & 0.050 & 0.094 & 0.050 & 0.104 & 0.050 & 0.088 & 0.202 & 0.542 \\
1000 & 0.043 & 0.088 & 0.043 & 0.102 & 0.043 & 0.084 & 0.203 & 0.552 \\
\bottomrule
\end{tabular}
\end{table}

\begin{table}[!htbp]
\caption{\label{Tab:DGP_Continuous} DGP2: stylised continuous design}
\centering
\begin{tabular}{@{}lcccccccc@{}}
\toprule
& \multicolumn{2}{c}{ DiD, $\Phi$ -  Normal} & \multicolumn{2}{c}{DiD, $\Phi$ -  Unif.} & \multicolumn{2}{c}{DiD, $\Phi$ -  Identity} & \multicolumn{2}{c}{ CiC } \\ \cmidrule(lr){2-3} \cmidrule(lr){4-5} \cmidrule(lr){6-7} \cmidrule(lr){8-9}
$N$ & $\mathcal{L}_2(\widehat{\DTT})$ & 10\% Rej. & $\mathcal{L}_2(\widehat{\DTT})$ & 10\% Rej. & $\mathcal{L}_2(\widehat{\DTT})$ & 10\% Rej. & $\mathcal{L}_2(\widehat{\DTT})$ & 10\% Rej. \\ \midrule
& \multicolumn{8}{l}{$U\sim \mathrm{ALD}(0,1,0.10)$} \\ \cmidrule{2-9}
200 & 0.118 & 0.104 & 0.109 & 0.144 & 0.110 & 0.104 & 0.110 & 0.036 \\
400 & 0.079 & 0.044 & 0.078 & 0.156 & 0.078 & 0.092 & 0.079 & 0.058 \\
600 & 0.065 & 0.076 & 0.065 & 0.180 & 0.065 & 0.112 & 0.066 & 0.066 \\
800 & 0.056 & 0.092 & 0.056 & 0.164 & 0.056 & 0.106 & 0.056 & 0.070 \\
1000 & 0.050 & 0.080 & 0.049 & 0.130 & 0.049 & 0.090 & 0.050 & 0.060 \\
\midrule
& \multicolumn{8}{l}{$U\sim \mathrm{ALD}(0,1,0.25)$} \\ \cmidrule{2-9}
200 & 0.118 & 0.100 & 0.109 & 0.146 & 0.110 & 0.108 & 0.110 & 0.040 \\
400 & 0.079 & 0.048 & 0.078 & 0.150 & 0.078 & 0.090 & 0.079 & 0.058 \\
600 & 0.065 & 0.076 & 0.065 & 0.174 & 0.065 & 0.102 & 0.066 & 0.064 \\
800 & 0.056 & 0.090 & 0.056 & 0.160 & 0.056 & 0.102 & 0.056 & 0.074 \\
1000 & 0.049 & 0.074 & 0.049 & 0.114 & 0.049 & 0.074 & 0.050 & 0.060 \\
\midrule
& \multicolumn{8}{l}{$U\sim \mathrm{ALD}(0,1,0.50)$} \\ \cmidrule{2-9}
200 & 0.119 & 0.114 & 0.109 & 0.138 & 0.109 & 0.100 & 0.109 & 0.046 \\
400 & 0.079 & 0.046 & 0.078 & 0.156 & 0.078 & 0.088 & 0.079 & 0.060 \\
600 & 0.066 & 0.064 & 0.065 & 0.176 & 0.065 & 0.112 & 0.065 & 0.062 \\
800 & 0.056 & 0.090 & 0.056 & 0.166 & 0.056 & 0.112 & 0.056 & 0.080 \\
1000 & 0.050 & 0.076 & 0.049 & 0.126 & 0.049 & 0.078 & 0.049 & 0.062 \\
\midrule
& \multicolumn{8}{l}{$U\sim \mathcal{N}(0,1)$} \\ \cmidrule{2-9}
200 & 0.120 & 0.082 & 0.110 & 0.168 & 0.111 & 0.128 & 0.111 & 0.048 \\
400 & 0.080 & 0.058 & 0.079 & 0.158 & 0.079 & 0.124 & 0.078 & 0.064 \\
600 & 0.064 & 0.060 & 0.063 & 0.156 & 0.063 & 0.110 & 0.064 & 0.052 \\
800 & 0.056 & 0.090 & 0.056 & 0.156 & 0.056 & 0.116 & 0.056 & 0.074 \\
1000 & 0.050 & 0.074 & 0.049 & 0.108 & 0.049 & 0.084 & 0.050 & 0.058 \\
\bottomrule
\end{tabular}
\end{table}

Several patterns emerge from \Cref{Tab:DGP_Empirical,Tab:DGP_Discrete,Tab:DGP_Continuous}. First, the proposed distributional DiD estimator performs well across all three designs, with $\mathcal{L}_2(\widehat{\DTT})$ generally declining as the sample size increases. This pattern is evident in the empirically calibrated design as well as in the stylised discrete and continuous designs, which is consistent with satisfactory finite-sample behaviour.

Secondly, the choice of working CDF matters for finite-sample performance, but the differences are moderate between the normal and identity links. In contrast, the uniform link tends to exhibit somewhat higher rejection frequencies, especially in the stylised designs. The identity link performs competitively throughout and often tracks the normal link closely. In the empirically calibrated design, all three versions of the proposed estimator deliver similar bias magnitudes, with the uniform link showing slightly greater over-rejection in some cases.

Finally, the competing Changes-in-Changes (CiC) estimator of \citet{athey-imbens-2006} performs comparatively less well in the discrete designs, particularly in DGP0 and DGP1, where rejection frequencies are often noticeably larger and the $\mathcal{L}_2$ error does not improve relative to the proposed procedure. In the continuous design, by contrast, CiC is more competitive, as might be expected given its natural suitability for continuous outcomes. Overall, the results suggest that the proposed distributional DiD estimator is robust across a range of designs, while the identity and normal links provide the most stable finite-sample performance.

\subsection{Over-identifying Restrictions Testing}\label{Sect:Sim_Spec_Test}

The data-generating process for the over-identifying restrictions test is calibrated directly to the empirical application and is therefore conditional on the observed data. Let $\widehat{\F}_{00}$ and $\widehat{\F}_{01}$ denote the empirical CDFs of the outcome for blocks without police presence in the pre-treatment and post-treatment periods, respectively, where the pre-treatment period pools months April through June and the post-treatment period pools months August through December. Likewise, let $\widehat{\F}_{10}$ and $\widehat{\F}_{11}$ denote the corresponding empirical CDFs for blocks with increased police presence. The support used in the simulations is the empirical support of the outcome, truncated above at $1.0$, and the associated quantile functions are obtained from these empirical CDFs.

Repeated cross-sections are then simulated for three groups, $G\in\{1,2,3\}$, over one pre-treatment period, $t=0$, and one post-treatment period, $t=1$. The treated group, $G=1$, is generated from $(\widehat{\F}_{10},\widehat{\F}_{11})$. The first control group, $G=2$, is generated from $(\widehat{\F}_{00},\widehat{\F}_{01})$ and therefore satisfies the null of valid over-identifying restrictions. The second control group, $G=3$, is used to introduce a controlled violation of the null. Its pre-treatment distribution is set equal to that of the valid control group, namely $\widehat{\F}_{30}=\widehat{\F}_{00}$, whereas its post-treatment distribution is defined on the empirical grid $\{y_\ell\}_{\ell=1}^L$ by
\[
\widehat{\F}_{31}^{(\eta)}(y_\ell)
=
(1-\eta)\widehat{\F}_{01}(y_\ell) + \eta \frac{\ell}{L},
\qquad \eta\in[0,1].
\]
Hence, when $\eta=0$, one has $\widehat{\F}_{31}^{(0)}=\widehat{\F}_{01}$, so that both control groups satisfy the same untreated transition and the null holds exactly. As $\eta$ increases, the post-treatment distribution of group $G=3$ is moved progressively away from the empirical benchmark $\widehat{\F}_{01}$, thereby generating increasingly pronounced violations of the functional over-identifying restrictions.

\begin{figure}[!htbp]
\centering 
\caption{Power Curves of the functional over-identifying restrictions tests}
\begin{subfigure}{0.45\textwidth}
\centering
\includegraphics[width=1\textwidth]{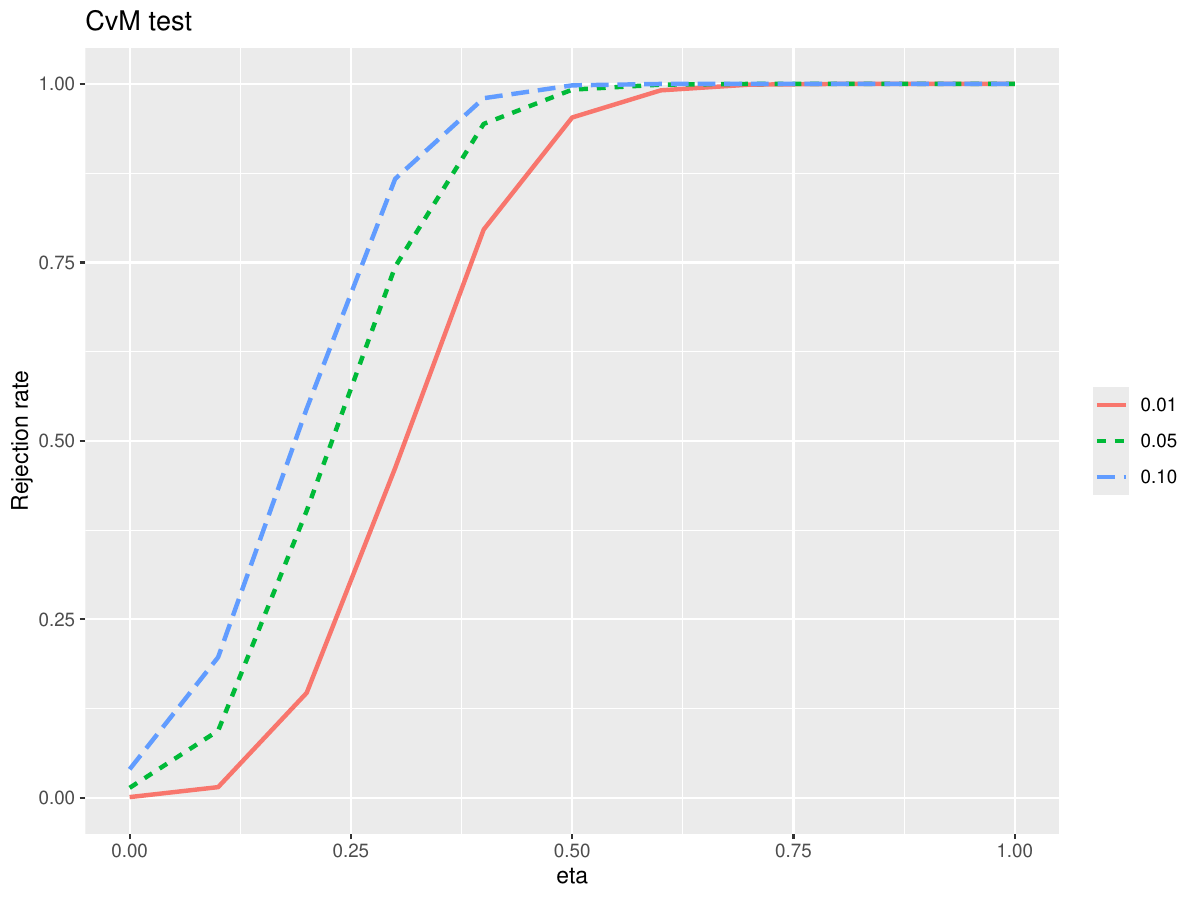}
\caption{ }
\end{subfigure}
\begin{subfigure}{0.45\textwidth}
\centering
\includegraphics[width=1\textwidth]{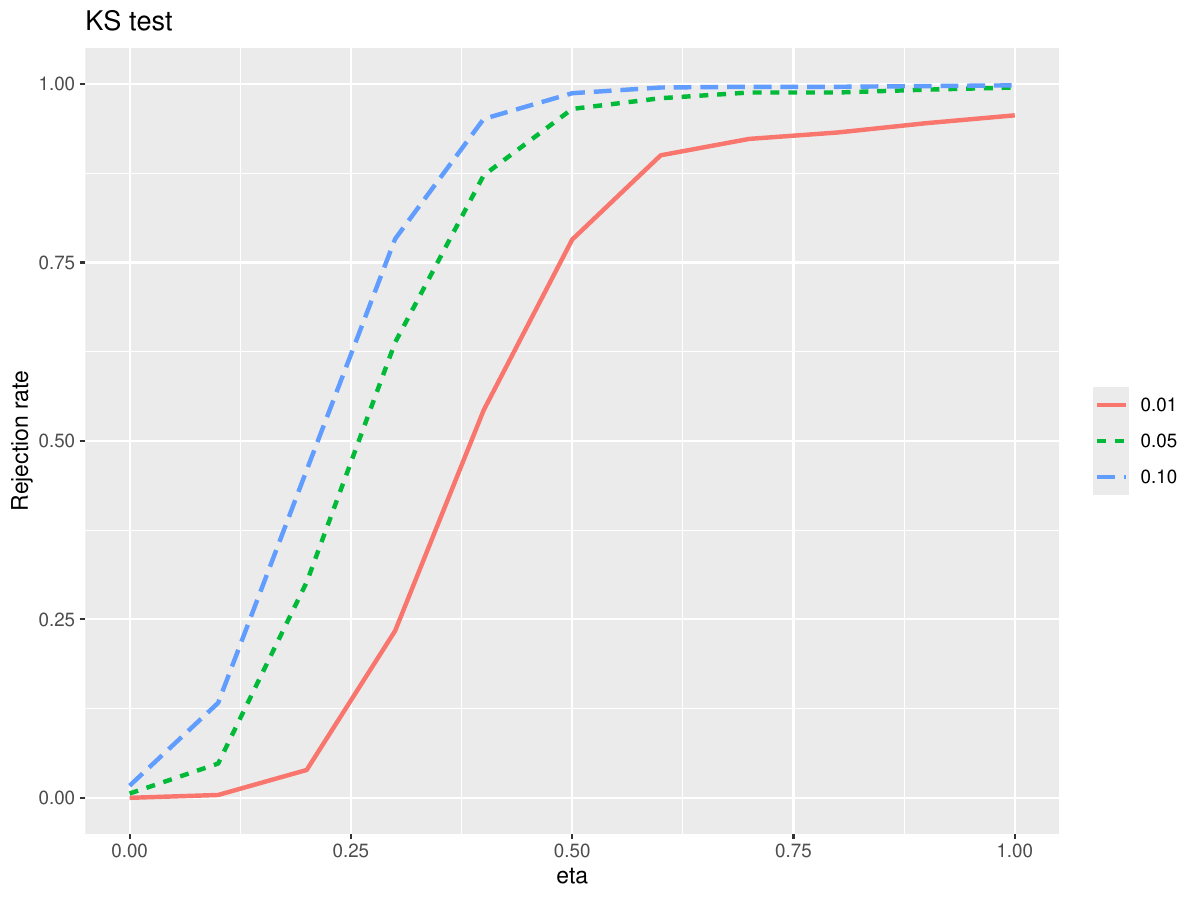}
\caption{ }
\end{subfigure}

{\footnotesize
\textit{Notes:} Both the CvM and KS tests are conducted at three nominal levels: $0.01$, $0.05$, and $0.10$. 999 empirical bootstrap replications are used for each of the 1000 Monte Carlo replications. The reported figure uses the standard normal link function for both tests.
}\label{Fig:powercurves}
\end{figure}

\Cref{Fig:powercurves} shows that both the CvM and KS tests have empirical rejection probabilities close to the nominal level when $\eta=0$, which is consistent with satisfactory size control under the null. As $\eta$ increases, the rejection probability rises for both tests, indicating non-trivial power against departures from the over-identifying restrictions. The increase is gradual for small values of $\eta$ and becomes more pronounced for larger deviations from the null. Overall, both statistics display the expected monotonic power pattern.

\newpage
\printbibliography
\end{refsection}
\end{document}